%% file: arxiv.tex
\documentclass[acmsmall,screen]{acmart}

\setcopyright{rightsretained}
\copyrightyear{2025}
\acmYear{2025}
\acmDOI{XXXXXXX.XXXXXXX}

\input{macros}

\begin{document}

\title[Black-box Context-free Grammar Inference for Readable \& Natural Grammars]{Black-box Context-free Grammar Inference for Readable \& Natural Grammars}
\author{Mohammad Rifat Arefin}
\affiliation{%
  \institution{University of Texas at Arlington}
  \city{Arlington}
  \country{USA}
}
\email{mxa7262@mavs.uta.edu}

\author{Shanto Rahman}
\affiliation{%
  \institution{University of Texas at Austin}
  \city{Austin}
  \country{USA}
}
\email{shanto.rahman@utexas.edu}

\author{Christoph Csallner}
\affiliation{%
  \institution{University of Texas at Arlington}
  \city{Arlington}
  \country{USA}
}
\email{csallner@uta.edu}

\begin{abstract}

Black-box context-free grammar inference is crucial for program analysis, reverse engineering, and security, yet existing tools such as \arvada{}, \treevada{}, and \kedavra{} struggle with scalability, readability, and accuracy on large, complex languages. We present \Tool{}, a novel LLM-guided grammar inference framework that extends \treevada{}’s parse tree recovery with three key innovations: bracket-guided bubble exploration, LLM-driven bubble generation and non-terminal labeling, and hierarchical delta debugging (HDD) for systematic tree simplification. Bracket-guided exploration leverages syntactic cues such as parentheses to propose well-structured grammar fragments, while LLM guidance produces meaningful non-terminal names and selects more promising merges. Finally, HDD incrementally reduces unnecessary rules, which makes the grammars both compact and interpretable.

In our experiments, we evaluate \Tool{} on a comprehensive benchmark suite ranging from small languages to larger ones such as \textit{lua}, \textit{c}, and \textit{mysql}. Our results show that \Tool{} consistently outperforms existing baselines in terms of \fscore{}. On average, \Tool{} achieves an \fscore{} of %
57\% for large languages, which is 25pp (percentage points) higher than the best-performing baseline, \treevada{} (32\%). In the case of interpretability, our generated grammars perform significantly better than those produced by existing approaches. Leveraging LLM-based node renaming and bubble exploration, \Tool{} produces rules with meaningful non-terminal names and compact structures that align more closely with human intuition. This allows developers and researchers to achieve higher accuracy with grammars that are easier to inspect, verify, and reason about in terms of their structure and semantics.

\end{abstract}

\maketitle

\npdecimalsign{.}

\section{Introduction}

Black-box context-free grammar inference from sample programs currently suffers from two key limitations. First, it is not clear if existing approaches scale to larger and more realistic languages. Second, existing approaches produce grammars that are difficult to consume for a human---even when these inferred grammars score very highly on traditional grammar inference metrics such as precision, recall, and F1 score.

Unclear scalability to larger programs and difficulties for human consumption are key roadblocks preventing black-box context-free grammar inference from benefiting software engineers and security analysts. While good performance on small languages is important, many languages commonly used in practice are large and complex. Besides the use-case of fuzzing and security analysis, another key use-case of grammar inference is reverse engineering rare and under-documented commercial languages, so readability is also very important.

Black-box grammar inference is a fundamentally hard problem, as we only have access to a black-box parser of an unknown language \textit{L} and a few sample \textit{L} programs. The sample programs may not cover all \textit{L} language features and will almost certainly not cover all possible combinations of these language features.

Recent advances in black-box context-free grammar inference include \arvada~\cite{kulkarni2021learning}, \treevada~\cite{arefin2024fast}, and \kedavra~\cite{li2024incremental}. \arvada{} addresses the problem by inferring grammars directly from complete input strings while exploring multiple generalization paths. However, this method suffers from low accuracy due to frequent over-generalization in complex grammars, slow processing speeds, and poor grammar readability. The more recent \treevada{} and \kedavra{} improve upon \arvada{} and are the most closely related approaches to our work. Despite these improvements, it is unclear how \treevada{} and \kedavra{} scale to larger languages and their inferred grammars are challenging for human consumption due to a larger number of inferred rules and their extensive use of non-descriptive node names.

Large language models (LLMs) have made significant advancements and show remarkable ability in many software engineering tasks. These models demonstrate robust performance in code understanding, enabling effective code summarization, refactoring, verification, and auto-completion. However, theoretical~\cite{Hahnlimitation} and empirical studies~\cite{bhattamishra2020ability, ebrahimi-etal-2020-self, delétang2023neuralnetworkschomskyhierarchy, zhang2024transformer} suggest that LLMs have limited ability to learn formal grammars (such as context-free grammars) from sample programs. This raises the question of whether these models merely mimic frequent patterns from their training data. Studies through probing techniques~\cite{ast-probe, whatdotheycapture} indicate that these models possess some structural information. These models struggle however in code syntax understanding tasks with large codebase~\cite{shen-etal-2022-benchmarking}.

In this paper we present \Tool{}, an automated grammar inference technique. The process begins with bubble exploration, where bracket-guided rules are used to identify initial structures. Next, \llm{} actively generates candidate bubbles, which are subsequently filtered using \treevada{} heuristics. Each validated bubble is incorporated into a candidate parse tree and checked against the oracle/black-box parser to gauge its likely correctness. Once checked, we apply hierarchical delta debugging to simplify the candidate tree. Finally, lexical inference is applied to enrich the grammar and increase diversity in the inferred rules. When comparing the tools on different-sized languages, \Tool{} demonstrates higher F1 scores than the competitors on large languages. \Tool{}-inferred grammars are also easier to read due to their descriptive non-terminal labels.

To summarize, the paper makes the following major contributions.
\begin{itemize}
    \item The paper presents \toolName{}, the first approach that scales black-box context-free grammar inference to larger languages. At the same time, \toolName{}-produced grammars are also more human readable.
    \item The paper compares \toolName{} empirically with its closest competitors (Barebone LLM, \treevada{}, and \kedavra{}), where for larger languages \Tool{} improves the best performing baseline, \treevada{}, by 25pp (percentage points).
    \item The \toolName{} source code, scripts, evaluation parameters, and training data are open-source and publicly available (\textbf{\url{https://github.com/rifatarefin/natgi}}).
\end{itemize}

\section{Background}

Context-free grammars are important because they form the foundation of programming language design, compiler construction, and parsing tools, ensuring that source code and structured data can be rigorously analyzed and processed. For grammar inference to be truly useful, the resulting grammars must be human-understandable. Readable grammars enable developers and researchers to inspect rules, validate whether they capture the intended language, and detect issues such as over-generalization or under-generalization.

\textbf{\textsc{\arvada{}}:}
\arvada{}~\cite{kulkarni2021learning} is the first tool to introduce recovering parse trees as a means to infer the target grammar. \arvada{}\ starts recovering parse trees from an initial flat state where all language tokens originate from the root node. Over the course of the algorithm, new hierarchies are added to the trees. \arvada{} uses the term \textit{bubbles} to denote syntactic units by grouping sibling nodes during parse tree reconstruction. 

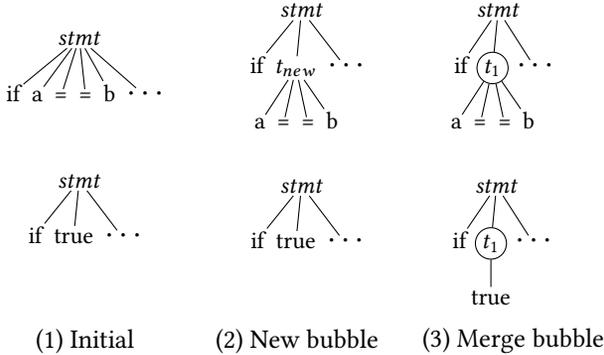
\begin{figure}[htbp]
\centering
\begin{tikzpicture}[>=Stealth, every node/.style={inner sep=1pt}]

\node (LeftTop) {
  \begin{forest}
    for tree={l sep=6pt, s sep=2pt, font=\small}
    [\textit{stmt}
      [if]
      [a]
      [{=}]
      [{=}]
      [b]
      [{\(\cdots\)}, font=\Large] %
    ]
  \end{forest}
};

\node (LeftBot) [below=.8cm of LeftTop] {
  \begin{forest}
    for tree={l sep=6pt, s sep=2pt, font=\small}
    [\textit{stmt}
      [if]
      [true]
      [{\(\cdots\)}, font=\Large]
    ]
  \end{forest}
};
\node (a) [below= of LeftBot.south] {(1) Initial};
\node (MidTop) [right= of LeftTop] {
  \begin{forest}
    for tree={l sep=6pt, s sep=2pt, font=\small}
    [\textit{stmt}
      [if]
      [$t_{new}$[a]
      [{=}]
      [{=}]
      [b]]
      [{\(\cdots\)}, font=\Large] %
    ]
  \end{forest}
};

\node (MidBot) [below=.5cm of MidTop] {
  \begin{forest}
    for tree={l sep=6pt, s sep=2pt, font=\small}
    [\textit{stmt}
      [if]
      [true]
      [{\(\cdots\)}, font=\Large]
    ]
  \end{forest}
};
\node (b) [below= of MidBot.south, right=1cm of a.east] {(2) New bubble};
\node (RightTop) [right= of MidTop] {
  \begin{forest}
    for tree={l sep=6pt, s sep=2pt, font=\small}
    [\textit{stmt}
      [if]
      [$t_1$, draw, circle[a]
      [{=}]
      [{=}]
      [b]]
      [{\(\cdots\)}, font=\Large] %
    ]
  \end{forest}
};

\node (RightBot) [below=.5cm of RightTop] {
  \begin{forest}
    for tree={l sep=6pt, s sep=2pt, font=\small}
    [\textit{stmt}
      [if]
      [$t_1$, draw, circle[true]]
      [{\(\cdots\)}, font=\Large]
    ]
  \end{forest}
};
\node (c) [below= of RightBot.south, right=.5cm of b.east] {(3) Merge bubble};
\end{tikzpicture}
\caption{Incremental parse tree construction: (1) parse tree segments at beginning, (2) node sequence $a==b$ bubbled-up under new node $t_{new}$, (3) $t_{new}$  and $true$ merged as $t_1$.}
\label{fig:bubble}
\Description{Pair of partial parse trees shown at three processing stages from start (left) to state after merge (right).}
\end{figure}

Figure~\ref{fig:bubble} explains how the \textit{bubbling} mechanism works. Two segments of parse tree in Figure~\ref{fig:bubble}.1 could be generalized further. Sibling sequence \texttt{a == b} is \textit{bubbled} up and a new internal node $t_{new}$ is introduced in Figure~\ref{fig:bubble}.2. Next, \arvada{} checks if $t_{new}$ can be replaced with any existing tree nodes and vice-versa. Here, replacing the tree node, \texttt{true}, with the subtree $t_{new}$ yields syntactically correct strings. The same holds in the reverse: substituting the subtree, $t_{new}$, with the node \texttt{true} also yields valid strings. 

This identical effect of swapping the pair of nodes indicates that the terminal sequence \texttt{a == b} and \texttt{true} might be alternations of the same grammar rule. At this point, $t_{new}$ and \texttt{true} are merged and relabeled as $t_1$, shown in Figure~\ref{fig:bubble}(c). The tree-induced grammar learns a new rule $t_1 \to a==b\; | \;true$. As soon as a new hierarchy is added to a tree, the tree-induced grammar learns a new rule. Tree nodes with the same label, but different children represent alternations in a production rule.

\arvada{}'s successor tool \treevada{} starts a from pre-structured trees instead of flat ones. \treevada{} uses \textit{brackets} a prior knowledge to pre-structure the parse trees, according to common programming languages~\cite{van2019lightweight}. We follow the \textit{Parse Tree Recovery} framework of \arvada{}/\treevada{} for grammar inference. \arvada{}/\treevada{} heuristically finds bubbles and uses \textsc{CheckBubble} to validate them before accepting them into the trees. \textsc{CheckBubble} operates in the following way:

\textbf{\textsc{CheckBubble}:} This procedure samples candidate strings by alternating the positions of a pair of nodes within the parse trees. If all candidate strings are passed by the oracle, \textsc{CheckBubble} returns true and the new structure is accepted. To avoid computational overhead, the number of sampled strings is capped at 100. However, this limitation may result in incorrect structures being adopted in tree reconstruction.

We rely on an LLM to form intermediate \textit{bubbles} at multiple steps, resulting in iterative reconstruction of the full parse trees. At each step, tree mutations are validated by \textsc{CheckBubble} procedure. Once the algorithm terminates, the grammar is read off from the final parse trees. Each parent and its children represent a production rule in the grammar, with the parent corresponding to the left-hand side non-terminal and the children forming the right-hand side sequence.

\section{Overview and Design}
\label{sec:methodology}

To infer a context-free grammar from sample programs via a black-box oracle, \Tool{} extends the earlier \treevada{} approach with three novel components. (1)~First, \Tool{}'s bracket-guided bubble exploration develops further \treevada{}'s ideas of bracket-induced tree pre-structuring, by treating each bracketed node sequence as a candidate bubble for potential merges even in places where the node sequence does not appear in brackets. (2)~Second, \Tool{}'s LLM-guided bubble exploration leverages LLM's wide background knowledge and pattern matching abilities to identify candidate bubbles. (3)~Finally, Tool{}'s hierarchical delta debugging (HDD) minimizes its parse trees to identify subtrees the parser accepts as programs and thereby generalize the inferred language.

\begin{figure}[h!t]
\centering
\includegraphics[width=\linewidth]{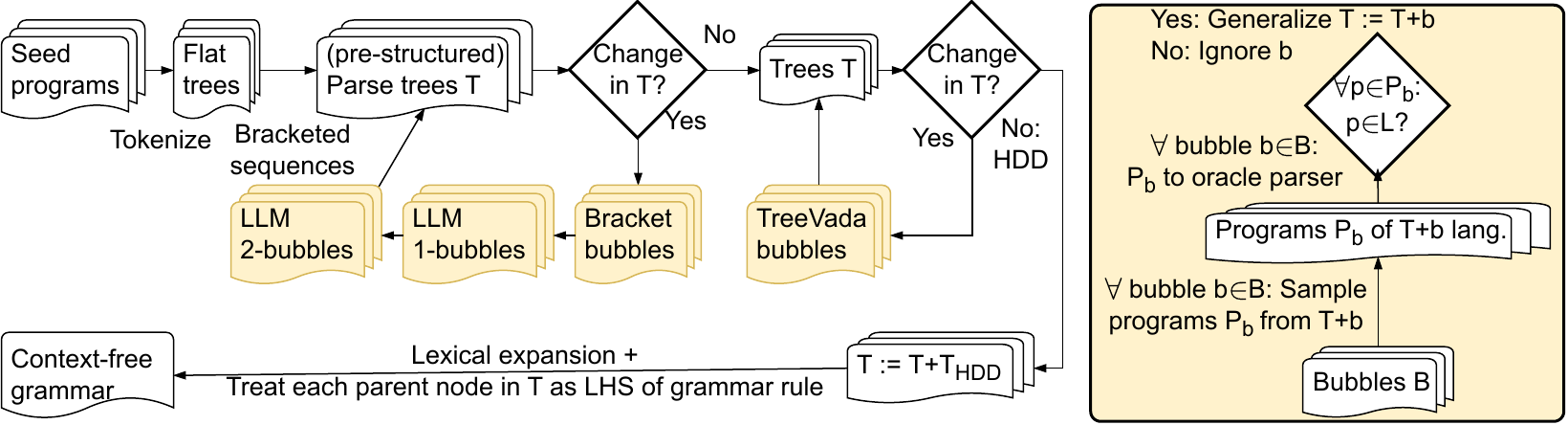}
\caption{Overview: After pre-processing \Tool{} iteratively generalizes its seed programs' parse trees (and thus grammar), first via brackets and LLM queries, then via \treevada{}, before adding sub-trees via hierarchical delta debugging (HDD). \Tool{} generalizes its trees \textit{T} with a candidate bubble \textit{b} (detailed in right sidebar) if its black-box parser accepts all programs sampled from the tentatively generalized trees (\textit{b} merged into \textit{T}).
}
\Description{Flow diagram showing the proposed tool's main processing stages as blocks and order of processing via arrows.}
\label{fig:tool_overview}
\end{figure}

Figure~\ref{fig:tool_overview} gives an overview of \Tool{}'s workflow.
Just as in related grammar inference tools, \Tool{}'s input is a set of ``seed'' programs of an unknown language, and its output is its context-free approximation of the grammar of the programs' programming language. Following \treevada{} (and the earlier \arvada{}), \Tool{} initially tokenizes each input program's text file. 

\Tool{} follows \treevada{}'s tokenization scheme, with the following two modifications. First, while \treevada{} separates upper- and lower-case letters into two classes, \Tool{} treats them as a single class. For example, the 8-character sequence \term{myFooBar} becomes three tokens in \treevada{} but one token in \Tool{}. Second, \Tool{} also adds to \treevada{}'s tokenization a scheme to remove redundant whitespace tokens. Specifically, \Tool{} iteratively removes one whitespace at a time and checks if the resulting program still passes the black-box oracle parser.

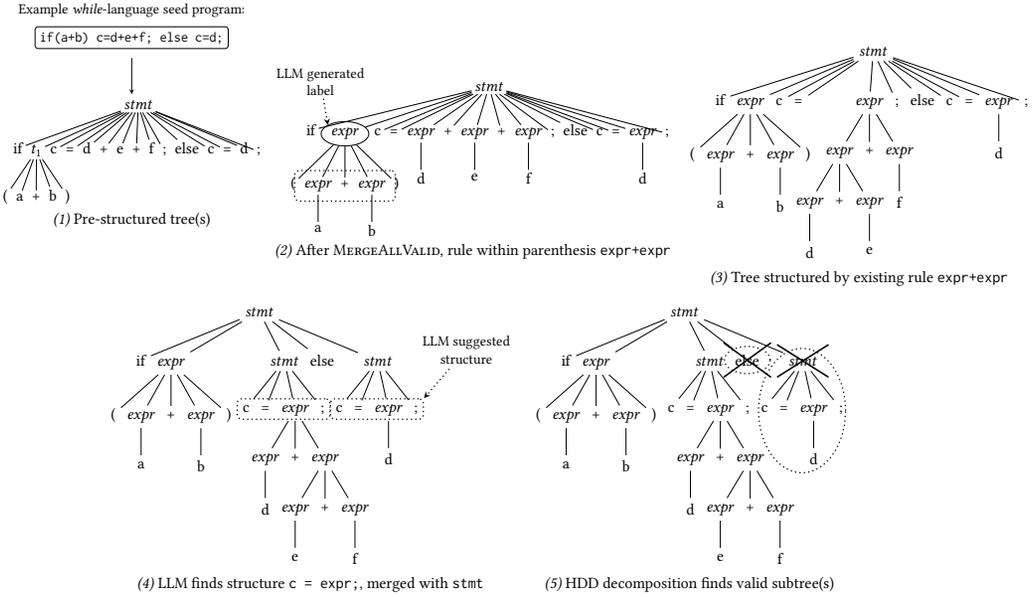
\begin{figure}[h!t]
\centering
\resizebox{\textwidth}{!}{
\begin{tikzpicture}[>=stealth, node distance=1mm]
\node (label) [font=\small] {Example \textit{while}-language seed program:};
\node (prog) [draw, rounded corners=2pt, inner sep=3pt, 
                below=2.5pt of label.south, 
              font=\ttfamily\small] {if(a+b) c=d+e+f; else c=d;};
\node (tree1) [below=1cm of prog.south,
                inner sep=1.8pt] {
\begin{forest}
for tree={l sep=.6cm, s sep=.5mm}
[\textit{stmt}, 
    [if]
    [$t_1$
        [(][a][+][b][)]
    ]
    [c]
    [{=}]
    [d]
    [+]
    [e]
    [+]
    [f]
    [;]
    [else]
    [c]
    [{=}]
    [d]
    [;]
]
\end{forest}
};
\draw[->, shorten <=2mm] (prog.south) -- (tree1.north)
      node[midway, right=2pt, font=\small, align=left] {};
\node (tree1cap) [anchor=south] at ([yshift=-.5cm]tree1.south) {\textit{(1)} Pre-structured tree(s)};
\node (tree2) [right=of tree1,
                inner sep=1.8pt] {
\begin{forest}
for tree={l sep=.6cm, s sep=.5mm}
[\textit{stmt}
    [if]
    [\textit{expr}, draw, ellipse, name=condnode
        [(]
        [\textit{expr}, name=L[a]]
        [+]
        [\textit{expr}, name=R[b]][)]
    ]
    [c]
    [{=}]
    [\textit{expr}[d]]
    [+]
    [\textit{expr}[e]]
    [+]
    [\textit{expr}[f]]
    [;]
    [else]
    [c]
    [{=}]
    [\textit{expr}[d]]
    [;]
]
\node[align=center, anchor=south east, font=\small] at ([xshift=.5cm, yshift=.5cm]condnode.north) (callout) {LLM generated\\label};
\draw[dotted,thick,->] (callout) -- (condnode.north west);
\node[draw,dotted,thick,rounded corners=3pt,inner sep=4pt,fit=(L)(R)] {};
\end{forest}
};
\node (tree2cap) [anchor=south] at ([yshift=-.5cm]tree2.south) {\textit{(2)} After \textsc{MergeAllValid}, rule within parenthesis \texttt{expr+expr}};
\node (tree3) [right=of tree2] {
\begin{forest}
for tree={l sep=.6cm, s sep=.1mm}
[\textit{stmt}
    [if]
    [\textit{expr}
        [(]
        [\textit{expr}[a]]
        [+]
        [\textit{expr}[b]][)]
    ]
    [c]
    [{=}]
    [\textit{expr}
    [\textit{expr}[\textit{expr}[d]]
    [+]
    [\textit{expr}[e]]]
    [+]
    [\textit{expr}[f]]]
    [;]
    [else]
    [c]
    [{=}]
    [\textit{expr}[d]]
    [;]
]
\end{forest}
};
\node (tree3cap) [anchor=south] at ([yshift=-.5cm]tree3.south) {\textit{(3)} Tree structured by existing rule \texttt{expr+expr}};

\node (tree4) [below=1.4cm of tree1cap, xshift=4cm] {
\begin{forest}
for tree={l sep=.6cm, s sep=.1mm}
[\textit{stmt}
    [if]
    [\textit{expr}
        [(]
        [\textit{expr}[a]]
        [+]
        [\textit{expr}[b]][)]
    ]
    [\textit{stmt} [c, name=L1]
    [{=}]
    [\textit{expr}[\textit{expr}[d]]
    [+]
    [\textit{expr}[\textit{expr}[e]]
    [+]
    [\textit{expr}[f]]]]
    [;, name=R1]]
    [else]
    [\textit{stmt} [c, name=L2]
    [{=}]
    [\textit{expr}[d]]
    [;, name=R2]]
]
\node[draw,dotted,thick,rounded corners=3pt,inner sep=0pt,fit=(L1)(R1)] {};
\node[draw,dotted,thick,rounded corners=3pt,inner sep=0pt,fit=(L2)(R2),name=labelnode] {};
\node[align=center, font=\small] at ([xshift=2cm, yshift=1.5cm]labelnode.north) (callout) {LLM suggested\\structure};
\draw[dotted,thick,->] (callout) -- (labelnode.north east);
\end{forest}
};
\node (tree4cap) [anchor=south] at ([yshift=-.5cm]tree4.south) {\textit{(4)} LLM finds structure \texttt{c = expr;}, merged with \texttt{stmt}};
\node (tree5) [right= of tree4] {
\begin{forest}
for tree={l sep=.6cm, s sep=.1mm}
[\textit{stmt}
    [if]
    [\textit{expr}
        [(]
        [\textit{expr}[a]]
        [+]
        [\textit{expr}[b]][)]
    ]
    [\textit{stmt} [c]
    [{=}]
    [\textit{expr}[\textit{expr}[d]]
    [+]
    [\textit{expr}[\textit{expr}[e]]
    [+]
    [\textit{expr}[f]]]]
    [;]]
    [else, tikz={\node [draw, ellipse, dotted, thick, inner sep=-1pt,fit to=tree]{};}, name=else]
    [\textit{stmt}, tikz={\node [draw, ellipse, dotted, thick, inner sep=-10pt,fit to=tree]{};}, name=stmt [c]
    [{=}]
    [\textit{expr}[d]]
    [;]]
]
\node[cross out,
      line width=1pt,
      draw,
      fit=(else)]{};
\node[cross out,
    line width=1pt,
    draw,
    fit=(stmt)]{};
\end{forest}
};
\node (tree5cap) [anchor=south] at ([yshift=-.5cm]tree5.south) {\textit{(5)} HDD decomposition finds valid subtree(s)};
\end{tikzpicture}}
\caption{Key \Tool{} processing steps
on a toy seed program (top left):
Tokenization and bracket-based tree pre-structuring~(1),
merging the new parent node $t_1$ with existing nodes \& renaming $t_1$ via LLM suggestion~(2),
leveraging the bracketed sequence \textit{expr}+\textit{expr} as a successful bubble~(3),
merging LLM-suggested bubble c=\textit{expr} with \textit{stmt}~(4), and
pruning the tree via hierarchical delta debugging~(5).
}
\Description{Sequence of 5 parse trees, ordered from the pre-structured initial tree (left top) to the result of HDD decomposition (right bottom).}
\label{fig:workflow_by_example}
\end{figure}

Still following \arvada{} and \treevada{}, \Tool{} then treats each seed program as a flat parse tree. \Tool{} then moves (following \treevada{}) each bracketed node sequence under a new child node. Figure~\ref{fig:workflow_by_example} demonstrates key \Tool{} processing steps using an example program in the Figure~\ref{fig:four-grammars-for-human-understanding} \textit{while} language. After tokenizing the program and treating it as a flat parse tree, \Tool{} moves the bracketed sequence \term{(a+b)} under a new non-terminal node $t_1$ (Figure~\ref{fig:workflow_by_example}.1). At this point this pre-structured tree thus implies the new grammar rule $t_1$ → \term{(a+b)}.

\subsection{LLM-guided Non-terminal Renaming (\ailabel{})} 

LLMs are vulnerable to token replacement tricks~\cite{orvalho2025largelanguagemodelsrobust}. When replacing code tokens with meaningless or misleading tokens, their performance significantly drops on downstream tasks~\cite{zhang2025unseenhorizonsunveilingreal}. Since we want to recover parse trees iteratively, each added (non-terminal) parent node should represent the semantic meaning of its particular sub-tree. In addition to improving LLM reasoning in subsequent LLM-based tree restructuring steps, creating descriptive non-terminal labels also improves human grammar readability.

Whenever our oracle approves merging a new bubble parent node (with temporary label $b$) with an existing non-terminal tree node $a$, \toolName{} collects both node's complete subtrees via an in-order traversal (minus the node itself). To find a descriptive non-terminal label, \toolName{} passes the resulting two node sequences as part of the Figure~\ref{fig:prompt_label} prompt to an LLM. 

\begin{figure}[h!t]
\begin{tcolorbox}[colback=gray!5!white, colframe=gray!75!black]
\begin{Verbatim}[fontsize=\footnotesize]
You are an AI assistant. You will label the internal nodes in the parse tree of any arbitrary 
program. The root node of the parse tree has label stmt. You will be given a pair of 
substrings derivable from the same non-terminal symbol. Your job is to label the non-terminal 
symbol. Look at some example substring pairs and their labels:

# Examples:
    - '(n+n)','n'
    -> numexpr, this non-terminal can derive both the substrings (n+n) and n.
    ...
\end{Verbatim}
\end{tcolorbox}
\caption{LLM prompt for generating descriptive non-terminal labels.}
\Description{Prompt template for LLM query to get non-terminal name.}
\label{fig:prompt_label}
\end{figure}

In the pre-processing phase of our running example, \Tool{} tries to merge each non-terminal node pair of the Figure~\ref{fig:workflow_by_example}.1 pre-structured parse tree. The oracle agrees that we can merge the temporarily-named $t_1$ parent node with the (implicit and hence not shown) non-terminal $a$ (which has the terminal \term{a} as its only child). An in-order traversal yields the subtree representation strings \term{(a+b)} and \term{a}, for which the LLM (in our experiments GPT-4o-mini) suggests the descriptive label \textit{expr}. Similar merges with \textit{b}, \textit{d}, \textit{e}, and \textit{f} yield the post-merge Figure~\ref{fig:workflow_by_example}.2, which implies the new grammar rule 
\textit{expr} → \term{(}\textit{expr} \term{+} \textit{expr}\term{)} | \term{a} | \term{b} | \term{d} | \term{e} | \term{f}.

\subsection{Bracket-guided Bubble Selection} 
\label{sec:structure_infer}

\treevada{} built on what it called a ``soft'' assumption that many languages use \term{[]()\{\}} brackets as (recursive) nesting constructs. In the pre-processing phase, \treevada{} and \Tool{} can often merge the resulting added parent nodes with existing tree nodes (e.g., yielding Figure~\ref{fig:workflow_by_example}.2). While it treats the node sequence from opening-bracket to closing-bracket as a single unit, this tree pre-structuring does not exploit the fact that this bracket pair also creates a sequence of nodes between the brackets that can be meaningful on its own. Specifically, what appears bracketed here may appear elsewhere in the seed programs in a different context, i.e., within a different type of brackets or not bracketed at all.

To exploit this observation, \toolName{} treats each within-brackets sequence as a candidate bubble. In our running example, \toolName{} treats the sequence within the \term{()} brackets, i.e., \textit{expr} \term{+} \textit{expr}, as such a candidate bubble. As our oracle approves merging \textit{expr} \term{+} \textit{expr} with \textit{expr}, \Tool{} recursively applies this newly learnt rule, yielding the Figure~\ref{fig:workflow_by_example}.3 parse tree. At this point the parse tree implies the grammar rule
\textit{expr} → 
\term{(}\textit{expr} \term{+} \textit{expr}\term{)} | 
\textit{expr} \term{+} \textit{expr} | 
\term{a} | \term{b} | \term{d} | \term{e} | \term{f}.

While earlier approaches such as \arvada{} and \treevada{} may at some point also propose the same \textit{expr} \term{+} \textit{expr} bubble, these earlier approaches would do so as a result of their regular (expensive) bubble ranking process. Besides such extra runtime, earlier tools may also first get confused by one of their heuristics, get stuck in their exploration, and thus never apply such a within-brackets bubble candidate.

\subsection{LLM-Guided Bubble Exploration}

Once existing bracket sequences are applied to the trees cannot make any changes to the trees, we ask an LLM to further structure them. We prompt the LLM flat tree layers as text sequence. Prior work on LLMs' structural awareness~\cite{ast-probe, whatdotheycapture, hewitt2019structural} suggests that the dependency tree of an input sequence is encoded in the LLM's hidden representation. That makes an LLM a perfect tool for adding structure to partially complete parse trees.

The prompt begins with a natural language instruction directing the LLM to iteratively build the parse trees (Figure~\ref{fig:prompt}). Given an input consisting of a sequence of tree levels (encoded as lists enclosed in square brackets), we instruct the model to propose node groupings (bubbles) that could iteratively build higher levels of the parse tree. The bubbles will introduce new nodes into the trees, where the resulting parent–child relationships represent tree-induced grammar rules.

\begin{figure}[h!t]
\begin{tcolorbox}[colback=gray!5!white, colframe=gray!75!black]
\begin{Verbatim}[fontsize=\footnotesize]
You are an AI assistant. You will help to build parse trees from flat tree levels. 
You will combine adjacent nodes from a tree level. Those nodes will be placed under a new 
parent node at that position. You will suggest node groups in a way that will build the 
complete parse tree iteratively.

# Input: A list of tree levels separated by square brackets.
# Output: A list of unique groups as json.

# Workflow:
    - Break an input level into smaller independent units.
    - Think what smallest units represent a language construct, free of recursion.
    ...
\end{Verbatim}
\end{tcolorbox}
\caption{Prompt in generating bubble candidates. Full prompt available in the artifact.
}
\Description{Prompt template for LLM queries.}
\label{fig:prompt}
\end{figure}
To get high-quality response we set several constraints in the prompt~\cite{gpt4techniq}. Figure~\ref{fig:prompt} shows the workflow sequence we used in our experiments. To add meaningful structure to the trees, we direct the model to avoid overly long groupings and to focus on common language constructs. We instruct the LLM to find small, recursion-free groups and limit the suggestion list to 20 bubbles. Each prompt also stores the prior tree state, if available, to maintain context across invocations. 

For our experiments we query the GPT-4o LLM using a temperature of 0 and a fixed seed of 101 for deterministic output. Despite this ``deterministic'' setting we observed occasional non-deterministic behavior. We get a list of candidate bubbles from the LLM. Using the same oracle check as during bracket-guided exploration, we evaluate each bubble before accepting (and thus merging) it into the trees.

\subsubsection{LLM 1 \& 2 Bubbles}

First, we sort the list of bubbles obtained from the specialized prompt by bubble-length. We call those as 1-bubble candidates. \textsc{CheckBubble} verifies these candidate 1-bubbles one by one if these could be merged with any existing tree nodes. Whenever, \textsc{CheckBubble} returns true, the new structure is accepted into the trees, the applied bubble is updated with the new AI-generated label introduced by the merge, and the bubble is tried again. \toolName{} thus incrementally integrates recursive structures from the input into the parse trees. In Figure~\ref{fig:workflow_by_example}.4, the LLM suggested 1-bubble \fbox{\term{c} \term{=} \textit{expr} \term{;}} is accepted and merged with existing node \texttt{stmt}.

While the 1-bubbles are tried to merge with any existing tree nodes, we prompt the LLM to generate another set of bubble pairs, which could essentially be merged with each other. We refer to this second set of candidates as 2-bubbles. Once 1-bubbles no longer yield changes in the trees, \toolName{} starts trying these 2-bubbles. 2-bubbles become useful at this stage because no further mergeable nodes remain in the trees; instead, we must select two sibling sequences simultaneously to advance the tree structuring process. Similar to 1-bubble, \toolName{} repeatedly evaluates 2-bubbles in a loop until no further changes occur in the trees. The three bubble mechanisms — \textit{bracket-guided bubbles}, \textit{1-bubbles}, and \textit{2-bubbles} are evaluated within an additional loop. If no structural changes are observed after five consecutive attempts, the process terminates and we get \toolNameWithoutTreevada{} structured parse trees. In \toolName{} we apply \treevada{} heuristics to further refine the trees in case LLM stops processing the trees early. The overall grammar inference procedure is summarized in Algorithm~\ref{algo:NatGI}.

\begin{algorithm}
\caption{\toolNameWithoutTreevada{} and \toolName{} Overview}
\begin{algorithmic}[1]
\REQUIRE Seed inputs $\mathcal{S}$, Oracle $\mathcal{O}$
\ENSURE Grammar $\mathcal{G}$, covering the language of $\mathcal{S}$
\STATE tokens $\gets$ \textsc{Pretokenization}($\mathcal{S}$)
\STATE prestructured\_trees $\gets$ \textsc{CreateNa\"\i veTrees}(tokens)
\STATE best\_trees $\gets$ \textsc{MergeAllValid}(prestructured\_trees, $\mathcal{O}$)

\WHILE{best\_trees update}
  \STATE best\_trees $\gets$ \textsc{CheckBubble}(In-bracket\_bubbles, $\mathcal{O}$)
  \STATE best\_trees $\gets$ \textsc{CheckBubble}(LLM\_1\_bubbles, $\mathcal{O}$)
  \STATE best\_trees $\gets$ \textsc{CheckBubble}(LLM\_2\_bubbles, $\mathcal{O}$)
\ENDWHILE

\STATE best\_trees $\gets$ \textsc{TreeVadaHeuristics}(best\_trees, $\mathcal{O}$) 

\STATE all\_trees $\gets$ \textsc{HDD\_Decompose}(best\_trees, $\mathcal{O}$)
\STATE $\mathcal{G} \gets$ \textsc{InducedGrammar}(all\_trees)
\STATE $\mathcal{G} \gets$ \textsc{ExpandTokens}(all\_trees, $\mathcal{G}$, $\mathcal{O}$)
\RETURN $\mathcal{G}$

\end{algorithmic}
\label{algo:NatGI}
\end{algorithm}

\subsection{Iterative Tree Generalization via \treevada{} Heuristics} %
LLMs have known limitations such as hallucinations~\cite{xu2024hallucination}, token perturbations~\cite{lin2025single}, and difficulty in handling large codebases~\cite{shen-etal-2022-benchmarking}. To compensate for any early termination of LLM guided tree-structuring, we apply \treevada{}'s heuristics to the trees obtained in the previous step. \treevada{} extends prior tool \arvada{}'s heuristics. \arvada{} introduced left/right context and frequency as measures for bubble ranking. \treevada{} further enhances this scheme by incorporating a depth and length-aware bubble ranking scheme. \treevada{} heuristics provide a deterministic set of bubbles, ranked according to their likelihood of getting merged. \toolName{} evaluates these bubbles using \textsc{CheckBubble} before accepting them into the parse trees.

It's worth noting than \toolNameWithoutTreevada{} doesn't have this \treevada{} heuristic step on top of LLM-structured trees. LLM-structured trees tend to be more natural, whereas \treevada{} heuristics make the parse trees overly complex, resulting in a longer and less-readable grammar (Figure~\ref{list:grammar_treevada}). \toolName{} gets the best of both worlds, \toolName{} compensates for any missing LLM step while minimally hurting the grammar readability (Figure~\ref{fig:four-grammars-for-human-understanding}-right).

\subsection{HDD for Program Decomposition} %

The main \toolName{} loop stops when the LLM (and \treevada{} heuristics) cannot structure the parse trees anymore. From this point onward, we decompose each parse tree into smaller fragments to enable the resulting grammar to capture a broader set of language rules. For example, Figure~\ref{fig:workflow_by_example}.4 shows the final parse tree. The tree-induced grammar would only recognize languages of the format \term{if} \texttt{cond} \texttt{stmt} \term{else} \texttt{stmt}. However, the language could also allow only \term{if} statements without the \term{else} part. Figure~\ref{fig:workflow_by_example}.(4) shows the pruned parse tree which also resembles a valid program.

To find such decomposed language via effective pruning, we use the classical \textit{Delta Debugging}~\cite{hddzeller} approach. Since our data type is hierarchical in nature, we use Hierarchical Delta Debugging~\cite{hdd}. Delta debugging iteratively reduces the input as long as the test continues to fail. In our case, the failing criterion is defined as follows: 50 strings sampled from the pruned tree (the same way we sample strings during \textsc{CheckBubble}) must be accepted by the black-box parser. As long as this condition is satisfied, we continue pruning the tree. We store all intermediate trees that meet this criterion. The final grammar extracted from the trees supports a broader range of rules compared to the one induced solely from the seed program’s parse trees.

\subsection{Lexical Inference}
Once \toolName{} recovers the parse tree structure of the seed inputs, the leaf nodes are limited to those tokens existing in the seeds. However, the seeds represent a tiny fraction of the allowed tokens at each leaf location. Like~\cite{kulkarni2021learning, arefin2024fast}, we perform lexical expansion of these tokens to enable the inferred grammar to accommodate a broader range of terminal symbols. We incrementally try to generalize each token with a larger character group, for example we try to expand a single digit to a numbers and check if that still satisfies the parser. If the parser accepts the new string, the resulting grammar permits numbers at that location. Similarly, we try to expand single characters to multiple characters, alpha-numerals, etc. For each expansion, we sample 50 strings like before, and check if all 50 strings are accepted by the black-box parser. However, the set of lexical expansion rules are fixed and \toolName{} cannot learn complex regular expressions for the tokens. The overall learning concludes with the extraction of rules from the recovered parse trees.

\section{Grammar Readability}

\lstdefinelanguage{Grammar}{
  morekeywords={start,stmt,numexpr, boolexpr, n1, n2, n0, n3, n4, t0, t1, stmt_1, stmt_2, boolexpr_, expr, expr_1, expr_2, expr_stmt, t9869, t9562, t7115, t4197, t9770, t4806, t9228, t8994, t6716, t5816, t6146, t5491, t3582, t3830, t1408, t363, t612, t489, t3953, t3325, t2781, t1144, t9342, simple_stmt, assignment, compound_stmt, if_stmt, while_stmt, condition, stmt_13, complex_stmt, conditional_stmt, stmt_4, bool_expr_1, boolstmt, stmt_11, stmt_7, stmt_3, boolstmt_2, complex_numexpr_2, complex_numexpr, complex_expr, unary_expr, unary_op, stmt_15 },    %
  keywordstyle= \normalcolor\itshape, %
  sensitive=true
}

\begin{figure}[h!t]
  \centering
  \begin{subfigure}[t]{0.25\textwidth}
\begin{lstlisting}[basicstyle=\color{teal}\ttfamily\scriptsize,
language=Grammar,
escapeinside={(*@}{@*)}]
start(*@\normalcolor:@*) stmt
stmt(*@\normalcolor:@*) stmt ; stmt
(*@\normalcolor|@*) L = numexpr
(*@\normalcolor|@*) while boolexpr 
    do stmt
(*@\normalcolor|@*) if boolexpr then stmt
    else stmt
(*@\normalcolor|@*) skip
boolexpr(*@\normalcolor:@*) ~boolexpr
(*@\normalcolor|@*) boolexpr & boolexpr
(*@\normalcolor|@*) numexpr == numexpr
(*@\normalcolor|@*) true
(*@\normalcolor|@*) false
numexpr(*@\normalcolor:@*)  ( numexpr + numexpr )
(*@\normalcolor|@*) L
(*@\normalcolor|@*) n
\end{lstlisting}
    \caption{Golden grammar of the \textit{while} language~\cite{kulkarni2021learning}.}
  \end{subfigure}
  \hfill
  \begin{subfigure}[t]{0.20\textwidth}
\begin{lstlisting}[basicstyle=\color{teal}\ttfamily\scriptsize,
language=Grammar,
escapeinside={(*@}{@*)}]
n1(*@\normalcolor:@*) skip
(*@\normalcolor|@*) n1 ; n1
(*@\normalcolor|@*) L = n2
(*@\normalcolor|@*) if n3 n4 then n1 
    else n1
(*@\normalcolor|@*) while n3 n4 do n1
n0(*@\normalcolor:@*) (*@\textvisiblespace@*)
(*@\normalcolor|@*) n0 t0
n2(*@\normalcolor:@*) n
(*@\normalcolor|@*) ( n2 + n2 )
(*@\normalcolor|@*) L
n3(*@\normalcolor:@*) n0
(*@\normalcolor|@*) n3 n4 & n3
n4(*@\normalcolor:@*) t1
(*@\normalcolor|@*) n2 == n2
t0(*@\normalcolor:@*) ~(*@\normalcolor+@*)
t1(*@\normalcolor:@*) true
(*@\normalcolor|@*) false
\end{lstlisting}
\caption{\kedavra{}: The + in t0's rule (t0:  \textasciitilde+) is positive closure (so \textasciitilde{} can appear 1..n times).}
\end{subfigure}
\hfill
\begin{subfigure}[t]{0.25\textwidth}
\begin{lstlisting}[basicstyle=\color{teal}\ttfamily\scriptsize,
language=Grammar,
escapeinside={(*@}{@*)}]
start(*@\normalcolor:@*) stmt
stmt(*@\normalcolor:@*) stmt ; stmt
(*@\normalcolor|@*) L = numexpr
(*@\normalcolor|@*) L = L
(*@\normalcolor|@*) while boolexpr 
    do stmt
(*@\normalcolor|@*) if boolexpr then stmt 
    else stmt
(*@\normalcolor|@*) skip
boolexpr(*@\normalcolor:@*) ~boolexpr
(*@\normalcolor|@*) boolexpr & boolexpr
(*@\normalcolor|@*) L == numexpr
(*@\normalcolor|@*) numexpr == numexpr
(*@\normalcolor|@*) numexpr == L
(*@\normalcolor|@*) L == L
(*@\normalcolor|@*) true
(*@\normalcolor|@*) false
numexpr(*@\normalcolor:@*) ( L + L )
(*@\normalcolor|@*) ( L + numexpr )
(*@\normalcolor|@*) ( numexpr + L )
(*@\normalcolor|@*) ( numexpr + numexpr )
(*@\normalcolor|@*) n
\end{lstlisting}
\caption{\toolNameWithoutTreevada{}}
\end{subfigure}
\hfill
\begin{subfigure}[t]{0.23\textwidth}
\begin{lstlisting}[basicstyle=\color{teal}\ttfamily\scriptsize,
language=Grammar,
escapeinside={(*@}{@*)}]
start(*@\normalcolor:@*) stmt
stmt(*@\normalcolor:@*) stmt_2 stmt
(*@\normalcolor|@*) skip
(*@\normalcolor|@*) L = expr_2
stmt_2(*@\normalcolor:@*) stmt ;
(*@\normalcolor|@*) stmt_1 boolexpr 
    then stmt else
(*@\normalcolor|@*) while boolexpr do
stmt_1(*@\normalcolor:@*) if
(*@\normalcolor|@*) stmt_2 stmt_1
boolexpr(*@\normalcolor:@*) true
(*@\normalcolor|@*) false
(*@\normalcolor|@*) boolexpr_ boolexpr
(*@\normalcolor|@*) expr_stmt expr_2
boolexpr_(*@\normalcolor:@*) boolexpr & 
(*@\normalcolor|@*) ~
expr_2(*@\normalcolor:@*) numexpr
(*@\normalcolor|@*) L
numexpr(*@\normalcolor:@*) ( expr )
(*@\normalcolor|@*) n
expr(*@\normalcolor:@*) expr_1 numexpr
(*@\normalcolor|@*) expr_1 L
expr_1(*@\normalcolor:@*) L+
(*@\normalcolor|@*) numexpr +
expr_stmt(*@\normalcolor:@*) L ==
(*@\normalcolor|@*) numexpr ==
    \end{lstlisting}
    \caption{\toolName{}}
  \end{subfigure}

\caption{Golden grammar of the \textit{while} language (left) and grammars inferred by \kedavra{}, \toolNameWithoutTreevada{}, and \toolName{}. The inferred grammars all achieved perfect scores on the commonly used \precision{}, \recall{}, and \fscore{} grammar quality metrics. Yet human readability differs markedly. Standalone whitespace is marked as~\texttt{\textvisiblespace} (whitespace before or after terminals is not shown).
}
\Description{Golden grammar (left) compared with grammars inferred by (in order) Kedavra and two versions of the proposed tool (right).}
\label{fig:four-grammars-for-human-understanding}
\end{figure}

\begin{figure}[h!t]
\centering
\begin{subfigure}[t]{0.24\textwidth}
\begin{lstlisting}[basicstyle=\color{teal}\ttfamily\scriptsize,
language=Grammar,
escapeinside={(*@}{@*)}]
start(*@\normalcolor:@*) t0
t0(*@\normalcolor:@*) t9869 t0
(*@\normalcolor|@*) skip
(*@\normalcolor|@*) L t1144
t9869(*@\normalcolor:@*) t9562 t6146
(*@\normalcolor|@*) t9770
(*@\normalcolor|@*) t9869 t9869
t9562(*@\normalcolor:@*) t7115 t6716
(*@\normalcolor|@*) t9562 t9869
t7115(*@\normalcolor:@*) t4197 t9342
(*@\normalcolor|@*) t9869 t7115
t4197(*@\normalcolor:@*) t9869 t4197
(*@\normalcolor|@*) if
\end{lstlisting}
\end{subfigure}
\hfill
\begin{subfigure}[t]{0.24\textwidth}
\begin{lstlisting}[basicstyle=\color{teal}\ttfamily\scriptsize,
language=Grammar,
escapeinside={(*@}{@*)}]
t9770(*@\normalcolor:@*) t4806 ;
(*@\normalcolor|@*) t3582 t9342 t3830
(*@\normalcolor|@*) t9869 t9770
t4806(*@\normalcolor:@*) t0
(*@\normalcolor|@*) t9869 t4806
t9342(*@\normalcolor:@*)  t9228
(*@\normalcolor|@*) t9342 & t9342
(*@\normalcolor|@*) t3953 = t1144
(*@\normalcolor|@*) t2781 t9228
t9228(*@\normalcolor:@*) t8994
(*@\normalcolor|@*) ~ t9228
t8994(*@\normalcolor:@*) true
(*@\normalcolor|@*) false
(*@\normalcolor|@*) t363 = t1144
\end{lstlisting}
\end{subfigure}
\hfill
\begin{subfigure}[t]{0.24\textwidth}
\begin{lstlisting}[basicstyle=\color{teal}\ttfamily\scriptsize,
language=Grammar,
escapeinside={(*@}{@*)}]
t6716(*@\normalcolor:@*) t5816
(*@\normalcolor|@*) t6716 t9869
t5816(*@\normalcolor:@*) then
(*@\normalcolor|@*) & t9342 t5816
(*@\normalcolor|@*) t6716 t9770
t6146(*@\normalcolor:@*) t4806 t5491
(*@\normalcolor|@*) t6146 t9869
t5491(*@\normalcolor:@*) else
(*@\normalcolor|@*) t5491 t9869
t3582(*@\normalcolor:@*) while
(*@\normalcolor|@*) t9869 t3582
t3830(*@\normalcolor:@*) do
(*@\normalcolor|@*) & t9342 t3830
t1144(*@\normalcolor:@*) = t1408
\end{lstlisting}
\end{subfigure}
\hfill
\begin{subfigure}[t]{0.24\textwidth}
\begin{lstlisting}[basicstyle=\color{teal}\ttfamily\scriptsize,
language=Grammar,
escapeinside={(*@}{@*)}]
t1408(*@\normalcolor:@*)  t363
(*@\normalcolor|@*)  L
t363(*@\normalcolor:@*) ( t612 )
(*@\normalcolor|@*) n
t612(*@\normalcolor:@*) t489 t363
(*@\normalcolor|@*) t489 L
t489(*@\normalcolor:@*) L +
(*@\normalcolor|@*) t363 +
t3953(*@\normalcolor:@*) t1408
(*@\normalcolor|@*) t3325
(*@\normalcolor|@*) t9342 & t3953
t3325(*@\normalcolor:@*) t2781 L
(*@\normalcolor|@*) t2781 t363
t2781(*@\normalcolor:@*) ~
(*@\normalcolor|@*) t2781 ~
\end{lstlisting}
\end{subfigure}
\caption{Grammar \treevada{} inferred for the Figure~\ref{fig:four-grammars-for-human-understanding} \textit{while} langauge. As the Figure~\ref{fig:four-grammars-for-human-understanding} tool-inferred grammars, \treevada{}'s grammar also got perfect precision, recall, and F1 scores.
}
\Description{Grammar inferred by TreeVada showing many machine-generated identifiers that are hard to read.}
\label{list:grammar_treevada}
\end{figure}

\begin{figure}[h!t]
\centering
\begin{subfigure}[t]{0.48\textwidth}
\begin{lstlisting}[basicstyle=\color{teal}\ttfamily\scriptsize,
language=Grammar,
escapeinside={(*@}{@*)}]
start(*@\normalcolor:@*) stmt
stmt(*@\normalcolor:@*) simple_stmt
(*@\normalcolor|@*) compound_stmt
simple_stmt(*@\normalcolor:@*) skip
(*@\normalcolor|@*) assignment
assignment(*@\normalcolor:@*) L = expr
compound_stmt(*@\normalcolor:@*) if_stmt
(*@\normalcolor|@*) while_stmt
if_stmt(*@\normalcolor:@*) if condition then stmt else stmt
\end{lstlisting}
\end{subfigure}
\hfill
\begin{subfigure}[t]{0.48\textwidth}
\begin{lstlisting}[basicstyle=\color{teal}\ttfamily\scriptsize,
language=Grammar,
escapeinside={(*@}{@*)}]
while_stmt(*@\normalcolor:@*) while condition do stmt
condition(*@\normalcolor:@*) expr == expr
(*@\normalcolor|@*) expr
(*@\normalcolor|@*) condition & condition
(*@\normalcolor|@*) ~ condition
expr(*@\normalcolor:@*) n
(*@\normalcolor|@*) L
(*@\normalcolor|@*) expr + expr
(*@\normalcolor|@*) ( expr )
\end{lstlisting}
\end{subfigure}
\caption{A Barebone-LLM inferred grammar (\precision{} 0.25, \recall{} 0.01, \fscore{} 0.02) for the Figure~\ref{fig:four-grammars-for-human-understanding} \textit{while} langauge.}
\label{list:grammar_bb_llm}
\end{figure}

Black-box grammar inference tools such as \arvada{}, \treevada{}, \kedavra{}, and \toolName{} aim to recover parse trees that closely approximate the parse trees derived from the language's ground truth (``golden grammar''). The closer the tree approximation, the closer would the quality of the tree-induced grammar be to the target ground-truth grammar. The various tools' tree structuring heuristics however often fail to capture the target language's actual parse trees. Figure~\ref{fig:four-grammars-for-human-understanding} shows the ground truth (``golden grammar'') of the \textit{while} language used in earlier black-box grammar inference work~\cite{kulkarni2021learning,arefin2024fast}. Figures~\ref{fig:four-grammars-for-human-understanding} and \ref{list:grammar_treevada} also show the tool-inferred grammars. For the \textit{while} language, the \treevada{}, \kedavra{}, \toolNameWithoutTreevada{}, and \toolName{} inferred grammars in Figures~\ref{fig:four-grammars-for-human-understanding} and \ref{list:grammar_treevada} all scored a perfect 100\% on the commonly-used grammar quality metrics precision, recall, and F1~score.
Due to a limited number of string sampling for oracle check during \textsc{CheckBubble} step, many wrong structures are accepted during the tree recovery phase. The resulting grammar becomes has more number of rules and alternatives than the ground truth. On the other hand, \toolName{} uses an expert LLM to try new structures. The structural information contained within it enables us to locate correct structures without trying too many incorrect ones. This prevents the algorithm from wrongly accepting structures that do not belong to the language.

\section{Evaluation Setup}
\label{sec:eval}

At a high level, we would first like to compare \Tool{}'s performance against its most closely related competitors, i.e., \treevada{} and \kedavra{}, especially on larger than toy and small-scale languages. We would also like to better understand the main contributors to \Tool's performance. We thus ask the following research questions.

\begin{description}

\item \RQResultOnLargeProgram{}: \RQResultOnLargeProgramFull{} 

\item \RQGrammarUnderstand{}: \RQGrammarUnderstandFull{}

\item \RQToolComponents{}: \RQToolComponentsFull{}

\end{description}

\subsection{Evaluation Dataset}

For our evaluation, we adopt the setup of prior work, collecting both seed inputs and black-box parsers. The dataset is divided into two categories: (1) small, synthetic languages (upper half of Table~\ref{table:seed_stat}) and (2) larger, realistic languages (lower half of Table~\ref{table:seed_stat}), designed to better reflect practical use cases. The languages are ordered by the number of non-terminals (``column NT''), and Table~\ref{table:seed_stat} summarizes the detailed statistics of both the grammars and the corresponding seed programs.

The first set of languages originates from prior work, including \arvada~\cite{kulkarni2021learning}, \treevada~\cite{arefin2024fast}, and \kedavra{}. In our evaluation, we use the R1 variant for \textit{\Use{turtle_programname}}, \textit{\Use{while_programname}}, \textit{\Use{lisp_programname}}, \textit{\Use{xml_programname}}, \textit{\Use{json_programname}}, \textit{\Use{tinyc_programname}}, \textit{\Use{tiny_programname}}, and \textit{\Use{curl_programname}}. For \textit{c-500}, which corresponds to the same \textit{tinyc} language, we adopt the R5 variant. We emphasize that R1 and R5 are randomly generated seeds rather than hand-crafted variants. In addition, we include \Use{minic_programname}, introduced in the \kedavra~\cite{li2024incremental} study. Finally, we extend the dataset with four languages from the \textsc{Glade} replication study\cite{bendrissou2022synthesizing}. Among those, one (\textit{tiny} language) belongs to the small group, and the rest are large programs. These languages—\textit{c, lua, and mysql}—introduce more realistic and challenging scenarios for grammar inference due to their structural complexity. As shown in Table~\ref{table:seed_stat}, the number of non-terminals (NT) in these languages is substantially higher compared to the others.
 We reuse both seed inputs and the black-box parsers from the respective source. Table~\ref{table:seed_stat} (right) shows the seed program properties, measured in terms of character and token counts after applying the NatGI pretokenization scheme. \textit{Branch factor} stands for the average number of children per node in the bracket-based pre-structured parse trees.

\nprounddigits{1}
\input{tables/seed_stat}

\subsection{Evaluation Metrics}

Table~\ref{table:seed_stat} shows grammar complexity metrics computed via the gMetrics tool~\cite{Crepinsek2010} (which adds metrics to the earlier SynC tool~\cite{Power2004}, which in turn adds to earlier grammar metrics~\cite{Csuhaj-Varju1993}). At a high level, all else being equal, higher metric values imply larger grammar size and complexity. The most basic metrics are the number of terminals (T) and non-terminals (NT). The average right-hand side length (RHS) divides the total number of terminal and non-terminal occurrences on all rules' right-hand sides by the number of non-terminals (NT). The final basic metric, the average McCabe cyclomatic complexity (MCC), is inspired by programs' control-flow complexity---it divides the total number of ``branch'' operators (alternative rule \term{|}, optional term \term{?}, and closures \term{*} and \term{+}) of all rules by the number of non-terminals (NT). 

Two metrics generate from the grammar a LR parser~\cite{Aho2006} and count its properties, e.g., the LR automaton's states (LRS). The basic lookahead metric (LAT) adds for each terminal the number of LR automaton states that do not lead to an error when that terminal is in the lookahead, divided by the number of terminals (T). The final metric captures a language property directly from the grammar rules. The maximum number of different terminal pairs starting with a given terminal (LTPSM) iteratively expands all grammar rules to compute all possible terminal pairs.

Based on comparing 16~languages including DSLs and general-purpose languages (incl. several version of Java) on 21~metrics, gMetrics proposes to categorize grammars based on three metrics (LRS, LAT, and LTPSM)~\cite[Table~6]{Crepinsek2010}, i.e., into four grammar categories:
Tiny (``toy grammars''),
Small (``mainly DSL grammars''),
Intermediate (``3rd generation general-purpose language''), and
Big (``modern object-oriented languages''), giving ranges for each category-metric combination (e.g., LTPSM over 70 indicates a Big grammar). The cut-offs are LRS (100, 1k, 3k), LAT (20, 200, 500), and LTPSM (6, 50, 70).
To simplify this categorization, we combine Tiny and Small into ``low complexity'' (Lo) and Intermediate and Big into ``high complexity'' (Hi) languages. Our three metrics agree for all our languages on which category (Lo or Hi) a language belongs to.

While gMetrics only provided partial results for mysql, we place mysql into the Hi complexity category as it has an order of magnitude more terminals and more than three times the non-terminals of the second-highest scoring language (c), which aligns with our manual inspection of the grammar file. Only curl doesn't have a golden grammar available, we categorize it under Lo, because it's a regular language.

To evaluate how close the inferred grammar captures the intended language, we use the same metrics as closely related work, i.e., \precision{}, \recall{} and \fscore{}. Let $G_{\text{true}}$ denote the reference grammar that defines the target language, and $G_{\text{inf}}$ be the grammar inferred by the system. We denote by $L(G)$ the language accepted by a grammar $G$, i.e., the set of all strings generated by it. We define:
\[
\text{Precision} = \frac{|L(G_{\text{inf}}) \cap L(G_{\text{true}})|}{|L(G_{\text{inf}})|}
\]
\[
\text{Recall} = \frac{|L(G_{\text{inf}}) \cap L(G_{\text{true}})|}{|L(G_{\text{true}})|}
\]
For \precision{}, we generate 1k~languages from $G_{\text{inf}}$ and check how many of these are accepted by the black-box parser. For \recall{}, we reuse the held-out test set from the previous studies. We check how many of these valid languages are accepted by $G_{\text{inf}}$. To better capture over- or under-generalization, we use the \fscore{} (i.e., the harmonic mean of \precision{} and \recall{}). Here, $\text{F1} = 2 \cdot \frac{\text{Precision} \cdot \text{Recall}}{\text{Precision} + \text{Recall}}$.

\subsection{Runtime Environment}

We ran our evaluation on an Ubuntu machine equipped with an intel Xeon W-2245 8-core CPU running at 3.90GHz and 128 GB of RAM to run our experiments and measure runtimes. We sequentially ran different experiments one after another, and our implementation does not incorporate multi-threading.

\subsection{Baselines}

We compare \toolNameWithoutTreevada{} and \toolName{} against the following baselines:

\subsubsection{Barebone LLM}

\begin{figure}
\begin{tcolorbox}[colback=gray!5!white, colframe=gray!75!black]
\begin{Verbatim}[fontsize=\footnotesize]
You will derive a context-free grammar (CFG) in Backus-Naur Form (BNF) for the given example 
programs:

<program 1> <program 2> <program 3> ...

Ensure that,
- The Production rules must be enclosed within <production-rules> and </production-rules> tags.
- The start rule should be <stmt>.
...
\end{Verbatim}
\end{tcolorbox}
\caption{LLM prompt for zero-shot grammar inference (Barebone LLM).}
\Description{Prompt template for LLM query to get non-terminal name.}
\label{fig:prompt_bb_llm}
\end{figure}

We perform zero-shot grammar inference by prompting the LLM with only the seed inputs. We instruct the LLM to extract production rules in Backus-Naur Form (BNF), with strict formatting: the rules are enclosed in <production-rules> tags, whitespace are explicitly tokenized, terminals are quoted, and no explanatory text is added. We use OpenAI's powerful reasoning model (o4-mini) for this task. Since, the model infers grammar structure solely from the seed inputs, the resulting grammar is often restricted to the terminals present in the seeds. For fair comparison, we expand the grammar by doing an additional lexical inference step. The final grammar's performance is shown in Table~\ref{table:accuracy}. Our prompt template is shown in Figure~\ref{fig:prompt_bb_llm}.

\subsubsection{\treevada{}}

We compare \toolNameWithoutTreevada{} and \toolName{} with our second baseline, \treevada{}~\cite{arefin2024fast}. \treevada{} relies on parse tree recovery strategy for black-box grammar inference. It extended on prior tool \arvada{}~\cite{kulkarni2021learning} to heuristically structure parse trees in a deterministic way. \treevada{} introduced bracket-based pre-structuring trick for faster grammar inference.

\subsubsection{\kedavra{}}

\kedavra{}~\cite{li2024incremental} is our third baseline that proposes a different strategy than \arvada{} and \treevada{}. Unlike processing the entire input strings, \kedavra{} breaks the inputs into simpler and shorter inputs. It proposed an incremental inference algorithm that infers grammar from those decomposed inputs.

\input{tables/paper-main-result-table}

\nprounddigits{1}

\section{Results}

\subsection{\RQResultOnLargeProgram{}: Effectiveness on Bigger Languages}%

Table~\ref{table:accuracy} reports the \precision{}, \recall{}, and \fscore{} of grammar inference achieved by \barebonesLLM{}, \treevadaOrg{}, \kedavra{}, \toolNameWithoutTreevadaPlain{}, and \Tool{}. As expected, \barebonesLLM{} shows the weakest performance due to its zero-shot nature—operating without any feedback and relying solely on a single seed program. In contrast, \treevada{}, \kedavra{}, \toolNameWithoutTreevada{}, and \Tool{} all achieve significantly higher scores across most benchmarks. On average, \toolNameWithoutTreevada{} uses 22.3 and 38.9 LLM calls and \toolName{} uses 24.9 and 35.0 LLM calls for Lo and Hi category languages, respectively.

While \kedavra{} performs well on many smaller languages, it notably underperforms on the \textit{curl} benchmark, achieving only a \Use{curl_Kedavra_f1} \fscore{}, compared to \Use{curl_treevada_false_hdd_true_LLM_true_f1} and \Use{curl_treevada_true_hdd_true_LLM_true_f1} by \toolNameWithoutTreevada{} and \Tool{}, respectively. More critically, \kedavra{} fails to produce any output for large languages such as \textit{c}, \textit{lua}, and \textit{mysql}, even with a generous 24-hour timeout. In contrast, both \toolNameWithoutTreevada{} and \Tool{} perform robustly on these large-scale benchmarks, achieving \fscore{}s of \Use{c_treevada_true_hdd_true_LLM_true_f1}, \Use{lua_treevada_false_hdd_true_LLM_true_f1}, and \Use{mysql_treevada_true_hdd_true_LLM_true_f1} for \textit{c}, \textit{lua}, and \textit{mysql}, respectively. The average \fscore{} on the high-complexity languages is \Use{Avg-Big-Program_treevada_true_hdd_true_LLM_true_f1}, which is 25pp higher and corresponds to a \fscoreImprovementOverBestBaselineForBigProgram{}\% relative improvement over the best baseline (\treevadaOrg{}). These results highlight the consistent and scalable performance of \Tool{} in handling more-complex languages. 

In terms of runtime, we observe significant variation across tools, especially as the complexity of the input languages increases. \barebonesLLM{} is generally fast due to its zero-shot nature, but this comes at the cost of poor grammar quality. \treevadaOrg{} demonstrates moderate runtimes on small to medium languages (e.g., \Use{tiny_treevada_original_build_time}s for \textit{tiny}) but incurs substantial overhead on larger languages such as \textit{c-500} and \textit{mysql}, taking over \Use{mysql_treevada_original_build_time}s. \kedavra{}, while efficient on small examples (e.g., \Use{tiny_Kedavra_build_time}s for \textit{tiny}), fails to complete on large languages like \textit{c}, \textit{lua}, and \textit{mysql}, timing out despite a generous 24-hour window. In contrast, \Tool{} maintains robust runtime performance across all benchmarks, completing even the largest tasks (e.g., \textit{mysql}) in under \Use{mysql_treevada_true_hdd_true_LLM_true_build_time}s, without timing out or degrading in quality. This highlights \Tool{}’s scalability and reliability for practical grammar inference in real-world settings.

\begin{framed}
\noindent
\emph{
\textbf{\RQResultOnLargeProgram{}: }
For small languages, \Tool{} performs comparably to other baselines. In contrast, for large languages, \Tool{} improves the average \fscore{} of the best baseline (\treevada{}) by 25pp, corresponding to a \fscoreImprovementOverBestBaselineForBigProgram{}\% relative improvement.
}
\end{framed}

\subsection{\RQGrammarUnderstand{}: Readability of the Tool-inferred Grammar}

To evaluate the human interpretability and readability of the inferred grammars, we conduct a qualitative comparison of parse tree structures generated by different tools. Figure~\ref{fig:four-grammars-for-human-understanding} presents grammars produced by \kedavra{}, \toolNameWithoutTreevada{}, and \Tool{}, alongside a manually constructed golden grammar. The golden grammar reflects a clean, intuitive representation of program structure, with semantically meaningful non-terminals like \textit{stmt}, \textit{boolexpr}, and \textit{numexpr}. These labels directly map to common language constructs, making it easy for humans to understand how different parts of the program are organized and interpreted.

In contrast, the grammar inferred by \kedavra{} demonstrates poor human readability. It uses arbitrary and abstract non-terminal labels such as $n1$, $n2$, and $n3$, which do not convey any semantic meaning about the underlying constructs they represent. This makes it difficult to trace program logic or reason about grammar correctness. On the other hand, both \toolNameWithoutTreevada{} and \Tool{} produce grammars that are significantly more structured and semantically meaningful. They introduce non-terminals that directly reflect logical blocks (e.g., \textit{if} and \textit{boolexpr}), making the parse trees much more interpretable and aligned with human expectations. This improved interpretability is especially critical in debugging, grammar refinement, and educational contexts, where understanding the structural role of each token is essential.

\begin{framed}
\noindent
\emph{
\textbf{\RQGrammarUnderstand{}:}
While \kedavra{} generates grammars with arbitrary labels that obscure logic, \Tool{} yields interpretable, semantically aligned non-terminals, which (all else being equal) make parse trees more human-readable and useful for debugging and refinement.
}
\end{framed}

\subsection{\RQToolComponents{}: Impact of Different \Tool{} Components} 

Table~\ref{table:first_and_nd_ablation} and Table~\ref{table:third_and_fourth_ablation} show the impact of different components of \Tool{}. We systematically examine the effect of disabling each component and report how it impacts grammar quality, runtime, and oracle calls.

\MyPara{Impact of \ailabel{}}
Table~\ref{table:first_and_nd_ablation} shows the results of \Tool{} when we disable the AI label meaning that the LLM-based node renaming component is disabled. By default, \Tool{} incorporates node renaming using \llm{}, which aims to enhance grammar readability and consistency. The results indicate that removing this component generally leads to a drop in grammar quality, as reflected by lower \fscore{} values in most cases compared to the \Tool{}. %
The average \fscore{} drops for small and large languages are \avgfscoreDropInSmallProgramInFirstAblation{}, and \avgfscoreDropInBigProgramInFirstAblation{}, respectively. Although this finding shows that the drop is not that significant on average, on average we find runtime impacts. 

\input{tables/first_and_second_ablation}

We find that build time increases for many languages when we do not consider \ailabel{}. Without the AI label, build time increases for many languages, with average increases of \avgbuildTimeDropInSmallProgramInFirstAblation{} for small languages and \avgbuildTimeDropInBigProgramInFirstAblation{} for large languages. Similarly, oracle calls increase on average by \avgOracleCallIncreaseInSmallProgramInFirstAblation{} for small languages and \avgOracleCallIncreaseInBigProgramInFirstAblation{} for large languages. The number of \llm{} calls also grows, with increases of \avgLLMCallIncreaseInSmallProgramInFirstAblation{} and \avgLLMCallIncreaseInBigProgramInFirstAblation{} for small and large languages, respectively.

\input{tables/third_and_fourth_ablation}

\MyPara{Impact of \hddAbb{}}
In this experiment, we evaluate \Tool{} with \hddAbb{} (\hdd{}) disabled. Table~\ref{table:first_and_nd_ablation} summarizes the impact of removing \hdd{} from the pipeline. In this configuration, both LLM-based renaming and \treevada{} remain enabled, while hdd is only disabled. The results show that disabling \hdd{} has minimal effect on small languages but causes substantial degradation for \largeProg{}s. In particular, \fscore{} drops by \LuaavgfscoreDropWhenHDDIsDisabled{}, \cavgfscoreDropWhenHDDIsDisabled{}, and \mysqlavgfscoreDropWhenHDDIsDisabled{} for \textit{lua}, \textit{c}, and \textit{mysql}, respectively, with an average \fscore{} drops of \avgfscoreDropInBigProgramInSecondAblation{}. These findings underscore the critical role of \hdd{} in improving grammar accuracy, especially for larger and more complex languages.

\MyPara{Impact of Bracket-guided Bubbles}
Table~\ref{table:third_and_fourth_ablation} shows the impact of disabling bracket-guided bubble exploration in \Tool{}. By default, \Tool{} leverages brackets in the input languages to induce partial structure in parse trees, reflecting the common use of brackets in programming languages to denote nesting and hierarchy. As described in Section~\ref{sec:structure_infer}, this semi-structuring step helps the system make earlier and more reliable generalizations about grammar rules, particularly when bracket-enclosed patterns recur across the program.

Eliminating this structural guidance results in a significant drop in grammar quality, as measured by \fscore{}. Most languages suffer substantial losses—ranging from no reduction for \textit{json} to as much as 77pp for \textit{c-500}. On average, \fscore{} decreases by \avgfscoreDropInSmallProgramInThirdAblation{} for small languages and by \avgfscoreDropInBigProgramInThirdAblation{} for large languages. These results underscore the critical role of even minimal syntactic cues, such as brackets, in supporting accurate grammar induction by providing essential scaffolding for parse tree generalization and grammar recovery.

Beyond accuracy, efficiency is also affected. The average \runtime{} increases by \avgbuildTimeDropInSmallProgramInThirdAblation{} for small languages and by \avgbuildTimeDropInBigProgramInThirdAblation{} for large languages. Similarly, oracle calls rise by \avgOracleCallIncreaseInSmallProgramInThirdAblation{} and \avgOracleCallIncreaseInBigProgramInThirdAblation{}, while the number of \llm{} calls grows by \avgLLMCallIncreaseInSmallProgramInThirdAblation{} and \avgLLMCallIncreaseInBigProgramInThirdAblation{} for small and large languages, respectively.

\MyPara{Impact of \llm{}-guided Bubble Exploration}
Table~\ref{table:third_and_fourth_ablation} also shows the results of disabling the \llm{}-guided tree generation and reverting to the heuristic-based tree construction strategy used by \treevada{}. This shows that the average \fscore{} drops is very minimal, \avgfscoreDropInSmallProgramInFourthAblation{}. 
\begin{figure}[h!t]
\centering
\begin{subfigure}[t]{0.35\textwidth}
\begin{lstlisting}[basicstyle=\color{teal}\ttfamily\scriptsize,
language=Grammar,
escapeinside={(*@}{@*)}]
start(*@\normalcolor:@*) stmt
stmt_13(*@\normalcolor:@*) complex_stmt stmt_11
(*@\normalcolor|@*) stmt_1 
(*@\normalcolor|@*) stmt_13 stmt_13
complex_stmt(*@\normalcolor:@*) 
  conditional_stmt stmt_15
(*@\normalcolor|@*) complex_stmt stmt_13
conditional_stmt(*@\normalcolor:@*) 
  stmt_2 boolexpr_
(*@\normalcolor|@*) stmt_13 conditional_stmt
stmt_2(*@\normalcolor:@*) stmt_13 stmt_2
(*@\normalcolor|@*) if
stmt_1(*@\normalcolor:@*) stmt_4 ;
(*@\normalcolor|@*) stmt_3 boolexpr_ boolstmt_2
(*@\normalcolor|@*) stmt_13 stmt_1
stmt_4(*@\normalcolor:@*) stmt 
(*@\normalcolor|@*) stmt_13 stmt_4
boolexpr_(*@\normalcolor:@*)  bool_expr_1
(*@\normalcolor|@*) boolexpr_ & boolexpr_
(*@\normalcolor|@*) complex_expr = expr_1 
(*@\normalcolor|@*) unary_op bool_expr_1
\end{lstlisting}
\end{subfigure}
\hfill
\begin{subfigure}[t]{0.28\textwidth}
\begin{lstlisting}[basicstyle=\color{teal}\ttfamily\scriptsize,
language=Grammar,
escapeinside={(*@}{@*)}]
bool_expr_1(*@\normalcolor:@*) boolexpr 
(*@\normalcolor|@*) ~ bool_expr_1
boolexpr(*@\normalcolor:@*) true
(*@\normalcolor|@*) false
(*@\normalcolor|@*) numexpr = expr_1
boolstmt(*@\normalcolor:@*) then
(*@\normalcolor|@*) & boolexpr_ boolstmt
(*@\normalcolor|@*) stmt_15 stmt_1
stmt_11(*@\normalcolor:@*) stmt_4 stmt_7
(*@\normalcolor|@*) stmt_11 stmt_13
stmt_7(*@\normalcolor:@*) else
(*@\normalcolor|@*) stmt_7 stmt_13
stmt_3(*@\normalcolor:@*) while
(*@\normalcolor|@*) stmt_13 stmt_3
boolstmt_2(*@\normalcolor:@*) do
(*@\normalcolor|@*) & boolexpr_ boolstmt_2
expr_1(*@\normalcolor:@*) = complex_numexpr_2
complex_numexpr_2(*@\normalcolor:@*)  numexpr
(*@\normalcolor|@*) L
\end{lstlisting}
\end{subfigure}
\hfill
\begin{subfigure}[t]{0.31\textwidth}
\begin{lstlisting}[basicstyle=\color{teal}\ttfamily\scriptsize,
language=Grammar,
escapeinside={(*@}{@*)}]
numexpr(*@\normalcolor:@*) ( complex_numexpr )
(*@\normalcolor|@*) n
complex_numexpr(*@\normalcolor:@*) expr numexpr
(*@\normalcolor|@*) expr L
expr(*@\normalcolor:@*) L +
(*@\normalcolor|@*) numexpr +
complex_expr(*@\normalcolor:@*) 
  complex_numexpr_2 
(*@\normalcolor|@*) unary_expr 
(*@\normalcolor|@*) boolexpr_ & complex_expr
unary_expr(*@\normalcolor:@*) unary_op L
(*@\normalcolor|@*) unary_op numexpr
unary_op(*@\normalcolor:@*) ~
(*@\normalcolor|@*) unary_op ~
stmt(*@\normalcolor:@*) stmt_13 stmt
(*@\normalcolor|@*) skip
(*@\normalcolor|@*) L expr_1
(*@\normalcolor|@*) complex_stmt stmt_11 stmt
stmt_15(*@\normalcolor:@*) stmt_15 stmt_13
(*@\normalcolor|@*) boolstmt
\end{lstlisting}
\end{subfigure}

\caption{\toolName{} w/o LLM-guided bubble exploration: inferred \textit{while} grammar. Despite achieving a perfect \fscore{}, this grammar is less readable (with over 2x the rules/non-terminals (25 vs 11) and MCC (32 vs 14)) than the corresponding Figure~\ref{fig:four-grammars-for-human-understanding} (d) \toolName{}-inferred grammar.}
\label{list:grammar_nollm_ablation}
\end{figure}

However, the more impact is in the quality of the generated grammars. For example, Figure~\ref{list:grammar_nollm_ablation} shows the output of \toolName{} without LLM-guided bubble exploration. This grammar is significantly larger than the one produced by \Tool{} in Figure~\ref{fig:four-grammars-for-human-understanding} (d). Moreover, other evaluation metrics confirm this difference: the generated grammar achieves an MCC score of 32, compared to 14 with \Tool{}, representing the increase of more than twice. Similarly, the number of rules or non-terminals also doubles, with 25 in the ablated version versus 11 in \Tool{}.

\begin{framed}
\noindent
\emph{
\textbf{\RQToolComponents{}: }
Without LLM-based renaming, grammars lose readability and efficiency; without HDD, accuracy collapses on large programs; without bracket-guided bubbles, even minimal syntactic cues vanish and quality plummets, and without LLM-guided bubble exploration, grammars bloat in size and complexity.
}
\end{framed}

\section{Threats to Validity}
\label{sec:threats}

The \textit{syntactic hypothesis}~\cite{ast-probe, hewitt2019structural} states that the model's hidden states encode a latent parse tree of the input token sequence. We assume that the LLM has been pre-trained on programs written in the target language, such that its \textit{syntactic subspace} encodes sufficient information about the language's structure. \toolName{} would perform best with real seeds, because the LLM can fully leverage it's capacity to identify correct structure. Since most of the programming languages share common structure, the LLMs knowledge can be useful for new or previously unseen language. However, if the seeds are artificially generated under strict constraints and lack naturalness, the LLM might struggle to find structure from those.

\section{Related Work}
\label{sec:related}

Inferring context-free grammars is fundamental for reverse engineering, code understanding, protocol specification, and automated test generation~\cite{stevenson2014survey}. For this problem, \textit{active learning}~\cite{ANGLUIN198787} is a popular technique that has been used for decades. In this setting, the learner has access to a Minimally Adequate Teacher (MAT) that responds to the learner's queries about the target language. Grammar inference tools such as Glade~\cite{bastani2017synthesizing}, Arvada~\cite{kulkarni2021learning}, TreeVada~\cite{arefin2024fast}, and Kedavra~\cite{li2024incremental} adopt this active learning setting where a black-box oracle is available as a MAT. The black-box oracle is usually the parser of the language which can only answer \textit{membership queries} (whether a string belongs to the language). However, the number of membership queries needed grows quickly if the language class is beyond regular~\cite{querycomplexity}.

Traditional white-box and grey-box methods leverage parser internals but are often impractical in real-world settings where only closed-source or remote parsers are available. This has motivated a series of black-box approaches that rely solely on example programs and oracle feedback.

\arvada{} is the pioneer tool for black-box grammar inference problem that starts from flat trees and iteratively performs bubble-and-merge generalizations~\cite{kulkarni2021learning}. While it achieved significant improvements over earlier tools like GLADE, \arvada{} remains limited by non-determinism, high runtime complexity, and susceptibility to over-generalization. Its reliance on repeated runs to explore generalization sequences makes results hard to reproduce and inefficient on larger or more complex languages.
\treevada{} addressed several of \arvada{}’s shortcomings by pre-structuring input programs according to bracket-implied nesting rules, recursively applying generalizations, and eliminating sources of non-determinism~\cite{arefin2024fast}. This design reduced runtime complexity to as low as it can and produced faster, higher-quality grammars with reproducibility in a single run. However, \treevada{} still struggles with readability when grammars become large and remains constrained by its reliance on syntactic nesting heuristics.
More recently, \kedavra{} introduced an incremental grammar inference approach that decomposes input programs into smaller fragments and infers grammars in stages~\cite{li2024incremental}. This approach is intended to improve \fscore{} compared to \arvada{} and \treevada{}. However, our evaluation indicates that \kedavra{} does not consistently achieve these goals. Although it performs well on small programs, it fails on higher-complexity languages, often timing out after 24 hours. Furthermore, the grammars it produces are typically larger and less interpretable, with structural redundancies that obscure readability. These findings show that while \kedavra{} explores an interesting direction toward incremental inference, its practical benefits remain limited.

\section{Conclusions}
\label{sec:conclusion}
We propose \Tool{}, an LLM-guided framework for black-box context-free grammar inference that effectively scales to larger languages. By combining bracket-guided bubble exploration, LLM-driven labeling, and hierarchical delta debugging, \Tool{} improves both accuracy and readability compared to prior approaches. On average, it improves \fscore{} by up to 25pp over \treevada{} while inferring grammars that are more compact and semantically meaningful. Unlike existing techniques, \Tool{} successfully handles challenging languages such as lua, c and mysql without failure or timeout, demonstrating robustness and scalability. These advances make \Tool{} a practical solution for real-world applications in software engineering, security, and reverse engineering, offering both accuracy and interpretability in grammar inference.

\bibliographystyle{ACM-Reference-Format}
\bibliography{ref}
\end{document}

%% file: macros.tex
\usepackage{algorithmic}

\usepackage{textcomp}
\usepackage{xcolor}
\usepackage{minted}
\usepackage{subcaption}

\usepackage{balance}
\usepackage{listings}
\usepackage{xspace}
\usepackage{subcaption}
\usepackage{booktabs}
\usepackage{framed}
\usepackage{forest}
\usepackage{tcolorbox} %
\usepackage{subcaption}
\usetikzlibrary{positioning, arrows.meta, shapes.arrows, shapes.misc}
\usepackage{algorithm}

\definecolor{Gray}{gray}{0.9}
\definecolor{cadmiumgreen}{rgb}{0.0, 0.42, 0.24}
\definecolor{Green}{rgb}{0.6,1,0.8}
\definecolor{Red}{rgb}{0.99, 0.76, 0.8}

\newcommand{\Comment}[1]{}
\definecolor{dkgreen}{rgb}{0,0.6,0}
\definecolor{gray}{rgb}{0.5,0.5,0.5}
\definecolor{mauve}{rgb}{0.58,0,0.82}

\usepackage{numprint}
\npthousandsep{,}

\newcommand{\barebonesLLM}{LLM\textsubscript{BB}}

\newcommand{\treevada}{\textsc{TreeVada}}
\newcommand{\treevadaOrg}{\treevada\textsubscript{orig}}
\newcommand{\kedavra}{\textsc{Kedavra}}
\newcommand{\arvada}{\textsc{Arvada}}
\newcommand{\ailabel}{AI Non-terminal Labeling}
\newcommand{\MyPara}[1]{\noindent\textbf{#1}.}
\newcommand{\hddAbb}{Hierarchical Delta-Debugging}
\newcommand{\hdd}{HDD}
\newcommand{\runtime}{runtime}
\newcommand{\toolNamePlain}{NatGI}%
\newcommand{\toolNameWithoutTreevadaPlain}{\toolNamePlain\textsubscript{Base}}
\newcommand{\toolName}{\textsc{\toolNamePlain}}
\newcommand{\toolNameWithoutTreevada}{\textsc{\toolNameWithoutTreevadaPlain}}
\newcommand{\Tool}{\toolName}
\newcommand{\llm}{LLM}
\newcommand{\fscore}{F1 score}
\newcommand{\precision}{precision}
\newcommand{\recall}{recall}

\newcommand{\Def}[2]{\expandafter\newcommand\csname rmk-#1\endcsname{#2}}
\newcommand{\Use}[1]{\csname rmk-#1\endcsname}
\input{data/def}

\newcommand{\RQResultOnLargeProgram}{RQ1}
\newcommand{\RQResultOnLargeProgramFull}{How does \Tool{} compare with \treevada{} and \kedavra{} on medium to large languages when measured via commonly used grammar inference metrics (i.e., precision, recall, F1 score, and runtime)?}

\newcommand{\RQGrammarUnderstand}{RQ2}
\newcommand{\RQGrammarUnderstandFull}{How does \Tool{} compare with \treevada{} and \kedavra{} in terms of the readability of the tool-inferred grammars?}

\newcommand{\RQToolComponents}{RQ3}
\newcommand{\RQToolComponentsFull}{How do \Tool's main components contribute to its grammar inference performance as measured via commonly used grammar inference metrics (i.e., precision, recall, F1 score, and runtime)?}

\newcommand{\largeProg}{large language}
\newcommand{\fscoreImprovementOverBestBaselineForBigProgram}{78} %
\newcommand{\avgfscoreDropInSmallProgramInFirstAblation}{2pp}
\newcommand{\avgfscoreDropInBigProgramInFirstAblation}{1pp}
\newcommand{\avgbuildTimeDropInSmallProgramInFirstAblation}{79s}
\newcommand{\avgbuildTimeDropInBigProgramInFirstAblation}{428s}
\newcommand{\avgOracleCallIncreaseInSmallProgramInFirstAblation}{9.7}
\newcommand{\avgOracleCallIncreaseInBigProgramInFirstAblation}{10.7}

\newcommand{\avgLLMCallIncreaseInSmallProgramInFirstAblation}{1.7}
\newcommand{\avgLLMCallIncreaseInBigProgramInFirstAblation}{6.74}

\newcommand{\avgfscoreDropInBigProgramInSecondAblation}{35pp}

\newcommand{\avgfscoreDropInSmallProgramInThirdAblation}{16pp}
\newcommand{\avgfscoreDropInBigProgramInThirdAblation}{3pp}

\newcommand{\avgbuildTimeDropInSmallProgramInThirdAblation}{466s}
\newcommand{\avgbuildTimeDropInBigProgramInThirdAblation}{403s}

\newcommand{\avgOracleCallIncreaseInSmallProgramInThirdAblation}{23.3s}
\newcommand{\avgOracleCallIncreaseInBigProgramInThirdAblation}{64.2s}

\newcommand{\avgLLMCallIncreaseInSmallProgramInThirdAblation}{3.5s}
\newcommand{\avgLLMCallIncreaseInBigProgramInThirdAblation}{7.9s}
\newcommand{\LuaavgfscoreDropWhenHDDIsDisabled}{35pp}
\newcommand{\cavgfscoreDropWhenHDDIsDisabled}{12pp}
\newcommand{\mysqlavgfscoreDropWhenHDDIsDisabled}{59pp}

\newcommand{\avgfscoreDropInSmallProgramInFourthAblation}{1pp}

\newcommand{\term}[1]{\textcolor{teal}{\texttt{#1}}}

%% file: data/def.tex
\Def{turtle_ID}{M1}
\Def{turtle_programname}{turtle}
\Def{turtle_program_seed_no}{35}
\Def{turtle_avg_char}{41.09}
\Def{turtle_max_char}{126}
\Def{turtle_avg_tokens}{27.00}
\Def{turtle_max_tokens}{82}
\Def{turtle_branching_factor}{8.28}
\Def{turtle_barebone_precision}{.01}
\Def{turtle_barebone_recall}{.75}
\Def{turtle_barebone_f1}{.02}
\Def{turtle_barebone_build_time}{.00}
\Def{turtle_barebone_oracle_calls}{.00}
\Def{turtle_barebone_llm_calls}{.78}
\Def{turtle_Kedavra_precision}{1.00}
\Def{turtle_Kedavra_recall}{1.00}
\Def{turtle_Kedavra_f1}{1.00}
\Def{turtle_Kedavra_build_time}{78}
\Def{turtle_Kedavra_oracle_calls}{19.0}
\Def{turtle_Kedavra_llm_calls}{-}
\Def{turtle_treevada_false_hdd_false_LLM_true_precision}{.97}
\Def{turtle_treevada_false_hdd_false_LLM_true_recall}{1.00}
\Def{turtle_treevada_false_hdd_false_LLM_true_f1}{.98}
\Def{turtle_treevada_false_hdd_false_LLM_true_build_time}{66}
\Def{turtle_treevada_false_hdd_false_LLM_true_oracle_calls}{17.6}
\Def{turtle_treevada_false_hdd_false_LLM_true_llm_calls}{16.10}
\Def{turtle_treevada_false_hdd_true_LLM_true_precision}{1.00}
\Def{turtle_treevada_false_hdd_true_LLM_true_recall}{.97}
\Def{turtle_treevada_false_hdd_true_LLM_true_f1}{.98}
\Def{turtle_treevada_false_hdd_true_LLM_true_build_time}{66}
\Def{turtle_treevada_false_hdd_true_LLM_true_oracle_calls}{17.6}
\Def{turtle_treevada_false_hdd_true_LLM_true_llm_calls}{16.10}
\Def{turtle_treevada_true_hdd_false_LLM_true_precision}{1.00}
\Def{turtle_treevada_true_hdd_false_LLM_true_recall}{1.00}
\Def{turtle_treevada_true_hdd_false_LLM_true_f1}{1.00}
\Def{turtle_treevada_true_hdd_false_LLM_true_build_time}{99} %
\Def{turtle_treevada_true_hdd_false_LLM_true_oracle_calls}{18.5}
\Def{turtle_treevada_true_hdd_false_LLM_true_llm_calls}{16.30}
\Def{turtle_treevada_true_hdd_true_LLM_true_precision}{1.00}
\Def{turtle_treevada_true_hdd_true_LLM_true_recall}{1.00}
\Def{turtle_treevada_true_hdd_true_LLM_true_f1}{1.00}
\Def{turtle_treevada_true_hdd_true_LLM_true_build_time}{99}
\Def{turtle_treevada_true_hdd_true_LLM_true_oracle_calls}{18.5}
\Def{turtle_treevada_true_hdd_true_LLM_true_llm_calls}{16.30}
\Def{turtle_treevada_original_precision}{.95}
\Def{turtle_treevada_original_recall}{1.00}
\Def{turtle_treevada_original_f1}{.97}
\Def{turtle_treevada_original_build_time}{48}
\Def{turtle_treevada_original_oracle_calls}{16.7}
\Def{turtle_first_ablation_precision}{.94}
\Def{turtle_first_ablation_recall}{1.00}
\Def{turtle_first_ablation_f1}{.97}
\Def{turtle_first_ablation_build_time}{89}
\Def{turtle_first_ablation_oracle_calls}{18.8}
\Def{turtle_first_ablation_llm_calls}{17.00}
\Def{turtle_third_ablation_precision}{.98}
\Def{turtle_third_ablation_recall}{1.00}
\Def{turtle_third_ablation_f1}{.99}
\Def{turtle_third_ablation_build_time}{79}
\Def{turtle_third_ablation_oracle_calls}{17.3}
\Def{turtle_third_ablation_llm_calls}{19.90}
\Def{turtle_fourth_ablation_precision}{1.00}
\Def{turtle_fourth_ablation_recall}{1.00}
\Def{turtle_fourth_ablation_f1}{1.00}
\Def{turtle_fourth_ablation_build_time}{77}
\Def{turtle_fourth_ablation_oracle_calls}{19.6}
\Def{turtle_fourth_ablation_llm_calls}{.00}
\Def{turtle_grammar_lrs}{66}
\Def{turtle_grammar_stat_term}{10}
\Def{turtle_grammar_stat_non_term}{4}
\Def{turtle_grammar_rhs_length}{6.25}
\Def{turtle_grammar_mcc}{1.50}
\Def{turtle_grammar_lat}{8.64}
\Def{turtle_grammar_ltpsm}{8}
\Def{while_ID}{M2}
\Def{while_programname}{while}
\Def{while_program_seed_no}{30}
\Def{while_avg_char}{171.43}
\Def{while_max_char}{793}
\Def{while_avg_tokens}{110.70}
\Def{while_max_tokens}{534}
\Def{while_branching_factor}{24.39}
\Def{while_barebone_precision}{.30}
\Def{while_barebone_recall}{.25}
\Def{while_barebone_f1}{.18}
\Def{while_barebone_build_time}{.00}
\Def{while_barebone_oracle_calls}{.00}
\Def{while_barebone_llm_calls}{3.00}
\Def{while_Kedavra_precision}{1.00}
\Def{while_Kedavra_recall}{1.00}
\Def{while_Kedavra_f1}{1.00}
\Def{while_Kedavra_build_time}{464}
\Def{while_Kedavra_oracle_calls}{123.8}
\Def{while_Kedavra_llm_calls}{-}
\Def{while_treevada_false_hdd_false_LLM_true_precision}{1.00}
\Def{while_treevada_false_hdd_false_LLM_true_recall}{1.00}
\Def{while_treevada_false_hdd_false_LLM_true_f1}{1.00}
\Def{while_treevada_false_hdd_false_LLM_true_build_time}{339}
\Def{while_treevada_false_hdd_false_LLM_true_oracle_calls}{30.9}
\Def{while_treevada_false_hdd_false_LLM_true_llm_calls}{19.80}
\Def{while_treevada_false_hdd_true_LLM_true_precision}{1.00}
\Def{while_treevada_false_hdd_true_LLM_true_recall}{1.00}
\Def{while_treevada_false_hdd_true_LLM_true_f1}{1.00}
\Def{while_treevada_false_hdd_true_LLM_true_build_time}{339}
\Def{while_treevada_false_hdd_true_LLM_true_oracle_calls}{30.9}
\Def{while_treevada_false_hdd_true_LLM_true_llm_calls}{19.80}
\Def{while_treevada_true_hdd_false_LLM_true_precision}{1.00}
\Def{while_treevada_true_hdd_false_LLM_true_recall}{1.00}
\Def{while_treevada_true_hdd_false_LLM_true_f1}{1.00}
\Def{while_treevada_true_hdd_false_LLM_true_build_time}{1,430} %
\Def{while_treevada_true_hdd_false_LLM_true_oracle_calls}{38.3}
\Def{while_treevada_true_hdd_false_LLM_true_llm_calls}{23.20}
\Def{while_treevada_true_hdd_true_LLM_true_precision}{1.00}
\Def{while_treevada_true_hdd_true_LLM_true_recall}{1.00}
\Def{while_treevada_true_hdd_true_LLM_true_f1}{1.00}
\Def{while_treevada_true_hdd_true_LLM_true_build_time}{1,430}
\Def{while_treevada_true_hdd_true_LLM_true_oracle_calls}{38.3}
\Def{while_treevada_true_hdd_true_LLM_true_llm_calls}{23.20}
\Def{while_treevada_original_precision}{1.00}
\Def{while_treevada_original_recall}{1.00}
\Def{while_treevada_original_f1}{1.00}
\Def{while_treevada_original_build_time}{1,030}
\Def{while_treevada_original_oracle_calls}{34.2}
\Def{while_first_ablation_precision}{1.00}
\Def{while_first_ablation_recall}{1.00}
\Def{while_first_ablation_f1}{1.00}
\Def{while_first_ablation_build_time}{1,469}
\Def{while_first_ablation_oracle_calls}{40.7}
\Def{while_first_ablation_llm_calls}{34.50}
\Def{while_third_ablation_precision}{.79}
\Def{while_third_ablation_recall}{.96}
\Def{while_third_ablation_f1}{.86}
\Def{while_third_ablation_build_time}{456}
\Def{while_third_ablation_oracle_calls}{32.3}
\Def{while_third_ablation_llm_calls}{27.10}
\Def{while_fourth_ablation_precision}{1.00}
\Def{while_fourth_ablation_recall}{1.00}
\Def{while_fourth_ablation_f1}{1.00}
\Def{while_fourth_ablation_build_time}{1,420}
\Def{while_fourth_ablation_oracle_calls}{47.1}
\Def{while_fourth_ablation_llm_calls}{.00}
\Def{while_grammar_lrs}{124}
\Def{while_grammar_stat_term}{20}
\Def{while_grammar_stat_non_term}{4}
\Def{while_grammar_rhs_length}{13.00}
\Def{while_grammar_mcc}{2.50}
\Def{while_grammar_lat}{10.14}
\Def{while_grammar_ltpsm}{16}
\Def{lisp_ID}{M3}
\Def{lisp_programname}{lisp}
\Def{lisp_program_seed_no}{30}
\Def{lisp_avg_char}{79.20}
\Def{lisp_max_char}{428}
\Def{lisp_avg_tokens}{75.50}
\Def{lisp_max_tokens}{411}
\Def{lisp_branching_factor}{4.08}
\Def{lisp_barebone_precision}{.19}
\Def{lisp_barebone_recall}{.89}
\Def{lisp_barebone_f1}{.26}
\Def{lisp_barebone_build_time}{.00}
\Def{lisp_barebone_oracle_calls}{.00}
\Def{lisp_barebone_llm_calls}{37.50}
\Def{lisp_Kedavra_precision}{1.00}
\Def{lisp_Kedavra_recall}{1.00}
\Def{lisp_Kedavra_f1}{1.00}
\Def{lisp_Kedavra_build_time}{124}
\Def{lisp_Kedavra_oracle_calls}{15.9}
\Def{lisp_Kedavra_llm_calls}{-}
\Def{lisp_treevada_false_hdd_false_LLM_true_precision}{1.00}
\Def{lisp_treevada_false_hdd_false_LLM_true_recall}{1.00}
\Def{lisp_treevada_false_hdd_false_LLM_true_f1}{1.00}
\Def{lisp_treevada_false_hdd_false_LLM_true_build_time}{337}
\Def{lisp_treevada_false_hdd_false_LLM_true_oracle_calls}{44.0}
\Def{lisp_treevada_false_hdd_false_LLM_true_llm_calls}{13.50}
\Def{lisp_treevada_false_hdd_true_LLM_true_precision}{1.00}
\Def{lisp_treevada_false_hdd_true_LLM_true_recall}{1.00}
\Def{lisp_treevada_false_hdd_true_LLM_true_f1}{1.00}
\Def{lisp_treevada_false_hdd_true_LLM_true_build_time}{337}
\Def{lisp_treevada_false_hdd_true_LLM_true_oracle_calls}{44.0}
\Def{lisp_treevada_false_hdd_true_LLM_true_llm_calls}{13.50}
\Def{lisp_treevada_true_hdd_false_LLM_true_precision}{1.00}
\Def{lisp_treevada_true_hdd_false_LLM_true_recall}{1.00}
\Def{lisp_treevada_true_hdd_false_LLM_true_f1}{1.00}
\Def{lisp_treevada_true_hdd_false_LLM_true_build_time}{380} %
\Def{lisp_treevada_true_hdd_false_LLM_true_oracle_calls}{62.1}
\Def{lisp_treevada_true_hdd_false_LLM_true_llm_calls}{13.60}
\Def{lisp_treevada_true_hdd_true_LLM_true_precision}{1.00}
\Def{lisp_treevada_true_hdd_true_LLM_true_recall}{.99}
\Def{lisp_treevada_true_hdd_true_LLM_true_f1}{1.00}
\Def{lisp_treevada_true_hdd_true_LLM_true_build_time}{380}
\Def{lisp_treevada_true_hdd_true_LLM_true_oracle_calls}{62.1}
\Def{lisp_treevada_true_hdd_true_LLM_true_llm_calls}{13.60}
\Def{lisp_treevada_original_precision}{.97}
\Def{lisp_treevada_original_recall}{1.00}
\Def{lisp_treevada_original_f1}{.99}
\Def{lisp_treevada_original_build_time}{516}
\Def{lisp_treevada_original_oracle_calls}{58.3}
\Def{lisp_first_ablation_precision}{.90}
\Def{lisp_first_ablation_recall}{1.00}
\Def{lisp_first_ablation_f1}{.95}
\Def{lisp_first_ablation_build_time}{326}
\Def{lisp_first_ablation_oracle_calls}{52.7}
\Def{lisp_first_ablation_llm_calls}{13.30}
\Def{lisp_third_ablation_precision}{.99}
\Def{lisp_third_ablation_recall}{1.00}
\Def{lisp_third_ablation_f1}{1.00}
\Def{lisp_third_ablation_build_time}{341}
\Def{lisp_third_ablation_oracle_calls}{44.7}
\Def{lisp_third_ablation_llm_calls}{16.60}
\Def{lisp_fourth_ablation_precision}{.94}
\Def{lisp_fourth_ablation_recall}{1.00}
\Def{lisp_fourth_ablation_f1}{.97}
\Def{lisp_fourth_ablation_build_time}{536}
\Def{lisp_fourth_ablation_oracle_calls}{55.2}
\Def{lisp_fourth_ablation_llm_calls}{.00}
\Def{lisp_grammar_lrs}{84}
\Def{lisp_grammar_stat_term}{7}
\Def{lisp_grammar_stat_non_term}{5}
\Def{lisp_grammar_rhs_length}{5.00}
\Def{lisp_grammar_mcc}{1.40}
\Def{lisp_grammar_lat}{17.12}
\Def{lisp_grammar_ltpsm}{5}
\Def{xml_ID}{M4}
\Def{xml_programname}{xml}
\Def{xml_program_seed_no}{20}
\Def{xml_avg_char}{27.80}
\Def{xml_max_char}{78}
\Def{xml_avg_tokens}{23.20}
\Def{xml_max_tokens}{60}
\Def{xml_branching_factor}{23.20}
\Def{xml_barebone_precision}{.03}
\Def{xml_barebone_recall}{.31}
\Def{xml_barebone_f1}{.04}
\Def{xml_barebone_build_time}{.00}
\Def{xml_barebone_oracle_calls}{.00}
\Def{xml_barebone_llm_calls}{.83}
\Def{xml_Kedavra_precision}{1.00}
\Def{xml_Kedavra_recall}{1.00}
\Def{xml_Kedavra_f1}{1.00}
\Def{xml_Kedavra_build_time}{338}
\Def{xml_Kedavra_oracle_calls}{85.9}
\Def{xml_Kedavra_llm_calls}{-}
\Def{xml_treevada_false_hdd_false_LLM_true_precision}{1.00}
\Def{xml_treevada_false_hdd_false_LLM_true_recall}{.29}
\Def{xml_treevada_false_hdd_false_LLM_true_f1}{.45}
\Def{xml_treevada_false_hdd_false_LLM_true_build_time}{147}
\Def{xml_treevada_false_hdd_false_LLM_true_oracle_calls}{27.7}
\Def{xml_treevada_false_hdd_false_LLM_true_llm_calls}{14.80}
\Def{xml_treevada_false_hdd_true_LLM_true_precision}{.36}
\Def{xml_treevada_false_hdd_true_LLM_true_recall}{1.00}
\Def{xml_treevada_false_hdd_true_LLM_true_f1}{.53}
\Def{xml_treevada_false_hdd_true_LLM_true_build_time}{147}
\Def{xml_treevada_false_hdd_true_LLM_true_oracle_calls}{27.7}
\Def{xml_treevada_false_hdd_true_LLM_true_llm_calls}{14.80}
\Def{xml_treevada_true_hdd_false_LLM_true_precision}{1.00}
\Def{xml_treevada_true_hdd_false_LLM_true_recall}{.66}
\Def{xml_treevada_true_hdd_false_LLM_true_f1}{.76}
\Def{xml_treevada_true_hdd_false_LLM_true_build_time}{190} %
\Def{xml_treevada_true_hdd_false_LLM_true_oracle_calls}{25.7}
\Def{xml_treevada_true_hdd_false_LLM_true_llm_calls}{14.80}
\Def{xml_treevada_true_hdd_true_LLM_true_precision}{.72}
\Def{xml_treevada_true_hdd_true_LLM_true_recall}{1.00}
\Def{xml_treevada_true_hdd_true_LLM_true_f1}{.82}
\Def{xml_treevada_true_hdd_true_LLM_true_build_time}{190}
\Def{xml_treevada_true_hdd_true_LLM_true_oracle_calls}{25.7}
\Def{xml_treevada_true_hdd_true_LLM_true_llm_calls}{14.80}
\Def{xml_treevada_original_precision}{1.00}
\Def{xml_treevada_original_recall}{1.00}
\Def{xml_treevada_original_f1}{1.00}
\Def{xml_treevada_original_build_time}{65}
\Def{xml_treevada_original_oracle_calls}{12.5}
\Def{xml_first_ablation_precision}{.64}
\Def{xml_first_ablation_recall}{.85}
\Def{xml_first_ablation_f1}{.66}
\Def{xml_first_ablation_build_time}{222}
\Def{xml_first_ablation_oracle_calls}{26.5}
\Def{xml_first_ablation_llm_calls}{18.70}
\Def{xml_third_ablation_precision}{1.00}
\Def{xml_third_ablation_recall}{.41}
\Def{xml_third_ablation_f1}{.58}
\Def{xml_third_ablation_build_time}{197}
\Def{xml_third_ablation_oracle_calls}{30.6}
\Def{xml_third_ablation_llm_calls}{23.20}
\Def{xml_fourth_ablation_precision}{1.00}
\Def{xml_fourth_ablation_recall}{1.00}
\Def{xml_fourth_ablation_f1}{1.00}
\Def{xml_fourth_ablation_build_time}{140}
\Def{xml_fourth_ablation_oracle_calls}{20.4}
\Def{xml_fourth_ablation_llm_calls}{.00}
\Def{xml_grammar_lrs}{342}
\Def{xml_grammar_stat_term}{17}
\Def{xml_grammar_stat_non_term}{7}
\Def{xml_grammar_rhs_length}{6.86}
\Def{xml_grammar_mcc}{2.43}
\Def{xml_grammar_lat}{31.00}
\Def{xml_grammar_ltpsm}{8}
\Def{json_ID}{M5}
\Def{json_programname}{json}
\Def{json_program_seed_no}{30}
\Def{json_avg_char}{11.73}
\Def{json_max_char}{90}
\Def{json_avg_tokens}{8.20}
\Def{json_max_tokens}{72}
\Def{json_branching_factor}{3.77}
\Def{json_barebone_precision}{.04}
\Def{json_barebone_recall}{.41}
\Def{json_barebone_f1}{.07}
\Def{json_barebone_build_time}{.00}
\Def{json_barebone_oracle_calls}{.00}
\Def{json_barebone_llm_calls}{1.38}
\Def{json_Kedavra_precision}{1.00}
\Def{json_Kedavra_recall}{.96}
\Def{json_Kedavra_f1}{.98}
\Def{json_Kedavra_build_time}{39}
\Def{json_Kedavra_oracle_calls}{7.4}
\Def{json_Kedavra_llm_calls}{-}
\Def{json_treevada_false_hdd_false_LLM_true_precision}{.98}
\Def{json_treevada_false_hdd_false_LLM_true_recall}{.90}
\Def{json_treevada_false_hdd_false_LLM_true_f1}{.94}
\Def{json_treevada_false_hdd_false_LLM_true_build_time}{104}
\Def{json_treevada_false_hdd_false_LLM_true_oracle_calls}{12.4}
\Def{json_treevada_false_hdd_false_LLM_true_llm_calls}{15.30}
\Def{json_treevada_false_hdd_true_LLM_true_precision}{.90}
\Def{json_treevada_false_hdd_true_LLM_true_recall}{.99}
\Def{json_treevada_false_hdd_true_LLM_true_f1}{.94}
\Def{json_treevada_false_hdd_true_LLM_true_build_time}{104}
\Def{json_treevada_false_hdd_true_LLM_true_oracle_calls}{12.4}
\Def{json_treevada_false_hdd_true_LLM_true_llm_calls}{15.30}
\Def{json_treevada_true_hdd_false_LLM_true_precision}{.96}
\Def{json_treevada_true_hdd_false_LLM_true_recall}{.90}
\Def{json_treevada_true_hdd_false_LLM_true_f1}{.93}
\Def{json_treevada_true_hdd_false_LLM_true_build_time}{96} %
\Def{json_treevada_true_hdd_false_LLM_true_oracle_calls}{13.7}
\Def{json_treevada_true_hdd_false_LLM_true_llm_calls}{16.30}
\Def{json_treevada_true_hdd_true_LLM_true_precision}{.90}
\Def{json_treevada_true_hdd_true_LLM_true_recall}{.98}
\Def{json_treevada_true_hdd_true_LLM_true_f1}{.94}
\Def{json_treevada_true_hdd_true_LLM_true_build_time}{96}
\Def{json_treevada_true_hdd_true_LLM_true_oracle_calls}{13.7}
\Def{json_treevada_true_hdd_true_LLM_true_llm_calls}{16.30}
\Def{json_treevada_original_precision}{.96}
\Def{json_treevada_original_recall}{.90}
\Def{json_treevada_original_f1}{.93}
\Def{json_treevada_original_build_time}{12}
\Def{json_treevada_original_oracle_calls}{9.0}
\Def{json_first_ablation_precision}{.99}
\Def{json_first_ablation_recall}{.90}
\Def{json_first_ablation_f1}{.94}
\Def{json_first_ablation_build_time}{65}
\Def{json_first_ablation_oracle_calls}{13.7}
\Def{json_first_ablation_llm_calls}{15.70}
\Def{json_third_ablation_precision}{.99}
\Def{json_third_ablation_recall}{.90}
\Def{json_third_ablation_f1}{.94}
\Def{json_third_ablation_build_time}{82}
\Def{json_third_ablation_oracle_calls}{12.2}
\Def{json_third_ablation_llm_calls}{19.70}
\Def{json_fourth_ablation_precision}{.96}
\Def{json_fourth_ablation_recall}{.90}
\Def{json_fourth_ablation_f1}{.93}
\Def{json_fourth_ablation_build_time}{22}
\Def{json_fourth_ablation_oracle_calls}{12.0}
\Def{json_fourth_ablation_llm_calls}{.00}
\Def{json_grammar_lrs}{85}
\Def{json_grammar_stat_term}{12}
\Def{json_grammar_stat_non_term}{7}
\Def{json_grammar_rhs_length}{4.29}
\Def{json_grammar_mcc}{1.43}
\Def{json_grammar_lat}{13.23}
\Def{json_grammar_ltpsm}{8}
\Def{tinyc_ID}{M6}
\Def{tinyc_programname}{tinyc}
\Def{tinyc_program_seed_no}{25}
\Def{tinyc_avg_char}{96.16}
\Def{tinyc_max_char}{257}
\Def{tinyc_avg_tokens}{47.24}
\Def{tinyc_max_tokens}{126}
\Def{tinyc_branching_factor}{7.71}
\Def{tinyc_barebone_precision}{.21}
\Def{tinyc_barebone_recall}{.10}
\Def{tinyc_barebone_f1}{.08}
\Def{tinyc_barebone_build_time}{.00}
\Def{tinyc_barebone_oracle_calls}{.00}
\Def{tinyc_barebone_llm_calls}{5.62}
\Def{tinyc_Kedavra_precision}{1.00}
\Def{tinyc_Kedavra_recall}{1.00}
\Def{tinyc_Kedavra_f1}{1.00}
\Def{tinyc_Kedavra_build_time}{47}
\Def{tinyc_Kedavra_oracle_calls}{21.9}
\Def{tinyc_Kedavra_llm_calls}{-}
\Def{tinyc_treevada_false_hdd_false_LLM_true_precision}{.93}
\Def{tinyc_treevada_false_hdd_false_LLM_true_recall}{.40}
\Def{tinyc_treevada_false_hdd_false_LLM_true_f1}{.55}
\Def{tinyc_treevada_false_hdd_false_LLM_true_build_time}{384}
\Def{tinyc_treevada_false_hdd_false_LLM_true_oracle_calls}{63.7}
\Def{tinyc_treevada_false_hdd_false_LLM_true_llm_calls}{30.78}
\Def{tinyc_treevada_false_hdd_true_LLM_true_precision}{.56}
\Def{tinyc_treevada_false_hdd_true_LLM_true_recall}{.94}
\Def{tinyc_treevada_false_hdd_true_LLM_true_f1}{.69}
\Def{tinyc_treevada_false_hdd_true_LLM_true_build_time}{384}
\Def{tinyc_treevada_false_hdd_true_LLM_true_oracle_calls}{63.7}
\Def{tinyc_treevada_false_hdd_true_LLM_true_llm_calls}{30.78}
\Def{tinyc_treevada_true_hdd_false_LLM_true_precision}{.75}
\Def{tinyc_treevada_true_hdd_false_LLM_true_recall}{.75}
\Def{tinyc_treevada_true_hdd_false_LLM_true_f1}{.74}
\Def{tinyc_treevada_true_hdd_false_LLM_true_build_time}{646} %
\Def{tinyc_treevada_true_hdd_false_LLM_true_oracle_calls}{95.9}
\Def{tinyc_treevada_true_hdd_false_LLM_true_llm_calls}{48.10}
\Def{tinyc_treevada_true_hdd_true_LLM_true_precision}{.81}
\Def{tinyc_treevada_true_hdd_true_LLM_true_recall}{.70}
\Def{tinyc_treevada_true_hdd_true_LLM_true_f1}{.75}
\Def{tinyc_treevada_true_hdd_true_LLM_true_build_time}{646}
\Def{tinyc_treevada_true_hdd_true_LLM_true_oracle_calls}{95.9}
\Def{tinyc_treevada_true_hdd_true_LLM_true_llm_calls}{48.10}
\Def{tinyc_treevada_original_precision}{.86}
\Def{tinyc_treevada_original_recall}{.97}
\Def{tinyc_treevada_original_f1}{.91}
\Def{tinyc_treevada_original_build_time}{1,285}
\Def{tinyc_treevada_original_oracle_calls}{164.5}
\Def{tinyc_first_ablation_precision}{.66}
\Def{tinyc_first_ablation_recall}{.81}
\Def{tinyc_first_ablation_f1}{.72}
\Def{tinyc_first_ablation_build_time}{613}
\Def{tinyc_first_ablation_oracle_calls}{103.8}
\Def{tinyc_first_ablation_llm_calls}{36.90}
\Def{tinyc_third_ablation_precision}{.90}
\Def{tinyc_third_ablation_recall}{.58}
\Def{tinyc_third_ablation_f1}{.68}
\Def{tinyc_third_ablation_build_time}{379}
\Def{tinyc_third_ablation_oracle_calls}{57.9}
\Def{tinyc_third_ablation_llm_calls}{43.90}
\Def{tinyc_fourth_ablation_precision}{.49}
\Def{tinyc_fourth_ablation_recall}{.97}
\Def{tinyc_fourth_ablation_f1}{.65}
\Def{tinyc_fourth_ablation_build_time}{782}
\Def{tinyc_fourth_ablation_oracle_calls}{161.5}
\Def{tinyc_fourth_ablation_llm_calls}{.00}
\Def{tinyc_grammar_lrs}{265}
\Def{tinyc_grammar_stat_term}{15}
\Def{tinyc_grammar_stat_non_term}{9}
\Def{tinyc_grammar_rhs_length}{4.67}
\Def{tinyc_grammar_mcc}{1.44}
\Def{tinyc_grammar_lat}{59.69}
\Def{tinyc_grammar_ltpsm}{12}
\Def{c-500_ID}{M7}
\Def{c-500_programname}{c-500}
\Def{c-500_program_seed_no}{10}
\Def{c-500_avg_char}{514.00}
\Def{c-500_max_char}{593}
\Def{c-500_avg_tokens}{244.30}
\Def{c-500_max_tokens}{279}
\Def{c-500_branching_factor}{8.45}
\Def{c-500_barebone_precision}{.09}
\Def{c-500_barebone_recall}{.08}
\Def{c-500_barebone_f1}{.05}
\Def{c-500_barebone_build_time}{.00}
\Def{c-500_barebone_oracle_calls}{.00}
\Def{c-500_barebone_llm_calls}{.75}
\Def{c-500_Kedavra_precision}{1.00}
\Def{c-500_Kedavra_recall}{1.00}
\Def{c-500_Kedavra_f1}{1.00}
\Def{c-500_Kedavra_build_time}{68}
\Def{c-500_Kedavra_oracle_calls}{23.8}
\Def{c-500_Kedavra_llm_calls}{-}
\Def{c-500_treevada_false_hdd_false_LLM_true_precision}{.83}
\Def{c-500_treevada_false_hdd_false_LLM_true_recall}{.07}
\Def{c-500_treevada_false_hdd_false_LLM_true_f1}{.10}
\Def{c-500_treevada_false_hdd_false_LLM_true_build_time}{378}
\Def{c-500_treevada_false_hdd_false_LLM_true_oracle_calls}{70.5}
\Def{c-500_treevada_false_hdd_false_LLM_true_llm_calls}{19.20}
\Def{c-500_treevada_false_hdd_true_LLM_true_precision}{.14}
\Def{c-500_treevada_false_hdd_true_LLM_true_recall}{.83}
\Def{c-500_treevada_false_hdd_true_LLM_true_f1}{.19}
\Def{c-500_treevada_false_hdd_true_LLM_true_build_time}{378}
\Def{c-500_treevada_false_hdd_true_LLM_true_oracle_calls}{70.5}
\Def{c-500_treevada_false_hdd_true_LLM_true_llm_calls}{19.20}
\Def{c-500_treevada_true_hdd_false_LLM_true_precision}{.75}
\Def{c-500_treevada_true_hdd_false_LLM_true_recall}{.73}
\Def{c-500_treevada_true_hdd_false_LLM_true_f1}{.73}
\Def{c-500_treevada_true_hdd_false_LLM_true_build_time}{2,205}
\Def{c-500_treevada_true_hdd_false_LLM_true_oracle_calls}{189.0}
\Def{c-500_treevada_true_hdd_false_LLM_true_llm_calls}{16.90}
\Def{c-500_treevada_true_hdd_true_LLM_true_precision}{.89}
\Def{c-500_treevada_true_hdd_true_LLM_true_recall}{.77}
\Def{c-500_treevada_true_hdd_true_LLM_true_f1}{.82}
\Def{c-500_treevada_true_hdd_true_LLM_true_build_time}{2,227}
\Def{c-500_treevada_true_hdd_true_LLM_true_oracle_calls}{189.0}
\Def{c-500_treevada_true_hdd_true_LLM_true_llm_calls}{16.90}
\Def{c-500_treevada_original_precision}{.93}
\Def{c-500_treevada_original_recall}{.53}
\Def{c-500_treevada_original_f1}{.67}
\Def{c-500_treevada_original_build_time}{3,164}
\Def{c-500_treevada_original_oracle_calls}{175.5}
\Def{c-500_first_ablation_precision}{.67}
\Def{c-500_first_ablation_recall}{.84}
\Def{c-500_first_ablation_f1}{.71}
\Def{c-500_first_ablation_build_time}{3,058}
\Def{c-500_first_ablation_oracle_calls}{279.7}
\Def{c-500_first_ablation_llm_calls}{38.20}
\Def{c-500_third_ablation_precision}{.94}
\Def{c-500_third_ablation_recall}{.03}
\Def{c-500_third_ablation_f1}{.05}
\Def{c-500_third_ablation_build_time}{223}
\Def{c-500_third_ablation_oracle_calls}{55.7}
\Def{c-500_third_ablation_llm_calls}{12.10}
\Def{c-500_fourth_ablation_precision}{.81}
\Def{c-500_fourth_ablation_recall}{.38}
\Def{c-500_fourth_ablation_f1}{.52}
\Def{c-500_fourth_ablation_build_time}{4,834}
\Def{c-500_fourth_ablation_oracle_calls}{447.1}
\Def{c-500_fourth_ablation_llm_calls}{.00}
\Def{c-500_grammar_lrs}{265}
\Def{c-500_grammar_stat_term}{15}
\Def{c-500_grammar_stat_non_term}{9}
\Def{c-500_grammar_rhs_length}{4.67}
\Def{c-500_grammar_mcc}{1.44}
\Def{c-500_grammar_lat}{59.69}
\Def{c-500_grammar_ltpsm}{12}
\Def{tiny_ID}{M8}
\Def{tiny_programname}{tiny}
\Def{tiny_program_seed_no}{10}
\Def{tiny_avg_char}{39.70}
\Def{tiny_max_char}{76}
\Def{tiny_avg_tokens}{15.50}
\Def{tiny_max_tokens}{31}
\Def{tiny_branching_factor}{15.50}
\Def{tiny_barebone_precision}{.41}
\Def{tiny_barebone_recall}{.17}
\Def{tiny_barebone_f1}{.23}
\Def{tiny_barebone_build_time}{.00}
\Def{tiny_barebone_oracle_calls}{.00}
\Def{tiny_barebone_llm_calls}{78.00}
\Def{tiny_Kedavra_precision}{1.00}
\Def{tiny_Kedavra_recall}{.51}
\Def{tiny_Kedavra_f1}{.67}
\Def{tiny_Kedavra_build_time}{16}
\Def{tiny_Kedavra_oracle_calls}{5.9}
\Def{tiny_Kedavra_llm_calls}{-}
\Def{tiny_treevada_false_hdd_false_LLM_true_precision}{.85}
\Def{tiny_treevada_false_hdd_false_LLM_true_recall}{.21}
\Def{tiny_treevada_false_hdd_false_LLM_true_f1}{.31}
\Def{tiny_treevada_false_hdd_false_LLM_true_build_time}{167}
\Def{tiny_treevada_false_hdd_false_LLM_true_oracle_calls}{12.1}
\Def{tiny_treevada_false_hdd_false_LLM_true_llm_calls}{29.80}
\Def{tiny_treevada_false_hdd_true_LLM_true_precision}{.26}
\Def{tiny_treevada_false_hdd_true_LLM_true_recall}{.82}
\Def{tiny_treevada_false_hdd_true_LLM_true_f1}{.37}
\Def{tiny_treevada_false_hdd_true_LLM_true_build_time}{167}
\Def{tiny_treevada_false_hdd_true_LLM_true_oracle_calls}{12.1}
\Def{tiny_treevada_false_hdd_true_LLM_true_llm_calls}{29.80}
\Def{tiny_treevada_true_hdd_false_LLM_true_precision}{.93}
\Def{tiny_treevada_true_hdd_false_LLM_true_recall}{.59}
\Def{tiny_treevada_true_hdd_false_LLM_true_f1}{.72}
\Def{tiny_treevada_true_hdd_false_LLM_true_build_time}{229}
\Def{tiny_treevada_true_hdd_false_LLM_true_oracle_calls}{27.1}
\Def{tiny_treevada_true_hdd_false_LLM_true_llm_calls}{30.30}
\Def{tiny_treevada_true_hdd_true_LLM_true_precision}{.60}
\Def{tiny_treevada_true_hdd_true_LLM_true_recall}{.56}
\Def{tiny_treevada_true_hdd_true_LLM_true_f1}{.58}
\Def{tiny_treevada_true_hdd_true_LLM_true_build_time}{229}
\Def{tiny_treevada_true_hdd_true_LLM_true_oracle_calls}{27.1}
\Def{tiny_treevada_true_hdd_true_LLM_true_llm_calls}{30.30}
\Def{tiny_treevada_original_precision}{.95}
\Def{tiny_treevada_original_recall}{.68}
\Def{tiny_treevada_original_f1}{.79}
\Def{tiny_treevada_original_build_time}{53}
\Def{tiny_treevada_original_oracle_calls}{29.3}
\Def{tiny_first_ablation_precision}{.94}
\Def{tiny_first_ablation_recall}{.65}
\Def{tiny_first_ablation_f1}{.76}
\Def{tiny_first_ablation_build_time}{250}
\Def{tiny_first_ablation_oracle_calls}{31.6}
\Def{tiny_first_ablation_llm_calls}{30.50}
\Def{tiny_third_ablation_precision}{.97}
\Def{tiny_third_ablation_recall}{.27}
\Def{tiny_third_ablation_f1}{.41}
\Def{tiny_third_ablation_build_time}{218}
\Def{tiny_third_ablation_oracle_calls}{15.9}
\Def{tiny_third_ablation_llm_calls}{31.40}
\Def{tiny_fourth_ablation_precision}{.59}
\Def{tiny_fourth_ablation_recall}{.67}
\Def{tiny_fourth_ablation_f1}{.63}
\Def{tiny_fourth_ablation_build_time}{78}
\Def{tiny_fourth_ablation_oracle_calls}{29.6}
\Def{tiny_fourth_ablation_llm_calls}{.00}
\Def{tiny_grammar_lrs}{64}
\Def{tiny_grammar_stat_term}{10}
\Def{tiny_grammar_stat_non_term}{13}
\Def{tiny_grammar_rhs_length}{2.69}
\Def{tiny_grammar_mcc}{.69}
\Def{tiny_grammar_lat}{27.10}
\Def{tiny_grammar_ltpsm}{8}
\Def{minic_ID}{M9}
\Def{minic_programname}{minic}
\Def{minic_program_seed_no}{10}
\Def{minic_avg_char}{107.00}
\Def{minic_max_char}{368}
\Def{minic_avg_tokens}{45.70}
\Def{minic_max_tokens}{158}
\Def{minic_branching_factor}{5.33}
\Def{minic_barebone_precision}{.30}
\Def{minic_barebone_recall}{.69}
\Def{minic_barebone_f1}{.39}
\Def{minic_barebone_build_time}{.00}
\Def{minic_barebone_oracle_calls}{.00}
\Def{minic_barebone_llm_calls}{27.00}
\Def{minic_Kedavra_precision}{1.00}
\Def{minic_Kedavra_recall}{.48}
\Def{minic_Kedavra_f1}{.65}
\Def{minic_Kedavra_build_time}{2,708}
\Def{minic_Kedavra_oracle_calls}{31.6}
\Def{minic_Kedavra_llm_calls}{-}
\Def{minic_treevada_false_hdd_false_LLM_true_precision}{.96}
\Def{minic_treevada_false_hdd_false_LLM_true_recall}{.33}
\Def{minic_treevada_false_hdd_false_LLM_true_f1}{.47}
\Def{minic_treevada_false_hdd_false_LLM_true_build_time}{1,753}
\Def{minic_treevada_false_hdd_false_LLM_true_oracle_calls}{30.4}
\Def{minic_treevada_false_hdd_false_LLM_true_llm_calls}{32.60}
\Def{minic_treevada_false_hdd_true_LLM_true_precision}{.33}
\Def{minic_treevada_false_hdd_true_LLM_true_recall}{.94}
\Def{minic_treevada_false_hdd_true_LLM_true_f1}{.47}
\Def{minic_treevada_false_hdd_true_LLM_true_build_time}{1,753}
\Def{minic_treevada_false_hdd_true_LLM_true_oracle_calls}{30.4}
\Def{minic_treevada_false_hdd_true_LLM_true_llm_calls}{32.60}
\Def{minic_treevada_true_hdd_false_LLM_true_precision}{.95}
\Def{minic_treevada_true_hdd_false_LLM_true_recall}{.45}
\Def{minic_treevada_true_hdd_false_LLM_true_f1}{.61}
\Def{minic_treevada_true_hdd_false_LLM_true_build_time}{3,634}
\Def{minic_treevada_true_hdd_false_LLM_true_oracle_calls}{71.3}
\Def{minic_treevada_true_hdd_false_LLM_true_llm_calls}{36.60}
\Def{minic_treevada_true_hdd_true_LLM_true_precision}{.45}
\Def{minic_treevada_true_hdd_true_LLM_true_recall}{.84}
\Def{minic_treevada_true_hdd_true_LLM_true_f1}{.58}
\Def{minic_treevada_true_hdd_true_LLM_true_build_time}{3,634}
\Def{minic_treevada_true_hdd_true_LLM_true_oracle_calls}{71.3}
\Def{minic_treevada_true_hdd_true_LLM_true_llm_calls}{36.60}
\Def{minic_treevada_original_precision}{.63}
\Def{minic_treevada_original_recall}{.62}
\Def{minic_treevada_original_f1}{.63}
\Def{minic_treevada_original_build_time}{5,337}
\Def{minic_treevada_original_oracle_calls}{93.6}
\Def{minic_first_ablation_precision}{.75}
\Def{minic_first_ablation_recall}{.47}
\Def{minic_first_ablation_f1}{.57}
\Def{minic_first_ablation_build_time}{3,641}
\Def{minic_first_ablation_oracle_calls}{71.9}
\Def{minic_first_ablation_llm_calls}{30.50}
\Def{minic_third_ablation_precision}{.91}
\Def{minic_third_ablation_recall}{.23}
\Def{minic_third_ablation_f1}{.35}
\Def{minic_third_ablation_build_time}{2,299}
\Def{minic_third_ablation_oracle_calls}{43.6}
\Def{minic_third_ablation_llm_calls}{56.00}
\Def{minic_fourth_ablation_precision}{.99}
\Def{minic_fourth_ablation_recall}{.57}
\Def{minic_fourth_ablation_f1}{.72}
\Def{minic_fourth_ablation_build_time}{4,259}
\Def{minic_fourth_ablation_oracle_calls}{95.4}
\Def{minic_fourth_ablation_llm_calls}{.00}
\Def{minic_grammar_lrs}{355}
\Def{minic_grammar_stat_term}{22}
\Def{minic_grammar_stat_non_term}{14}
\Def{minic_grammar_rhs_length}{6.36}
\Def{minic_grammar_mcc}{2.00}
\Def{minic_grammar_lat}{68.04}
\Def{minic_grammar_ltpsm}{17}
\Def{curl_ID}{M10}
\Def{curl_programname}{curl}
\Def{curl_program_seed_no}{25}
\Def{curl_avg_char}{22.08}
\Def{curl_max_char}{39}
\Def{curl_avg_tokens}{14.80}
\Def{curl_max_tokens}{25}
\Def{curl_branching_factor}{14.80}
\Def{curl_barebone_precision}{.44}
\Def{curl_barebone_recall}{.61}
\Def{curl_barebone_f1}{.51}
\Def{curl_barebone_build_time}{.00}
\Def{curl_barebone_oracle_calls}{.00}
\Def{curl_barebone_llm_calls}{72.60}
\Def{curl_Kedavra_precision}{.96}
\Def{curl_Kedavra_recall}{.08}
\Def{curl_Kedavra_f1}{.14}
\Def{curl_Kedavra_build_time}{84}
\Def{curl_Kedavra_oracle_calls}{19.6}
\Def{curl_Kedavra_llm_calls}{-}
\Def{curl_treevada_false_hdd_false_LLM_true_precision}{.91}
\Def{curl_treevada_false_hdd_false_LLM_true_recall}{.71}
\Def{curl_treevada_false_hdd_false_LLM_true_f1}{.79}
\Def{curl_treevada_false_hdd_false_LLM_true_build_time}{294}
\Def{curl_treevada_false_hdd_false_LLM_true_oracle_calls}{74.6}
\Def{curl_treevada_false_hdd_false_LLM_true_llm_calls}{31.00}
\Def{curl_treevada_false_hdd_true_LLM_true_precision}{.78}
\Def{curl_treevada_false_hdd_true_LLM_true_recall}{.78}
\Def{curl_treevada_false_hdd_true_LLM_true_f1}{.78}
\Def{curl_treevada_false_hdd_true_LLM_true_build_time}{294}
\Def{curl_treevada_false_hdd_true_LLM_true_oracle_calls}{74.6}
\Def{curl_treevada_false_hdd_true_LLM_true_llm_calls}{31.00}
\Def{curl_treevada_true_hdd_false_LLM_true_precision}{.72}
\Def{curl_treevada_true_hdd_false_LLM_true_recall}{.80}
\Def{curl_treevada_true_hdd_false_LLM_true_f1}{.74}
\Def{curl_treevada_true_hdd_false_LLM_true_build_time}{283}
\Def{curl_treevada_true_hdd_false_LLM_true_oracle_calls}{74.6}
\Def{curl_treevada_true_hdd_false_LLM_true_llm_calls}{32.40}
\Def{curl_treevada_true_hdd_true_LLM_true_precision}{.82}
\Def{curl_treevada_true_hdd_true_LLM_true_recall}{.70}
\Def{curl_treevada_true_hdd_true_LLM_true_f1}{.74}
\Def{curl_treevada_true_hdd_true_LLM_true_build_time}{283}
\Def{curl_treevada_true_hdd_true_LLM_true_oracle_calls}{74.6}
\Def{curl_treevada_true_hdd_true_LLM_true_llm_calls}{32.40}
\Def{curl_treevada_original_precision}{.96}
\Def{curl_treevada_original_recall}{.62}
\Def{curl_treevada_original_f1}{.76}
\Def{curl_treevada_original_build_time}{103}
\Def{curl_treevada_original_oracle_calls}{45.1}
\Def{curl_first_ablation_precision}{.60}
\Def{curl_first_ablation_recall}{.90}
\Def{curl_first_ablation_f1}{.71}
\Def{curl_first_ablation_build_time}{271}
\Def{curl_first_ablation_oracle_calls}{73.8}
\Def{curl_first_ablation_llm_calls}{30.20}
\Def{curl_third_ablation_precision}{.75}
\Def{curl_third_ablation_recall}{.83}
\Def{curl_third_ablation_f1}{.79}
\Def{curl_third_ablation_build_time}{278}
\Def{curl_third_ablation_oracle_calls}{73.0}
\Def{curl_third_ablation_llm_calls}{34.10}
\Def{curl_fourth_ablation_precision}{.55}
\Def{curl_fourth_ablation_recall}{.81}
\Def{curl_fourth_ablation_f1}{.66}
\Def{curl_fourth_ablation_build_time}{160}
\Def{curl_fourth_ablation_oracle_calls}{53.1}
\Def{curl_fourth_ablation_llm_calls}{.00}
\Def{curl_grammar_lrs}{n/a}
\Def{curl_grammar_stat_term}{n/a}
\Def{curl_grammar_stat_non_term}{n/a}
\Def{curl_grammar_rhs_length}{n/a}
\Def{curl_grammar_mcc}{n/a}
\Def{curl_grammar_lat}{n/a}
\Def{curl_grammar_ltpsm}{n/a}
\Def{c_ID}{M11}
\Def{c_programname}{c}
\Def{c_program_seed_no}{10}
\Def{c_avg_char}{110.50}
\Def{c_max_char}{986}
\Def{c_avg_tokens}{72.90}
\Def{c_max_tokens}{665}
\Def{c_branching_factor}{11.47}
\Def{c_barebone_precision}{.08}
\Def{c_barebone_recall}{.94}
\Def{c_barebone_f1}{.15}
\Def{c_barebone_build_time}{.00}
\Def{c_barebone_oracle_calls}{.00}
\Def{c_barebone_llm_calls}{.00}
\Def{c_Kedavra_precision}{.00}
\Def{c_Kedavra_recall}{.00}
\Def{c_Kedavra_f1}{.00}
\Def{c_Kedavra_build_time}{86,400}
\Def{c_Kedavra_oracle_calls}{t/o}
\Def{c_Kedavra_llm_calls}{-}
\Def{c_treevada_false_hdd_false_LLM_true_precision}{.94}
\Def{c_treevada_false_hdd_false_LLM_true_recall}{.28}
\Def{c_treevada_false_hdd_false_LLM_true_f1}{.42}
\Def{c_treevada_false_hdd_false_LLM_true_build_time}{471}
\Def{c_treevada_false_hdd_false_LLM_true_oracle_calls}{59.2}
\Def{c_treevada_false_hdd_false_LLM_true_llm_calls}{29.90}
\Def{c_treevada_false_hdd_true_LLM_true_precision}{.34}
\Def{c_treevada_false_hdd_true_LLM_true_recall}{.93}
\Def{c_treevada_false_hdd_true_LLM_true_f1}{.49}
\Def{c_treevada_false_hdd_true_LLM_true_build_time}{471}
\Def{c_treevada_false_hdd_true_LLM_true_oracle_calls}{59.2}
\Def{c_treevada_false_hdd_true_LLM_true_llm_calls}{29.90}
\Def{c_treevada_true_hdd_false_LLM_true_precision}{.91}
\Def{c_treevada_true_hdd_false_LLM_true_recall}{.28}
\Def{c_treevada_true_hdd_false_LLM_true_f1}{.42}
\Def{c_treevada_true_hdd_false_LLM_true_build_time}{713} %
\Def{c_treevada_true_hdd_false_LLM_true_oracle_calls}{78.7}
\Def{c_treevada_true_hdd_false_LLM_true_llm_calls}{22.20}
\Def{c_treevada_true_hdd_true_LLM_true_precision}{.39}
\Def{c_treevada_true_hdd_true_LLM_true_recall}{.90}
\Def{c_treevada_true_hdd_true_LLM_true_f1}{.54}
\Def{c_treevada_true_hdd_true_LLM_true_build_time}{713}
\Def{c_treevada_true_hdd_true_LLM_true_oracle_calls}{78.7}
\Def{c_treevada_true_hdd_true_LLM_true_llm_calls}{22.20}
\Def{c_treevada_original_precision}{.88}
\Def{c_treevada_original_recall}{.23}
\Def{c_treevada_original_f1}{.36}
\Def{c_treevada_original_build_time}{360}
\Def{c_treevada_original_oracle_calls}{48.8}
\Def{c_first_ablation_precision}{.89}
\Def{c_first_ablation_recall}{.34}
\Def{c_first_ablation_f1}{.49}
\Def{c_first_ablation_build_time}{1,845}
\Def{c_first_ablation_oracle_calls}{90.4}
\Def{c_first_ablation_llm_calls}{35.11}
\Def{c_third_ablation_precision}{.92}
\Def{c_third_ablation_recall}{.34}
\Def{c_third_ablation_f1}{.49}
\Def{c_third_ablation_build_time}{510}
\Def{c_third_ablation_oracle_calls}{60.6}
\Def{c_third_ablation_llm_calls}{31.90}
\Def{c_fourth_ablation_precision}{.91}
\Def{c_fourth_ablation_recall}{.43}
\Def{c_fourth_ablation_f1}{.58}
\Def{c_fourth_ablation_build_time}{552}
\Def{c_fourth_ablation_oracle_calls}{55.5}
\Def{c_fourth_ablation_llm_calls}{.00}
\Def{lua_grammar_lrs}{2868}
\Def{lua_grammar_stat_term}{61}
\Def{lua_grammar_stat_non_term}{37}
\Def{lua_grammar_rhs_length}{5.87}
\Def{lua_grammar_mcc}{2.35}
\Def{lua_grammar_lat}{991.84}
\Def{lua_grammar_ltpsm}{52}
\Def{lua_ID}{M12}
\Def{lua_programname}{lua}
\Def{lua_program_seed_no}{10}
\Def{lua_avg_char}{111.70}
\Def{lua_max_char}{406}
\Def{lua_avg_tokens}{46.70}
\Def{lua_max_tokens}{158}
\Def{lua_branching_factor}{10.85}
\Def{lua_barebone_precision}{.27}
\Def{lua_barebone_recall}{.00}
\Def{lua_barebone_f1}{.01}
\Def{lua_barebone_build_time}{.00}
\Def{lua_barebone_oracle_calls}{.00}
\Def{lua_barebone_llm_calls}{2.33}
\Def{lua_Kedavra_precision}{.00}
\Def{lua_Kedavra_recall}{.00}
\Def{lua_Kedavra_f1}{.00}
\Def{lua_Kedavra_build_time}{86,400}
\Def{lua_Kedavra_oracle_calls}{t/o}
\Def{lua_Kedavra_llm_calls}{-}
\Def{lua_treevada_false_hdd_false_LLM_true_precision}{.44}
\Def{lua_treevada_false_hdd_false_LLM_true_recall}{.09}
\Def{lua_treevada_false_hdd_false_LLM_true_f1}{.15}
\Def{lua_treevada_false_hdd_false_LLM_true_build_time}{594}
\Def{lua_treevada_false_hdd_false_LLM_true_oracle_calls}{98.1}
\Def{lua_treevada_false_hdd_false_LLM_true_llm_calls}{56.40}
\Def{lua_treevada_false_hdd_true_LLM_true_precision}{.53}
\Def{lua_treevada_false_hdd_true_LLM_true_recall}{.53}
\Def{lua_treevada_false_hdd_true_LLM_true_f1}{.53}
\Def{lua_treevada_false_hdd_true_LLM_true_build_time}{594}
\Def{lua_treevada_false_hdd_true_LLM_true_oracle_calls}{98.1}
\Def{lua_treevada_false_hdd_true_LLM_true_llm_calls}{56.40}
\Def{lua_treevada_true_hdd_false_LLM_true_precision}{.46}
\Def{lua_treevada_true_hdd_false_LLM_true_recall}{.10}
\Def{lua_treevada_true_hdd_false_LLM_true_f1}{.17}
\Def{lua_treevada_true_hdd_false_LLM_true_build_time}{1,125} %
\Def{lua_treevada_true_hdd_false_LLM_true_oracle_calls}{208.7}
\Def{lua_treevada_true_hdd_false_LLM_true_llm_calls}{57.60}
\Def{lua_treevada_true_hdd_true_LLM_true_precision}{.53}
\Def{lua_treevada_true_hdd_true_LLM_true_recall}{.52}
\Def{lua_treevada_true_hdd_true_LLM_true_f1}{.52}
\Def{lua_treevada_true_hdd_true_LLM_true_build_time}{1,125}
\Def{lua_treevada_true_hdd_true_LLM_true_oracle_calls}{208.7}
\Def{lua_treevada_true_hdd_true_LLM_true_llm_calls}{57.60}
\Def{lua_treevada_original_precision}{.64}
\Def{lua_treevada_original_recall}{.09}
\Def{lua_treevada_original_f1}{.16}
\Def{lua_treevada_original_build_time}{535}
\Def{lua_treevada_original_oracle_calls}{105.2}
\Def{lua_first_ablation_precision}{.52}
\Def{lua_first_ablation_recall}{.54}
\Def{lua_first_ablation_f1}{.52}
\Def{lua_first_ablation_build_time}{1,114}
\Def{lua_first_ablation_oracle_calls}{197.9}
\Def{lua_first_ablation_llm_calls}{55.90}
\Def{lua_third_ablation_precision}{.50}
\Def{lua_third_ablation_recall}{.51}
\Def{lua_third_ablation_f1}{.50}
\Def{lua_third_ablation_build_time}{682}
\Def{lua_third_ablation_oracle_calls}{99.2}
\Def{lua_third_ablation_llm_calls}{58.30}
\Def{lua_fourth_ablation_precision}{.52}
\Def{lua_fourth_ablation_recall}{.53}
\Def{lua_fourth_ablation_f1}{.52}
\Def{lua_fourth_ablation_build_time}{958}
\Def{lua_fourth_ablation_oracle_calls}{182.5}
\Def{lua_fourth_ablation_llm_calls}{.00}
\Def{c_grammar_lrs}{4675}
\Def{c_grammar_stat_term}{108}
\Def{c_grammar_stat_non_term}{87}
\Def{c_grammar_rhs_length}{6.03}
\Def{c_grammar_mcc}{2.75}
\Def{c_grammar_lat}{647.93}
\Def{c_grammar_ltpsm}{106}
\Def{mysql_ID}{M13}
\Def{mysql_programname}{mysql}
\Def{mysql_program_seed_no}{10}
\Def{mysql_avg_char}{99.20}
\Def{mysql_max_char}{402}
\Def{mysql_avg_tokens}{39.10}
\Def{mysql_max_tokens}{178}
\Def{mysql_branching_factor}{19.37}
\Def{mysql_barebone_precision}{.30}
\Def{mysql_barebone_recall}{.25}
\Def{mysql_barebone_f1}{.24}
\Def{mysql_barebone_build_time}{.00}
\Def{mysql_barebone_oracle_calls}{.00}
\Def{mysql_barebone_llm_calls}{17.00}
\Def{mysql_Kedavra_precision}{.00}
\Def{mysql_Kedavra_recall}{.00}
\Def{mysql_Kedavra_f1}{.00}
\Def{mysql_Kedavra_build_time}{86,400}
\Def{mysql_Kedavra_oracle_calls}{t/o}
\Def{mysql_Kedavra_llm_calls}{-}
\Def{mysql_treevada_false_hdd_false_LLM_true_precision}{.80}
\Def{mysql_treevada_false_hdd_false_LLM_true_recall}{.01}
\Def{mysql_treevada_false_hdd_false_LLM_true_f1}{.02}
\Def{mysql_treevada_false_hdd_false_LLM_true_build_time}{1,638}
\Def{mysql_treevada_false_hdd_false_LLM_true_oracle_calls}{108.8}
\Def{mysql_treevada_false_hdd_false_LLM_true_llm_calls}{30.30}
\Def{mysql_treevada_false_hdd_true_LLM_true_precision}{.47}
\Def{mysql_treevada_false_hdd_true_LLM_true_recall}{.87}
\Def{mysql_treevada_false_hdd_true_LLM_true_f1}{.59}
\Def{mysql_treevada_false_hdd_true_LLM_true_build_time}{1,638}
\Def{mysql_treevada_false_hdd_true_LLM_true_oracle_calls}{108.8}
\Def{mysql_treevada_false_hdd_true_LLM_true_llm_calls}{30.30}
\Def{mysql_treevada_true_hdd_false_LLM_true_precision}{.69}
\Def{mysql_treevada_true_hdd_false_LLM_true_recall}{.03}
\Def{mysql_treevada_true_hdd_false_LLM_true_f1}{.06}
\Def{mysql_treevada_true_hdd_false_LLM_true_build_time}{2,595}
\Def{mysql_treevada_true_hdd_false_LLM_true_oracle_calls}{197.9}
\Def{mysql_treevada_true_hdd_false_LLM_true_llm_calls}{25.30}
\Def{mysql_treevada_true_hdd_true_LLM_true_precision}{.58}
\Def{mysql_treevada_true_hdd_true_LLM_true_recall}{.75}
\Def{mysql_treevada_true_hdd_true_LLM_true_f1}{.65}
\Def{mysql_treevada_true_hdd_true_LLM_true_build_time}{2,595}
\Def{mysql_treevada_true_hdd_true_LLM_true_oracle_calls}{197.9}
\Def{mysql_treevada_true_hdd_true_LLM_true_llm_calls}{25.30}
\Def{mysql_treevada_original_precision}{.65}
\Def{mysql_treevada_original_recall}{.32}
\Def{mysql_treevada_original_f1}{.43}
\Def{mysql_treevada_original_build_time}{1,390}
\Def{mysql_treevada_original_oracle_calls}{87.9}
\Def{mysql_first_ablation_precision}{.75}
\Def{mysql_first_ablation_recall}{.58}
\Def{mysql_first_ablation_f1}{.65}
\Def{mysql_first_ablation_build_time}{2,760}
\Def{mysql_first_ablation_oracle_calls}{229.2}
\Def{mysql_first_ablation_llm_calls}{32.50}
\Def{mysql_third_ablation_precision}{.86}
\Def{mysql_third_ablation_recall}{.52}
\Def{mysql_third_ablation_f1}{.63}
\Def{mysql_third_ablation_build_time}{1,854}
\Def{mysql_third_ablation_oracle_calls}{133.1}
\Def{mysql_third_ablation_llm_calls}{38.50}
\Def{mysql_fourth_ablation_precision}{.70}
\Def{mysql_fourth_ablation_recall}{.58}
\Def{mysql_fourth_ablation_f1}{.63}
\Def{mysql_fourth_ablation_build_time}{1,716}
\Def{mysql_fourth_ablation_oracle_calls}{134.9}
\Def{mysql_fourth_ablation_llm_calls}{.00}
\Def{mysql_grammar_lrs}{n/a}
\Def{mysql_grammar_stat_term}{1028}
\Def{mysql_grammar_stat_non_term}{319}
\Def{mysql_grammar_rhs_length}{n/a}
\Def{mysql_grammar_mcc}{n/a}
\Def{mysql_grammar_lat}{n/a}
\Def{mysql_grammar_ltpsm}{n/a}
\Def{Avg-Small-Program_ID}{M14}
\Def{Avg-Small-Program_programname}{Lo\textsubscript{avg}}
\Def{Avg-Small-Program_program_seed_no}{0}
\Def{Avg-Small-Program_avg_char}{.00}
\Def{Avg-Small-Program_max_char}{0}
\Def{Avg-Small-Program_avg_tokens}{.00}
\Def{Avg-Small-Program_max_tokens}{0}
\Def{Avg-Small-Program_branching_factor}{.00}
\Def{Avg-Small-Program_barebone_precision}{.20}
\Def{Avg-Small-Program_barebone_recall}{.43}
\Def{Avg-Small-Program_barebone_f1}{.18}
\Def{Avg-Small-Program_barebone_build_time}{.00}
\Def{Avg-Small-Program_barebone_oracle_calls}{.00}
\Def{Avg-Small-Program_barebone_llm_calls}{22.75}
\Def{Avg-Small-Program_Kedavra_precision}{1.00}
\Def{Avg-Small-Program_Kedavra_recall}{.80}
\Def{Avg-Small-Program_Kedavra_f1}{.84}
\Def{Avg-Small-Program_Kedavra_build_time}{397}
\Def{Avg-Small-Program_Kedavra_oracle_calls}{35.5}
\Def{Avg-Small-Program_Kedavra_llm_calls}{-}
\Def{Avg-Small-Program_treevada_false_hdd_false_LLM_true_precision}{.94}
\Def{Avg-Small-Program_treevada_false_hdd_false_LLM_true_recall}{.59}
\Def{Avg-Small-Program_treevada_false_hdd_false_LLM_true_f1}{.66}
\Def{Avg-Small-Program_treevada_false_hdd_false_LLM_true_build_time}{397}
\Def{Avg-Small-Program_treevada_false_hdd_false_LLM_true_oracle_calls}{38.4}
\Def{Avg-Small-Program_treevada_false_hdd_false_LLM_true_llm_calls}{22.29}
\Def{Avg-Small-Program_treevada_false_hdd_true_LLM_true_precision}{.59}
\Def{Avg-Small-Program_treevada_false_hdd_true_LLM_true_recall}{.93}
\Def{Avg-Small-Program_treevada_false_hdd_true_LLM_true_f1}{.69}
\Def{Avg-Small-Program_treevada_false_hdd_true_LLM_true_build_time}{397}
\Def{Avg-Small-Program_treevada_false_hdd_true_LLM_true_oracle_calls}{38.4}
\Def{Avg-Small-Program_treevada_false_hdd_true_LLM_true_llm_calls}{22.29}
\Def{Avg-Small-Program_treevada_true_hdd_false_LLM_true_precision}{.90}
\Def{Avg-Small-Program_treevada_true_hdd_false_LLM_true_recall}{.64}
\Def{Avg-Small-Program_treevada_true_hdd_false_LLM_true_f1}{.82}
\Def{Avg-Small-Program_treevada_true_hdd_false_LLM_true_build_time}{921}
\Def{Avg-Small-Program_treevada_true_hdd_false_LLM_true_oracle_calls}{61.6}
\Def{Avg-Small-Program_treevada_true_hdd_false_LLM_true_llm_calls}{24.85}
\Def{Avg-Small-Program_treevada_true_hdd_true_LLM_true_precision}{.75}
\Def{Avg-Small-Program_treevada_true_hdd_true_LLM_true_recall}{.86}
\Def{Avg-Small-Program_treevada_true_hdd_true_LLM_true_f1}{.82}
\Def{Avg-Small-Program_treevada_true_hdd_true_LLM_true_build_time}{921}
\Def{Avg-Small-Program_treevada_true_hdd_true_LLM_true_oracle_calls}{61.6}
\Def{Avg-Small-Program_treevada_true_hdd_true_LLM_true_llm_calls}{24.85}
\Def{Avg-Small-Program_treevada_original_precision}{.92}
\Def{Avg-Small-Program_treevada_original_recall}{.83}
\Def{Avg-Small-Program_treevada_original_f1}{.87}
\Def{Avg-Small-Program_treevada_original_build_time}{1161}
\Def{Avg-Small-Program_treevada_original_oracle_calls}{63.9}
\Def{Avg-Small-Program_first_ablation_precision}{.79}
\Def{Avg-Small-Program_first_ablation_recall}{.84}
\Def{Avg-Small-Program_first_ablation_f1}{.80}
\Def{Avg-Small-Program_first_ablation_build_time}{1,000}
\Def{Avg-Small-Program_first_ablation_oracle_calls}{71.3}
\Def{Avg-Small-Program_first_ablation_llm_calls}{26.55}
\Def{Avg-Small-Program_third_ablation_precision}{.88}
\Def{Avg-Small-Program_third_ablation_recall}{.58}
\Def{Avg-Small-Program_third_ablation_f1}{.66}
\Def{Avg-Small-Program_third_ablation_build_time}{455}
\Def{Avg-Small-Program_third_ablation_oracle_calls}{38.3}
\Def{Avg-Small-Program_third_ablation_llm_calls}{28.40}
\Def{Avg-Small-Program_fourth_ablation_precision}{.80}
\Def{Avg-Small-Program_fourth_ablation_recall}{.83}
\Def{Avg-Small-Program_fourth_ablation_f1}{.81}
\Def{Avg-Small-Program_fourth_ablation_build_time}{1,231}
\Def{Avg-Small-Program_fourth_ablation_oracle_calls}{94.1}
\Def{Avg-Small-Program_fourth_ablation_llm_calls}{.00}
\Def{Avg-Small-Program_grammar_lrs}{0}
\Def{Avg-Small-Program_grammar_stat_term}{0}
\Def{Avg-Small-Program_grammar_stat_non_term}{0}
\Def{Avg-Small-Program_grammar_rhs_length}{.00}
\Def{Avg-Small-Program_grammar_mcc}{.00}
\Def{Avg-Small-Program_grammar_lat}{.00}
\Def{Avg-Small-Program_grammar_ltpsm}{0}
\Def{Avg-Big-Program_ID}{M15}
\Def{Avg-Big-Program_programname}{Hi\textsubscript{avg}}
\Def{Avg-Big-Program_program_seed_no}{0}
\Def{Avg-Big-Program_avg_char}{.00}
\Def{Avg-Big-Program_max_char}{0}
\Def{Avg-Big-Program_avg_tokens}{.00}
\Def{Avg-Big-Program_max_tokens}{0}
\Def{Avg-Big-Program_branching_factor}{.00}
\Def{Avg-Big-Program_barebone_precision}{.22}
\Def{Avg-Big-Program_barebone_recall}{.40}
\Def{Avg-Big-Program_barebone_f1}{.13}
\Def{Avg-Big-Program_barebone_build_time}{.00}
\Def{Avg-Big-Program_barebone_oracle_calls}{.00}
\Def{Avg-Big-Program_barebone_llm_calls}{6.44}
\Def{Avg-Big-Program_Kedavra_precision}{.00}
\Def{Avg-Big-Program_Kedavra_recall}{.78}
\Def{Avg-Big-Program_Kedavra_f1}{.00}
\Def{Avg-Big-Program_Kedavra_build_time}{86,400}
\Def{Avg-Big-Program_Kedavra_oracle_calls}{t/o}
\Def{Avg-Big-Program_Kedavra_llm_calls}{-}
\Def{Avg-Big-Program_treevada_false_hdd_false_LLM_true_precision}{.73}
\Def{Avg-Big-Program_treevada_false_hdd_false_LLM_true_recall}{6.44}
\Def{Avg-Big-Program_treevada_false_hdd_false_LLM_true_f1}{.20}
\Def{Avg-Big-Program_treevada_false_hdd_false_LLM_true_build_time}{901}
\Def{Avg-Big-Program_treevada_false_hdd_false_LLM_true_oracle_calls}{88.7}
\Def{Avg-Big-Program_treevada_false_hdd_false_LLM_true_llm_calls}{38.87}
\Def{Avg-Big-Program_treevada_false_hdd_true_LLM_true_precision}{.45}
\Def{Avg-Big-Program_treevada_false_hdd_true_LLM_true_recall}{.78}
\Def{Avg-Big-Program_treevada_false_hdd_true_LLM_true_f1}{.54}
\Def{Avg-Big-Program_treevada_false_hdd_true_LLM_true_build_time}{901}
\Def{Avg-Big-Program_treevada_false_hdd_true_LLM_true_oracle_calls}{88.7}
\Def{Avg-Big-Program_treevada_false_hdd_true_LLM_true_llm_calls}{38.87}
\Def{Avg-Big-Program_treevada_true_hdd_false_LLM_true_precision}{.69}
\Def{Avg-Big-Program_treevada_true_hdd_false_LLM_true_recall}{.14}
\Def{Avg-Big-Program_treevada_true_hdd_false_LLM_true_f1}{.22}
\Def{Avg-Big-Program_treevada_true_hdd_false_LLM_true_build_time}{1,478}
\Def{Avg-Big-Program_treevada_true_hdd_false_LLM_true_oracle_calls}{161.8}
\Def{Avg-Big-Program_treevada_true_hdd_false_LLM_true_llm_calls}{35.03}
\Def{Avg-Big-Program_treevada_true_hdd_true_LLM_true_precision}{.50}
\Def{Avg-Big-Program_treevada_true_hdd_true_LLM_true_recall}{.72}
\Def{Avg-Big-Program_treevada_true_hdd_true_LLM_true_f1}{.57}
\Def{Avg-Big-Program_treevada_true_hdd_true_LLM_true_build_time}{1,478}
\Def{Avg-Big-Program_treevada_true_hdd_true_LLM_true_oracle_calls}{161.8}
\Def{Avg-Big-Program_treevada_true_hdd_true_LLM_true_llm_calls}{35.03}
\Def{Avg-Big-Program_treevada_original_precision}{.72}
\Def{Avg-Big-Program_treevada_original_recall}{.21}
\Def{Avg-Big-Program_treevada_original_f1}{.32}
\Def{Avg-Big-Program_treevada_original_build_time}{1154}
\Def{Avg-Big-Program_treevada_original_oracle_calls}{80.6}
\Def{Avg-Big-Program_first_ablation_precision}{.72}
\Def{Avg-Big-Program_first_ablation_recall}{.49}
\Def{Avg-Big-Program_first_ablation_f1}{.56}
\Def{Avg-Big-Program_first_ablation_build_time}{1,906}
\Def{Avg-Big-Program_first_ablation_oracle_calls}{172.5}
\Def{Avg-Big-Program_first_ablation_llm_calls}{41.17}
\Def{Avg-Big-Program_third_ablation_precision}{.76}
\Def{Avg-Big-Program_third_ablation_recall}{.46}
\Def{Avg-Big-Program_third_ablation_f1}{.54}
\Def{Avg-Big-Program_third_ablation_build_time}{1,015}
\Def{Avg-Big-Program_third_ablation_oracle_calls}{97.6}
\Def{Avg-Big-Program_third_ablation_llm_calls}{42.90}
\Def{Avg-Big-Program_fourth_ablation_precision}{.71}
\Def{Avg-Big-Program_fourth_ablation_recall}{.51}
\Def{Avg-Big-Program_fourth_ablation_f1}{.58}
\Def{Avg-Big-Program_fourth_ablation_build_time}{1,075}
\Def{Avg-Big-Program_fourth_ablation_oracle_calls}{124.3}
\Def{Avg-Big-Program_fourth_ablation_llm_calls}{.00}
\Def{Avg-Big-Program_grammar_lrs}{0}
\Def{Avg-Big-Program_grammar_stat_term}{0}
\Def{Avg-Big-Program_grammar_stat_non_term}{0}
\Def{Avg-Big-Program_grammar_rhs_length}{.00}
\Def{Avg-Big-Program_grammar_mcc}{.00}
\Def{Avg-Big-Program_grammar_lat}{.00}
\Def{Avg-Big-Program_grammar_ltpsm}{0}

%% file: tables/seed_stat.tex
\begin{table}[h!t]
\centering
\caption{Low-complexity (top) and high-complexity (bottom) languages with their grammar (left), language (middle), and input (seed) programs' (post pre-tokenization) complexity (right); 
T~=~terminals; 
NT~=~non-terminals; 
RHS~=~avg. right-hand side length; 
MCC~=~avg. right-hand side branches;
LRS~=~LR automaton states;
LAT~=~lookahead metric; 
LTPSM~=~max terminal pairs;
P~=~seed program count;
C\textsubscript{avg}/C\textsubscript{max}~=~avg/max characters; 
T\textsubscript{avg}/T\textsubscript{max}~=~avg/max tokens; 
B~=~branch factor from the pre-structured trees.}
\label{table:seed_stat}
\resizebox{\textwidth}{!}{%
\begin{tabular}{l|rrN{2}{1}N{1}{1}|rN{3}{1}r|r|N{3}{1}r|N{3}{1}r|N{2}{1}}
\toprule
 & \multicolumn{7}{c|}{Grammar \& Language} & \multicolumn{6}{c}{Seed Programs}  \\
 & \multicolumn{1}{c}{T} 
 & \multicolumn{1}{c}{NT} 
 & \multicolumn{1}{c}{RHS} 
 & \multicolumn{1}{c|}{MCC} 
 & \multicolumn{1}{c}{LRS} 
 & \multicolumn{1}{c}{LAT} 
 & \multicolumn{1}{c|}{LTPSM} 
 & \multicolumn{1}{c|}{P} 
 & \multicolumn{1}{c}{C\textsubscript{avg}} 
 & \multicolumn{1}{c|}{C\textsubscript{max}} 
 & \multicolumn{1}{c}{T\textsubscript{avg}} 
 & \multicolumn{1}{c|}{T\textsubscript{max}} 
 & \multicolumn{1}{c}{B} \\
\midrule
 \Use{turtle_programname}  & \Use{turtle_grammar_stat_term} & \Use{turtle_grammar_stat_non_term} & \Use{turtle_grammar_rhs_length} & \Use{turtle_grammar_mcc}  & \Use{turtle_grammar_lrs} & \Use{turtle_grammar_lat} & \Use{turtle_grammar_ltpsm} & \Use{turtle_program_seed_no}  & \Use{turtle_avg_char} & \Use{turtle_max_char} & \Use{turtle_avg_tokens} & \Use{turtle_max_tokens} & \Use{turtle_branching_factor}\\
 \Use{while_programname}  & \Use{while_grammar_stat_term} & \Use{while_grammar_stat_non_term} & \Use{while_grammar_rhs_length} & \Use{while_grammar_mcc}  & \Use{while_grammar_lrs} & \Use{while_grammar_lat} & \Use{while_grammar_ltpsm} & \Use{while_program_seed_no}  & \Use{while_avg_char} & \Use{while_max_char} & \Use{while_avg_tokens} & \Use{while_max_tokens} & \Use{while_branching_factor}\\
 \Use{lisp_programname}  & \Use{lisp_grammar_stat_term} & \Use{lisp_grammar_stat_non_term} & \Use{lisp_grammar_rhs_length} & \Use{lisp_grammar_mcc}  & \Use{lisp_grammar_lrs} & \Use{lisp_grammar_lat} & \Use{lisp_grammar_ltpsm} & \Use{lisp_program_seed_no}  & \Use{lisp_avg_char} & \Use{lisp_max_char} & \Use{lisp_avg_tokens} & \Use{lisp_max_tokens} & \Use{lisp_branching_factor}\\
 \Use{xml_programname}  & \Use{xml_grammar_stat_term} & \Use{xml_grammar_stat_non_term} & \Use{xml_grammar_rhs_length} & \Use{xml_grammar_mcc}  & \Use{xml_grammar_lrs} & \Use{xml_grammar_lat} & \Use{xml_grammar_ltpsm} & \Use{xml_program_seed_no}  & \Use{xml_avg_char} & \Use{xml_max_char} & \Use{xml_avg_tokens} & \Use{xml_max_tokens} & \Use{xml_branching_factor}\\
 \Use{json_programname}  & \Use{json_grammar_stat_term} & \Use{json_grammar_stat_non_term} & \Use{json_grammar_rhs_length} & \Use{json_grammar_mcc}  & \Use{json_grammar_lrs} & \Use{json_grammar_lat} & \Use{json_grammar_ltpsm} & \Use{json_program_seed_no}  & \Use{json_avg_char} & \Use{json_max_char} & \Use{json_avg_tokens} & \Use{json_max_tokens} & \Use{json_branching_factor}\\
 \Use{tinyc_programname}  & \Use{tinyc_grammar_stat_term} & \Use{tinyc_grammar_stat_non_term} & \Use{tinyc_grammar_rhs_length} & \Use{tinyc_grammar_mcc}  & \Use{tinyc_grammar_lrs} & \Use{tinyc_grammar_lat} & \Use{tinyc_grammar_ltpsm} & \Use{tinyc_program_seed_no}  & \Use{tinyc_avg_char} & \Use{tinyc_max_char} & \Use{tinyc_avg_tokens} & \Use{tinyc_max_tokens} & \Use{tinyc_branching_factor}\\
 \Use{c-500_programname}  & \Use{c-500_grammar_stat_term} & \Use{c-500_grammar_stat_non_term} & \Use{c-500_grammar_rhs_length} & \Use{c-500_grammar_mcc}  & \Use{c-500_grammar_lrs} & \Use{c-500_grammar_lat} & \Use{c-500_grammar_ltpsm} & \Use{c-500_program_seed_no}  & \Use{c-500_avg_char} & \Use{c-500_max_char} & \Use{c-500_avg_tokens} & \Use{c-500_max_tokens} & \Use{c-500_branching_factor}\\
 \Use{tiny_programname}  & \Use{tiny_grammar_stat_term} & \Use{tiny_grammar_stat_non_term} & \Use{tiny_grammar_rhs_length} & \Use{tiny_grammar_mcc}  & \Use{tiny_grammar_lrs} & \Use{tiny_grammar_lat} & \Use{tiny_grammar_ltpsm} & \Use{tiny_program_seed_no}  & \Use{tiny_avg_char} & \Use{tiny_max_char} & \Use{tiny_avg_tokens} & \Use{tiny_max_tokens} & \Use{tiny_branching_factor}\\
 \Use{minic_programname}  & \Use{minic_grammar_stat_term} & \Use{minic_grammar_stat_non_term} & \Use{minic_grammar_rhs_length} & \Use{minic_grammar_mcc}  & \Use{minic_grammar_lrs} & \Use{minic_grammar_lat} & \Use{minic_grammar_ltpsm} & \Use{minic_program_seed_no}  & \Use{minic_avg_char} & \Use{minic_max_char} & \Use{minic_avg_tokens} & \Use{minic_max_tokens} & \Use{minic_branching_factor}\\
 \Use{curl_programname}  & \Use{curl_grammar_stat_term} & \Use{curl_grammar_stat_non_term} & \Use{curl_grammar_rhs_length} & \Use{curl_grammar_mcc}  & \Use{curl_grammar_lrs} & \Use{curl_grammar_lat} & \Use{curl_grammar_ltpsm} & \Use{curl_program_seed_no}  & \Use{curl_avg_char} & \Use{curl_max_char} & \Use{curl_avg_tokens} & \Use{curl_max_tokens} & \Use{curl_branching_factor}\\
\midrule
\Use{lua_programname}  & \Use{lua_grammar_stat_term} & \Use{lua_grammar_stat_non_term} & \Use{lua_grammar_rhs_length} & \Use{lua_grammar_mcc}  & \Use{lua_grammar_lrs} & \Use{lua_grammar_lat} & \Use{lua_grammar_ltpsm} & \Use{lua_program_seed_no}  & \Use{lua_avg_char} & \Use{lua_max_char} & \Use{lua_avg_tokens} & \Use{lua_max_tokens} & \Use{lua_branching_factor}\\
 \Use{c_programname}  & \Use{c_grammar_stat_term} & \Use{c_grammar_stat_non_term} & \Use{c_grammar_rhs_length} & \Use{c_grammar_mcc}  & \Use{c_grammar_lrs} & \Use{c_grammar_lat} & \Use{c_grammar_ltpsm} & \Use{c_program_seed_no}  & \Use{c_avg_char} & \Use{c_max_char} & \Use{c_avg_tokens} & \Use{c_max_tokens} & \Use{c_branching_factor}\\
 \Use{mysql_programname}  & \Use{mysql_grammar_stat_term} & \Use{mysql_grammar_stat_non_term} & \Use{mysql_grammar_rhs_length} & \Use{mysql_grammar_mcc}  & \Use{mysql_grammar_lrs} & \Use{mysql_grammar_lat} & \Use{mysql_grammar_ltpsm} & \Use{mysql_program_seed_no}  & \Use{mysql_avg_char} & \Use{mysql_max_char} & \Use{mysql_avg_tokens} & \Use{mysql_max_tokens} & \Use{mysql_branching_factor}\\
\bottomrule
\end{tabular}
}
\end{table}

%% file: tables/paper-main-result-table.tex
\begin{table*}
\centering
\caption{Accuracy of LLM-based techniques (average of 10 runs) and \treevada{} and \kedavra{} baselines: 
\barebonesLLM~=~Barebone LLM;
p~=~precision;  
r~=~recall; 
f1~=~\fscore{}; 
t~=~time in seconds; 
t/o~=~timeout (24h); 
bold~=~best \fscore{};
Lo\textsubscript{avg}/Hi\textsubscript{avg}~=~average of low/high language complexity group.
}
\label{table:accuracy}
\resizebox{\textwidth}{!}{%
\begin{tabular}{l|r r r|r r r r| r r r r|r r r r|r r r r}
\toprule
 &  \multicolumn{3}{c|}{\barebonesLLM{}} & \multicolumn{4}{c|}{\treevadaOrg{}} & \multicolumn{4}{c|}{\kedavra{}} &  \multicolumn{4}{c|}{\toolNameWithoutTreevadaPlain{}}  & \multicolumn{4}{c}{\toolName{}} \\
\multicolumn{1}{c|}{} & \multicolumn{1}{c}{p} & \multicolumn{1}{c}{r} & \multicolumn{1}{c|}{f1} & \multicolumn{1}{c}{p} & \multicolumn{1}{c}{r} & \multicolumn{1}{c}{f1} & \multicolumn{1}{c|}{t[s]} & \multicolumn{1}{c}{p} & \multicolumn{1}{c}{r} & \multicolumn{1}{c}{f1} & \multicolumn{1}{c|}{t[s]} & \multicolumn{1}{c}{p} & \multicolumn{1}{c}{r} & \multicolumn{1}{c}{f1} & \multicolumn{1}{c|}{t[s]} & \multicolumn{1}{c}{p} & \multicolumn{1}{c}{r} & \multicolumn{1}{c}{f1} & \multicolumn{1}{c}{t[s]} \\
\midrule
 \Use{turtle_programname} & \Use{turtle_barebone_precision}  & \Use{turtle_barebone_recall} & \Use{turtle_barebone_f1}  & \Use{turtle_treevada_original_precision}  & \Use{turtle_treevada_original_recall} & \Use{turtle_treevada_original_f1} & \Use{turtle_treevada_original_build_time} & \Use{turtle_Kedavra_precision}  & \Use{turtle_Kedavra_recall} & \textbf{\Use{turtle_Kedavra_f1}} & \Use{turtle_Kedavra_build_time} & \Use{turtle_treevada_false_hdd_true_LLM_true_precision} & \Use{turtle_treevada_false_hdd_true_LLM_true_recall} & \Use{turtle_treevada_false_hdd_true_LLM_true_f1} & \Use{turtle_treevada_false_hdd_true_LLM_true_build_time} & \Use{turtle_treevada_true_hdd_true_LLM_true_precision} & \Use{turtle_treevada_true_hdd_true_LLM_true_recall} & \textbf{\Use{turtle_treevada_true_hdd_true_LLM_true_f1}} & \Use{turtle_treevada_true_hdd_true_LLM_true_build_time} \\
 \Use{while_programname} & \Use{while_barebone_precision}  & \Use{while_barebone_recall} & \Use{while_barebone_f1}  & \Use{while_treevada_original_precision}  & \Use{while_treevada_original_recall} & \textbf{\Use{while_treevada_original_f1}} & \Use{while_treevada_original_build_time} & \Use{while_Kedavra_precision}  & \Use{while_Kedavra_recall} & \textbf{\Use{while_Kedavra_f1}} & \Use{while_Kedavra_build_time} & \Use{while_treevada_false_hdd_true_LLM_true_precision} & \Use{while_treevada_false_hdd_true_LLM_true_recall} & \textbf{\Use{while_treevada_false_hdd_true_LLM_true_f1}} & \Use{while_treevada_false_hdd_true_LLM_true_build_time} & \Use{while_treevada_true_hdd_true_LLM_true_precision} & \Use{while_treevada_true_hdd_true_LLM_true_recall} & \textbf{\Use{while_treevada_true_hdd_true_LLM_true_f1}} & \Use{while_treevada_true_hdd_true_LLM_true_build_time} \\
 \Use{lisp_programname} & \Use{lisp_barebone_precision}  & \Use{lisp_barebone_recall} & \Use{lisp_barebone_f1}  & \Use{lisp_treevada_original_precision}  & \Use{lisp_treevada_original_recall} & \Use{lisp_treevada_original_f1} & \Use{lisp_treevada_original_build_time} & \Use{lisp_Kedavra_precision}  & \Use{lisp_Kedavra_recall} & \textbf{\Use{lisp_Kedavra_f1}} & \Use{lisp_Kedavra_build_time} & \Use{lisp_treevada_false_hdd_true_LLM_true_precision} & \Use{lisp_treevada_false_hdd_true_LLM_true_recall} & \textbf{\Use{lisp_treevada_false_hdd_true_LLM_true_f1}} & \Use{lisp_treevada_false_hdd_true_LLM_true_build_time} & \Use{lisp_treevada_true_hdd_true_LLM_true_precision} & \Use{lisp_treevada_true_hdd_true_LLM_true_recall} & \textbf{\Use{lisp_treevada_true_hdd_true_LLM_true_f1}} & \Use{lisp_treevada_true_hdd_true_LLM_true_build_time} \\
 \Use{xml_programname} & \Use{xml_barebone_precision}  & \Use{xml_barebone_recall} & \Use{xml_barebone_f1}  & \Use{xml_treevada_original_precision}  & \Use{xml_treevada_original_recall} & \textbf{\Use{xml_treevada_original_f1}} & \Use{xml_treevada_original_build_time} & \Use{xml_Kedavra_precision}  & \Use{xml_Kedavra_recall} & \textbf{\Use{xml_Kedavra_f1}} & \Use{xml_Kedavra_build_time} & \Use{xml_treevada_false_hdd_true_LLM_true_precision} & \Use{xml_treevada_false_hdd_true_LLM_true_recall} & \Use{xml_treevada_false_hdd_true_LLM_true_f1} & \Use{xml_treevada_false_hdd_true_LLM_true_build_time} & \Use{xml_treevada_true_hdd_true_LLM_true_precision} & \Use{xml_treevada_true_hdd_true_LLM_true_recall} & \Use{xml_treevada_true_hdd_true_LLM_true_f1} & \Use{xml_treevada_true_hdd_true_LLM_true_build_time} \\
 \Use{json_programname} & \Use{json_barebone_precision}  & \Use{json_barebone_recall} & \Use{json_barebone_f1}  & \Use{json_treevada_original_precision}  & \Use{json_treevada_original_recall} & \Use{json_treevada_original_f1} & \Use{json_treevada_original_build_time} & \Use{json_Kedavra_precision}  & \Use{json_Kedavra_recall} & \textbf{\Use{json_Kedavra_f1}} & \Use{json_Kedavra_build_time} & \Use{json_treevada_false_hdd_true_LLM_true_precision} & \Use{json_treevada_false_hdd_true_LLM_true_recall} & \Use{json_treevada_false_hdd_true_LLM_true_f1} & \Use{json_treevada_false_hdd_true_LLM_true_build_time} & \Use{json_treevada_true_hdd_true_LLM_true_precision} & \Use{json_treevada_true_hdd_true_LLM_true_recall} & \Use{json_treevada_true_hdd_true_LLM_true_f1} & \Use{json_treevada_true_hdd_true_LLM_true_build_time} \\
 \Use{tinyc_programname} & \Use{tinyc_barebone_precision}  & \Use{tinyc_barebone_recall} & \Use{tinyc_barebone_f1}  & \Use{tinyc_treevada_original_precision}  & \Use{tinyc_treevada_original_recall} & \Use{tinyc_treevada_original_f1} & \Use{tinyc_treevada_original_build_time} & \Use{tinyc_Kedavra_precision}  & \Use{tinyc_Kedavra_recall} & \textbf{\Use{tinyc_Kedavra_f1}} & \Use{tinyc_Kedavra_build_time} & \Use{tinyc_treevada_false_hdd_true_LLM_true_precision} & \Use{tinyc_treevada_false_hdd_true_LLM_true_recall} & \Use{tinyc_treevada_false_hdd_true_LLM_true_f1} & \Use{tinyc_treevada_false_hdd_true_LLM_true_build_time} & \Use{tinyc_treevada_true_hdd_true_LLM_true_precision} & \Use{tinyc_treevada_true_hdd_true_LLM_true_recall} & \Use{tinyc_treevada_true_hdd_true_LLM_true_f1} & \Use{tinyc_treevada_true_hdd_true_LLM_true_build_time} \\
 \Use{c-500_programname} & \Use{c-500_barebone_precision}  & \Use{c-500_barebone_recall} & \Use{c-500_barebone_f1}  & \Use{c-500_treevada_original_precision}  & \Use{c-500_treevada_original_recall} & \Use{c-500_treevada_original_f1} & \Use{c-500_treevada_original_build_time} & \Use{c-500_Kedavra_precision}  & \Use{c-500_Kedavra_recall} & \textbf{\Use{c-500_Kedavra_f1}} & \Use{c-500_Kedavra_build_time} & \Use{c-500_treevada_false_hdd_true_LLM_true_precision} & \Use{c-500_treevada_false_hdd_true_LLM_true_recall} & \Use{c-500_treevada_false_hdd_true_LLM_true_f1} & \Use{c-500_treevada_false_hdd_true_LLM_true_build_time} & \Use{c-500_treevada_true_hdd_true_LLM_true_precision} & \Use{c-500_treevada_true_hdd_true_LLM_true_recall} & \Use{c-500_treevada_true_hdd_true_LLM_true_f1} & \Use{c-500_treevada_true_hdd_true_LLM_true_build_time} \\
 \Use{tiny_programname} & \Use{tiny_barebone_precision}  & \Use{tiny_barebone_recall} & \Use{tiny_barebone_f1}  & \Use{tiny_treevada_original_precision}  & \Use{tiny_treevada_original_recall} & \textbf{\Use{tiny_treevada_original_f1}} & \Use{tiny_treevada_original_build_time} & \Use{tiny_Kedavra_precision}  & \Use{tiny_Kedavra_recall} & \Use{tiny_Kedavra_f1} & \Use{tiny_Kedavra_build_time} & \Use{tiny_treevada_false_hdd_true_LLM_true_precision} & \Use{tiny_treevada_false_hdd_true_LLM_true_recall} & \Use{tiny_treevada_false_hdd_true_LLM_true_f1} & \Use{tiny_treevada_false_hdd_true_LLM_true_build_time} & \Use{tiny_treevada_true_hdd_true_LLM_true_precision} & \Use{tiny_treevada_true_hdd_true_LLM_true_recall} & \Use{tiny_treevada_true_hdd_true_LLM_true_f1} & \Use{tiny_treevada_true_hdd_true_LLM_true_build_time} \\
 \Use{minic_programname} & \Use{minic_barebone_precision}  & \Use{minic_barebone_recall} & \Use{minic_barebone_f1}  & \Use{minic_treevada_original_precision}  & \Use{minic_treevada_original_recall} & \Use{minic_treevada_original_f1} & \Use{minic_treevada_original_build_time} & \Use{minic_Kedavra_precision}  & \Use{minic_Kedavra_recall} & \textbf{\Use{minic_Kedavra_f1}} & \Use{minic_Kedavra_build_time} & \Use{minic_treevada_false_hdd_true_LLM_true_precision} & \Use{minic_treevada_false_hdd_true_LLM_true_recall} & \Use{minic_treevada_false_hdd_true_LLM_true_f1} & \Use{minic_treevada_false_hdd_true_LLM_true_build_time} & \Use{minic_treevada_true_hdd_true_LLM_true_precision} & \Use{minic_treevada_true_hdd_true_LLM_true_recall} & \Use{minic_treevada_true_hdd_true_LLM_true_f1} & \Use{minic_treevada_true_hdd_true_LLM_true_build_time} \\
 \Use{curl_programname} & \Use{curl_barebone_precision}  & \Use{curl_barebone_recall} & \Use{curl_barebone_f1}  & \Use{curl_treevada_original_precision}  & \Use{curl_treevada_original_recall} & \Use{curl_treevada_original_f1} & \Use{curl_treevada_original_build_time} & \Use{curl_Kedavra_precision}  & \Use{curl_Kedavra_recall} & \Use{curl_Kedavra_f1} & \Use{curl_Kedavra_build_time} & \Use{curl_treevada_false_hdd_true_LLM_true_precision} & \Use{curl_treevada_false_hdd_true_LLM_true_recall} & \textbf{\Use{curl_treevada_false_hdd_true_LLM_true_f1}} & \Use{curl_treevada_false_hdd_true_LLM_true_build_time} & \Use{curl_treevada_true_hdd_true_LLM_true_precision} & \Use{curl_treevada_true_hdd_true_LLM_true_recall} & \Use{curl_treevada_true_hdd_true_LLM_true_f1} & \Use{curl_treevada_true_hdd_true_LLM_true_build_time} \\
\midrule
\Use{Avg-Small-Program_programname} & \Use{Avg-Small-Program_barebone_precision}  & \Use{Avg-Small-Program_barebone_recall} & \Use{Avg-Small-Program_barebone_f1}  & \Use{Avg-Small-Program_treevada_original_precision}  & \Use{Avg-Small-Program_treevada_original_recall} & \textbf{\Use{Avg-Small-Program_treevada_original_f1}} & \Use{Avg-Small-Program_treevada_original_build_time} & \Use{Avg-Small-Program_Kedavra_precision}  & \Use{Avg-Small-Program_Kedavra_recall} & \Use{Avg-Small-Program_Kedavra_f1} & \Use{Avg-Small-Program_Kedavra_build_time} & \Use{Avg-Small-Program_treevada_false_hdd_true_LLM_true_precision} & \Use{Avg-Small-Program_treevada_false_hdd_true_LLM_true_recall} & \Use{Avg-Small-Program_treevada_false_hdd_true_LLM_true_f1} & \Use{Avg-Small-Program_treevada_false_hdd_true_LLM_true_build_time} & \Use{Avg-Small-Program_treevada_true_hdd_true_LLM_true_precision} & \Use{Avg-Small-Program_treevada_true_hdd_true_LLM_true_recall} & \Use{Avg-Small-Program_treevada_true_hdd_true_LLM_true_f1} & \Use{Avg-Small-Program_treevada_true_hdd_true_LLM_true_build_time} \\
\midrule
\Use{lua_programname} & \Use{lua_barebone_precision}  & \Use{lua_barebone_recall} & \Use{lua_barebone_f1}  & \Use{lua_treevada_original_precision}  & \Use{lua_treevada_original_recall} & \Use{lua_treevada_original_f1} & \Use{lua_treevada_original_build_time} & \multicolumn{3}{c}{t/o} & \Use{lua_Kedavra_build_time} & \Use{lua_treevada_false_hdd_true_LLM_true_precision} & \Use{lua_treevada_false_hdd_true_LLM_true_recall} & \textbf{\Use{lua_treevada_false_hdd_true_LLM_true_f1}} & \Use{lua_treevada_false_hdd_true_LLM_true_build_time} & \Use{lua_treevada_true_hdd_true_LLM_true_precision} & \Use{lua_treevada_true_hdd_true_LLM_true_recall} & \Use{lua_treevada_true_hdd_true_LLM_true_f1} & \Use{lua_treevada_true_hdd_true_LLM_true_build_time} \\
 \Use{c_programname} & \Use{c_barebone_precision}  & \Use{c_barebone_recall} & \Use{c_barebone_f1}  & \Use{c_treevada_original_precision}  & \Use{c_treevada_original_recall} & \Use{c_treevada_original_f1} & \Use{c_treevada_original_build_time} &  \multicolumn{3}{c}{t/o}  & \Use{c_Kedavra_build_time} & \Use{c_treevada_false_hdd_true_LLM_true_precision} & \Use{c_treevada_false_hdd_true_LLM_true_recall} & \Use{c_treevada_false_hdd_true_LLM_true_f1} & \Use{c_treevada_false_hdd_true_LLM_true_build_time} & \Use{c_treevada_true_hdd_true_LLM_true_precision} & \Use{c_treevada_true_hdd_true_LLM_true_recall} & \textbf{\Use{c_treevada_true_hdd_true_LLM_true_f1}} & \Use{c_treevada_true_hdd_true_LLM_true_build_time} \\
 \Use{mysql_programname} & \Use{mysql_barebone_precision}  & \Use{mysql_barebone_recall} & \Use{mysql_barebone_f1}  & \Use{mysql_treevada_original_precision}  & \Use{mysql_treevada_original_recall} & \Use{mysql_treevada_original_f1} & \Use{mysql_treevada_original_build_time} & \multicolumn{3}{c}{t/o} & \Use{mysql_Kedavra_build_time} & \Use{mysql_treevada_false_hdd_true_LLM_true_precision} & \Use{mysql_treevada_false_hdd_true_LLM_true_recall} & \Use{mysql_treevada_false_hdd_true_LLM_true_f1} & \Use{mysql_treevada_false_hdd_true_LLM_true_build_time} & \Use{mysql_treevada_true_hdd_true_LLM_true_precision} & \Use{mysql_treevada_true_hdd_true_LLM_true_recall} & \textbf{\Use{mysql_treevada_true_hdd_true_LLM_true_f1}} & \Use{mysql_treevada_true_hdd_true_LLM_true_build_time} \\
 \midrule
\Use{Avg-Big-Program_programname} & \Use{Avg-Big-Program_barebone_precision}  & \Use{Avg-Big-Program_barebone_recall} & \Use{Avg-Big-Program_barebone_f1}  & \Use{Avg-Big-Program_treevada_original_precision}  & \Use{Avg-Big-Program_treevada_original_recall} & \Use{Avg-Big-Program_treevada_original_f1} & \Use{Avg-Big-Program_treevada_original_build_time} & \multicolumn{3}{c}{t/o} & \Use{Avg-Big-Program_Kedavra_build_time} & \Use{Avg-Big-Program_treevada_false_hdd_true_LLM_true_precision} & \Use{Avg-Big-Program_treevada_false_hdd_true_LLM_true_recall} & \Use{Avg-Big-Program_treevada_false_hdd_true_LLM_true_f1} & \Use{Avg-Big-Program_treevada_false_hdd_true_LLM_true_build_time} & \Use{Avg-Big-Program_treevada_true_hdd_true_LLM_true_precision} & \Use{Avg-Big-Program_treevada_true_hdd_true_LLM_true_recall} & \textbf{\Use{Avg-Big-Program_treevada_true_hdd_true_LLM_true_f1}} & \Use{Avg-Big-Program_treevada_true_hdd_true_LLM_true_build_time} \\ 
\bottomrule
\end{tabular}
}
\end{table*}

%% file: tables/first_and_second_ablation.tex
\begin{table}[h!t]
\centering
\caption{%
10-run average: \toolName{} without AI label \& \toolName{} without HDD; 
p~=~\precision{}; 
r~=~\recall{}; 
f1~=~\fscore{}; 
t~=~time in seconds;
o~=~oracle membership calls (in thousands);
llm~=~LLM bubble ranking calls.
While we did not capture exact results, the ``w/o HDD'' ablation runtime is about 10s shorter than \toolName{}.
}
\label{table:first_and_nd_ablation}
\resizebox{\columnwidth}{!}{%
\begin{tabular}{l|rrrr|rrrr|rN{2}{1}|rrr|rN{2}{1}}
\toprule
\multicolumn{1}{c|}{} 
& \multicolumn{4}{c|}{\Tool{}}
& \multicolumn{6}{c|}{\Tool{} w/o AI Non-terminal Labeling} & \multicolumn{5}{c}{\Tool{} w/o HDD}\\
 & \multicolumn{1}{c}{p} & \multicolumn{1}{c}{r} & \multicolumn{1}{c}{f1} & \multicolumn{1}{c|}{t[s]}
 & \multicolumn{1}{c}{p} & \multicolumn{1}{c}{r} & \multicolumn{1}{c}{f1} & \multicolumn{1}{c|}{t[s]} 
 & \multicolumn{1}{c}{o[k]} & \multicolumn{1}{c|}{llm} & \multicolumn{1}{c}{p} & \multicolumn{1}{c}{r} & \multicolumn{1}{c|}{f1} & \multicolumn{1}{c}{o[k]} & \multicolumn{1}{c}{llm} \\
\midrule
\Use{turtle_programname} & 
\Use{turtle_treevada_true_hdd_true_LLM_true_precision}  & 
\Use{turtle_treevada_true_hdd_true_LLM_true_recall} & 
\Use{turtle_treevada_true_hdd_true_LLM_true_f1} &
\Use{turtle_treevada_true_hdd_true_LLM_true_build_time} & 
\Use{turtle_first_ablation_precision}  & \Use{turtle_first_ablation_recall} & \Use{turtle_first_ablation_f1} & \Use{turtle_first_ablation_build_time} &            \Use{turtle_first_ablation_oracle_calls} & \Use{turtle_first_ablation_llm_calls}  & \Use{turtle_treevada_true_hdd_false_LLM_true_precision}  & \Use{turtle_treevada_true_hdd_false_LLM_true_recall} & \Use{turtle_treevada_true_hdd_false_LLM_true_f1} &     \Use{turtle_treevada_true_hdd_false_LLM_true_oracle_calls} & \Use{turtle_treevada_true_hdd_false_LLM_true_llm_calls} \\
\Use{while_programname} & 
\Use{while_treevada_true_hdd_true_LLM_true_precision}  & 
\Use{while_treevada_true_hdd_true_LLM_true_recall} & 
\Use{while_treevada_true_hdd_true_LLM_true_f1} &
\Use{while_treevada_true_hdd_true_LLM_true_build_time} & 
\Use{while_first_ablation_precision}  & \Use{while_first_ablation_recall} & \Use{while_first_ablation_f1} & \Use{while_first_ablation_build_time} &            \Use{while_first_ablation_oracle_calls} & \Use{while_first_ablation_llm_calls}  & \Use{while_treevada_true_hdd_false_LLM_true_precision}  & \Use{while_treevada_true_hdd_false_LLM_true_recall} & \Use{while_treevada_true_hdd_false_LLM_true_f1} &            \Use{while_treevada_true_hdd_false_LLM_true_oracle_calls} & \Use{while_treevada_true_hdd_false_LLM_true_llm_calls} \\
\Use{lisp_programname} & 
\Use{lisp_treevada_true_hdd_true_LLM_true_precision}  & 
\Use{lisp_treevada_true_hdd_true_LLM_true_recall} & 
\Use{lisp_treevada_true_hdd_true_LLM_true_f1} &
\Use{lisp_treevada_true_hdd_true_LLM_true_build_time} & 
\Use{lisp_first_ablation_precision}  & \Use{lisp_first_ablation_recall} & \Use{lisp_first_ablation_f1} & \Use{lisp_first_ablation_build_time} &            \Use{lisp_first_ablation_oracle_calls} & \Use{lisp_first_ablation_llm_calls}  & \Use{lisp_treevada_true_hdd_false_LLM_true_precision}  & \Use{lisp_treevada_true_hdd_false_LLM_true_recall} & \Use{lisp_treevada_true_hdd_false_LLM_true_f1} &            \Use{lisp_treevada_true_hdd_false_LLM_true_oracle_calls} & \Use{lisp_treevada_true_hdd_false_LLM_true_llm_calls} \\
\Use{xml_programname} & 
\Use{xml_treevada_true_hdd_true_LLM_true_precision}  & 
\Use{xml_treevada_true_hdd_true_LLM_true_recall} & 
\Use{xml_treevada_true_hdd_true_LLM_true_f1} &
\Use{xml_treevada_true_hdd_true_LLM_true_build_time} & 
\Use{xml_first_ablation_precision}  & \Use{xml_first_ablation_recall} & \Use{xml_first_ablation_f1} & \Use{xml_first_ablation_build_time} &            \Use{xml_first_ablation_oracle_calls} & \Use{xml_first_ablation_llm_calls}  & \Use{xml_treevada_true_hdd_false_LLM_true_precision}  & \Use{xml_treevada_true_hdd_false_LLM_true_recall} & \Use{xml_treevada_true_hdd_false_LLM_true_f1} &            \Use{xml_treevada_true_hdd_false_LLM_true_oracle_calls} & \Use{xml_treevada_true_hdd_false_LLM_true_llm_calls} \\
\Use{json_programname} &
\Use{json_treevada_true_hdd_true_LLM_true_precision}  & 
\Use{json_treevada_true_hdd_true_LLM_true_recall} & 
\Use{json_treevada_true_hdd_true_LLM_true_f1} &
\Use{json_treevada_true_hdd_true_LLM_true_build_time} & 
 \Use{json_first_ablation_precision}  & \Use{json_first_ablation_recall} & \Use{json_first_ablation_f1} & \Use{json_first_ablation_build_time} &            \Use{json_first_ablation_oracle_calls} & \Use{json_first_ablation_llm_calls}  & \Use{json_treevada_true_hdd_false_LLM_true_precision}  & \Use{json_treevada_true_hdd_false_LLM_true_recall} & \Use{json_treevada_true_hdd_false_LLM_true_f1} &           \Use{json_treevada_true_hdd_false_LLM_true_oracle_calls} & \Use{json_treevada_true_hdd_false_LLM_true_llm_calls} \\
\Use{tinyc_programname} &
\Use{tinyc_treevada_true_hdd_true_LLM_true_precision}  & 
\Use{tinyc_treevada_true_hdd_true_LLM_true_recall} & 
\Use{tinyc_treevada_true_hdd_true_LLM_true_f1} &
\Use{tinyc_treevada_true_hdd_true_LLM_true_build_time} & 
\Use{tinyc_first_ablation_precision}  & \Use{tinyc_first_ablation_recall} & \Use{tinyc_first_ablation_f1} & \Use{tinyc_first_ablation_build_time} &            \Use{tinyc_first_ablation_oracle_calls} & \Use{tinyc_first_ablation_llm_calls}  & \Use{tinyc_treevada_true_hdd_false_LLM_true_precision}  & \Use{tinyc_treevada_true_hdd_false_LLM_true_recall} & \Use{tinyc_treevada_true_hdd_false_LLM_true_f1} &          \Use{tinyc_treevada_true_hdd_false_LLM_true_oracle_calls} & \Use{tinyc_treevada_true_hdd_false_LLM_true_llm_calls} \\
\Use{c-500_programname} &
\Use{c-500_treevada_true_hdd_true_LLM_true_precision}  & 
\Use{c-500_treevada_true_hdd_true_LLM_true_recall} & 
\Use{c-500_treevada_true_hdd_true_LLM_true_f1} &
\Use{c-500_treevada_true_hdd_true_LLM_true_build_time} & 
\Use{c-500_first_ablation_precision}  & \Use{c-500_first_ablation_recall} & \Use{c-500_first_ablation_f1} & \Use{c-500_first_ablation_build_time} &            \Use{c-500_first_ablation_oracle_calls} & \Use{c-500_first_ablation_llm_calls}  & \Use{c-500_treevada_true_hdd_false_LLM_true_precision}  & \Use{c-500_treevada_true_hdd_false_LLM_true_recall} & \Use{c-500_treevada_true_hdd_false_LLM_true_f1} &            \Use{c-500_treevada_true_hdd_false_LLM_true_oracle_calls} & \Use{c-500_treevada_true_hdd_false_LLM_true_llm_calls} \\
\Use{tiny_programname} &
\Use{tiny_treevada_true_hdd_true_LLM_true_precision}  & 
\Use{tiny_treevada_true_hdd_true_LLM_true_recall} & 
\Use{tiny_treevada_true_hdd_true_LLM_true_f1} &
\Use{tiny_treevada_true_hdd_true_LLM_true_build_time} & 
\Use{tiny_first_ablation_precision}  & \Use{tiny_first_ablation_recall} & \Use{tiny_first_ablation_f1} & \Use{tiny_first_ablation_build_time} &            \Use{tiny_first_ablation_oracle_calls} & \Use{tiny_first_ablation_llm_calls}  & \Use{tiny_treevada_true_hdd_false_LLM_true_precision}  & \Use{tiny_treevada_true_hdd_false_LLM_true_recall} & \Use{tiny_treevada_true_hdd_false_LLM_true_f1} &            \Use{tiny_treevada_true_hdd_false_LLM_true_oracle_calls} & \Use{tiny_treevada_true_hdd_false_LLM_true_llm_calls} \\
\Use{minic_programname} &
\Use{minic_treevada_true_hdd_true_LLM_true_precision}  & 
\Use{minic_treevada_true_hdd_true_LLM_true_recall} & 
\Use{minic_treevada_true_hdd_true_LLM_true_f1} &
\Use{minic_treevada_true_hdd_true_LLM_true_build_time} & 
\Use{minic_first_ablation_precision}  & \Use{minic_first_ablation_recall} & \Use{minic_first_ablation_f1} & \Use{minic_first_ablation_build_time} &            \Use{minic_first_ablation_oracle_calls} & \Use{minic_first_ablation_llm_calls}  & \Use{minic_treevada_true_hdd_false_LLM_true_precision}  & \Use{minic_treevada_true_hdd_false_LLM_true_recall} & \Use{minic_treevada_true_hdd_false_LLM_true_f1} &          \Use{minic_treevada_true_hdd_false_LLM_true_oracle_calls} & \Use{minic_treevada_true_hdd_false_LLM_true_llm_calls} \\
\Use{curl_programname} &
\Use{curl_treevada_true_hdd_true_LLM_true_precision}  & 
\Use{curl_treevada_true_hdd_true_LLM_true_recall} & 
\Use{curl_treevada_true_hdd_true_LLM_true_f1} &
\Use{curl_treevada_true_hdd_true_LLM_true_build_time} & 
\Use{curl_first_ablation_precision}  & \Use{curl_first_ablation_recall} & \Use{curl_first_ablation_f1} & \Use{curl_first_ablation_build_time} &            \Use{curl_first_ablation_oracle_calls} & \Use{curl_first_ablation_llm_calls}  & \Use{curl_treevada_true_hdd_false_LLM_true_precision}  & \Use{curl_treevada_true_hdd_false_LLM_true_recall} & \Use{curl_treevada_true_hdd_false_LLM_true_f1} &           \Use{curl_treevada_true_hdd_false_LLM_true_oracle_calls} & \Use{curl_treevada_true_hdd_false_LLM_true_llm_calls} \\
\midrule
\Use{Avg-Small-Program_programname} &
\Use{Avg-Small-Program_treevada_true_hdd_true_LLM_true_precision}  & 
\Use{Avg-Small-Program_treevada_true_hdd_true_LLM_true_recall} & 
\Use{Avg-Small-Program_treevada_true_hdd_true_LLM_true_f1} &
\Use{Avg-Small-Program_treevada_true_hdd_true_LLM_true_build_time} & 
\Use{Avg-Small-Program_first_ablation_precision}  & \Use{Avg-Small-Program_first_ablation_recall} & \Use{Avg-Small-Program_first_ablation_f1} & \Use{Avg-Small-Program_first_ablation_build_time} &            \Use{Avg-Small-Program_first_ablation_oracle_calls} & \Use{Avg-Small-Program_first_ablation_llm_calls}  & \Use{Avg-Small-Program_treevada_true_hdd_false_LLM_true_precision}  & \Use{Avg-Small-Program_treevada_true_hdd_false_LLM_true_recall} & \Use{Avg-Small-Program_treevada_true_hdd_false_LLM_true_f1} &            \Use{Avg-Small-Program_treevada_true_hdd_false_LLM_true_oracle_calls} & \Use{Avg-Small-Program_treevada_true_hdd_false_LLM_true_llm_calls} \\
\midrule
\Use{lua_programname} &
\Use{lua_treevada_true_hdd_true_LLM_true_precision}  & 
\Use{lua_treevada_true_hdd_true_LLM_true_recall} & 
\Use{lua_treevada_true_hdd_true_LLM_true_f1} &
\Use{lua_treevada_true_hdd_true_LLM_true_build_time} &  \Use{lua_first_ablation_precision}  & \Use{lua_first_ablation_recall} & \Use{lua_first_ablation_f1} & \Use{lua_first_ablation_build_time} &            \Use{lua_first_ablation_oracle_calls} & \Use{lua_first_ablation_llm_calls}  & \Use{lua_treevada_true_hdd_false_LLM_true_precision}  & \Use{lua_treevada_true_hdd_false_LLM_true_recall} & \Use{lua_treevada_true_hdd_false_LLM_true_f1} &            \Use{lua_treevada_true_hdd_false_LLM_true_oracle_calls} & \Use{lua_treevada_true_hdd_false_LLM_true_llm_calls} \\
\Use{c_programname} &
\Use{c_treevada_true_hdd_true_LLM_true_precision}  & 
\Use{c_treevada_true_hdd_true_LLM_true_recall} & 
\Use{c_treevada_true_hdd_true_LLM_true_f1} &
\Use{c_treevada_true_hdd_true_LLM_true_build_time} & \Use{c_first_ablation_precision}  & \Use{c_first_ablation_recall} & \Use{c_first_ablation_f1} & \Use{c_first_ablation_build_time} &            \Use{c_first_ablation_oracle_calls} & \Use{c_first_ablation_llm_calls}  & \Use{c_treevada_true_hdd_false_LLM_true_precision}  & \Use{c_treevada_true_hdd_false_LLM_true_recall} & \Use{c_treevada_true_hdd_false_LLM_true_f1} &          \Use{c_treevada_true_hdd_false_LLM_true_oracle_calls} & \Use{c_treevada_true_hdd_false_LLM_true_llm_calls} \\
\Use{mysql_programname} &
\Use{mysql_treevada_true_hdd_true_LLM_true_precision}  & 
\Use{mysql_treevada_true_hdd_true_LLM_true_recall} & 
\Use{mysql_treevada_true_hdd_true_LLM_true_f1} &
\Use{mysql_treevada_true_hdd_true_LLM_true_build_time} &  \Use{mysql_first_ablation_precision}  & \Use{mysql_first_ablation_recall} & \Use{mysql_first_ablation_f1} & \Use{mysql_first_ablation_build_time} &            \Use{mysql_first_ablation_oracle_calls} & \Use{mysql_first_ablation_llm_calls}  & \Use{mysql_treevada_true_hdd_false_LLM_true_precision}  & \Use{mysql_treevada_true_hdd_false_LLM_true_recall} & \Use{mysql_treevada_true_hdd_false_LLM_true_f1} &     \Use{mysql_treevada_true_hdd_false_LLM_true_oracle_calls} & \Use{mysql_treevada_true_hdd_false_LLM_true_llm_calls} \\
\midrule
\Use{Avg-Big-Program_programname} &
\Use{Avg-Big-Program_treevada_true_hdd_true_LLM_true_precision}  & 
\Use{Avg-Big-Program_treevada_true_hdd_true_LLM_true_recall} & 
\Use{Avg-Big-Program_treevada_true_hdd_true_LLM_true_f1} &
\Use{Avg-Big-Program_treevada_true_hdd_true_LLM_true_build_time} & 
\Use{Avg-Big-Program_first_ablation_precision}  & \Use{Avg-Big-Program_first_ablation_recall} & \Use{Avg-Big-Program_first_ablation_f1} & \Use{Avg-Big-Program_first_ablation_build_time} &            \Use{Avg-Big-Program_first_ablation_oracle_calls} & \Use{Avg-Big-Program_first_ablation_llm_calls}  & \Use{Avg-Big-Program_treevada_true_hdd_false_LLM_true_precision}  & \Use{Avg-Big-Program_treevada_true_hdd_false_LLM_true_recall} & \Use{Avg-Big-Program_treevada_true_hdd_false_LLM_true_f1} &           \Use{Avg-Big-Program_treevada_true_hdd_false_LLM_true_oracle_calls} & \Use{Avg-Big-Program_treevada_true_hdd_false_LLM_true_llm_calls} \\
\bottomrule
\end{tabular}
}
\end{table}

%% file: tables/third_and_fourth_ablation.tex
\begin{table}[h!t]
\centering
\caption{%
10-run average: \toolName{} without bracket-guided bubbles \& \toolName{} with \treevada{} bubble ranking heuristic replacing LLM-guided bubble exploration; p~=~precision; 
r~=~recall; 
f1~=~\fscore{}; 
t~=~time in seconds;
llm~=~LLM bubble calls;
o~=~oracle membership calls (in thousands).
}
\label{table:third_and_fourth_ablation}
\resizebox{\columnwidth}{!}{%
\begin{tabular}{l|rrrr|rrrr|rN{2}{1}|rrrr|rN{2}{1}}
\toprule
\multicolumn{1}{c|}{} 
& \multicolumn{4}{c|}{\Tool{}}
 & \multicolumn{6}{c|}{\toolName{} w/o pre-existing} & \multicolumn{6}{c}{\toolName{} w/o LLM guided} \\
 \multicolumn{1}{c|}{} 
& \multicolumn{4}{c|}{} & \multicolumn{6}{c|}{bracket-guided bubbles} & \multicolumn{6}{c}{bubble exploration} \\
 & \multicolumn{1}{c}{p} & \multicolumn{1}{c}{r} & \multicolumn{1}{c}{f1} & \multicolumn{1}{c|}{t[s]}
 & \multicolumn{1}{c}{p} & \multicolumn{1}{c}{r} & \multicolumn{1}{c}{f1} & \multicolumn{1}{c|}{t[s]} 
 & \multicolumn{1}{c}{o[k]} & \multicolumn{1}{c|}{llm} & \multicolumn{1}{c}{p} & \multicolumn{1}{c}{r} & \multicolumn{1}{c}{f1} & \multicolumn{1}{c|}{t[s]} & \multicolumn{1}{c}{o[k]} & \multicolumn{1}{c}{llm} \\
\midrule
\Use{turtle_programname} & 
\Use{turtle_treevada_true_hdd_true_LLM_true_precision}  & 
\Use{turtle_treevada_true_hdd_true_LLM_true_recall} & 
\Use{turtle_treevada_true_hdd_true_LLM_true_f1} &
\Use{turtle_treevada_true_hdd_true_LLM_true_build_time} & 
\Use{turtle_third_ablation_precision}  & \Use{turtle_third_ablation_recall} & \Use{turtle_third_ablation_f1} & \Use{turtle_third_ablation_build_time} &            \Use{turtle_third_ablation_oracle_calls} & \Use{turtle_third_ablation_llm_calls} & \Use{turtle_fourth_ablation_precision}  & \Use{turtle_fourth_ablation_recall} & \Use{turtle_fourth_ablation_f1} & \Use{turtle_fourth_ablation_build_time} &            \Use{turtle_fourth_ablation_oracle_calls} & \Use{turtle_fourth_ablation_llm_calls}\\
\Use{while_programname} &
\Use{while_treevada_true_hdd_true_LLM_true_precision}  & 
\Use{while_treevada_true_hdd_true_LLM_true_recall} & 
\Use{while_treevada_true_hdd_true_LLM_true_f1} &
\Use{while_treevada_true_hdd_true_LLM_true_build_time} & 
\Use{while_third_ablation_precision}  & \Use{while_third_ablation_recall} & \Use{while_third_ablation_f1} & \Use{while_third_ablation_build_time} &            \Use{while_third_ablation_oracle_calls} & \Use{while_third_ablation_llm_calls} & \Use{while_fourth_ablation_precision}  & \Use{while_fourth_ablation_recall} & \Use{while_fourth_ablation_f1} & \Use{while_fourth_ablation_build_time} &            \Use{while_fourth_ablation_oracle_calls} & \Use{while_fourth_ablation_llm_calls}\\
\Use{lisp_programname} &
\Use{lisp_treevada_true_hdd_true_LLM_true_precision}  & 
\Use{lisp_treevada_true_hdd_true_LLM_true_recall} & 
\Use{lisp_treevada_true_hdd_true_LLM_true_f1} &
\Use{lisp_treevada_true_hdd_true_LLM_true_build_time} & 
\Use{lisp_third_ablation_precision}  & \Use{lisp_third_ablation_recall} & \Use{lisp_third_ablation_f1} & \Use{lisp_third_ablation_build_time} &            \Use{lisp_third_ablation_oracle_calls} & \Use{lisp_third_ablation_llm_calls} & \Use{lisp_fourth_ablation_precision}  & \Use{lisp_fourth_ablation_recall} & \Use{lisp_fourth_ablation_f1} & \Use{lisp_fourth_ablation_build_time} &            \Use{lisp_fourth_ablation_oracle_calls} & \Use{lisp_fourth_ablation_llm_calls}\\
\Use{xml_programname} &
\Use{xml_treevada_true_hdd_true_LLM_true_precision}  & 
\Use{xml_treevada_true_hdd_true_LLM_true_recall} & 
\Use{xml_treevada_true_hdd_true_LLM_true_f1} &
\Use{xml_treevada_true_hdd_true_LLM_true_build_time} & 
\Use{xml_third_ablation_precision}  & \Use{xml_third_ablation_recall} & \Use{xml_third_ablation_f1} & \Use{xml_third_ablation_build_time} &            \Use{xml_third_ablation_oracle_calls} & \Use{xml_third_ablation_llm_calls} & \Use{xml_fourth_ablation_precision}  & \Use{xml_fourth_ablation_recall} & \Use{xml_fourth_ablation_f1} & \Use{xml_fourth_ablation_build_time} &            \Use{xml_fourth_ablation_oracle_calls} & \Use{xml_fourth_ablation_llm_calls}\\
\Use{json_programname} &
\Use{json_treevada_true_hdd_true_LLM_true_precision}  & 
\Use{json_treevada_true_hdd_true_LLM_true_recall} & 
\Use{json_treevada_true_hdd_true_LLM_true_f1} &
\Use{json_treevada_true_hdd_true_LLM_true_build_time} & 
\Use{json_third_ablation_precision}  & \Use{json_third_ablation_recall} & \Use{json_third_ablation_f1} & \Use{json_third_ablation_build_time} &            \Use{json_third_ablation_oracle_calls} & \Use{json_third_ablation_llm_calls} & \Use{json_fourth_ablation_precision}  & \Use{json_fourth_ablation_recall} & \Use{json_fourth_ablation_f1} & \Use{json_fourth_ablation_build_time} &            \Use{json_fourth_ablation_oracle_calls} & \Use{json_fourth_ablation_llm_calls}\\
\Use{tinyc_programname} &
\Use{tinyc_treevada_true_hdd_true_LLM_true_precision}  & 
\Use{tinyc_treevada_true_hdd_true_LLM_true_recall} & 
\Use{tinyc_treevada_true_hdd_true_LLM_true_f1} &
\Use{tinyc_treevada_true_hdd_true_LLM_true_build_time} & 
\Use{tinyc_third_ablation_precision}  & \Use{tinyc_third_ablation_recall} & \Use{tinyc_third_ablation_f1} & \Use{tinyc_third_ablation_build_time} &            \Use{tinyc_third_ablation_oracle_calls} & \Use{tinyc_third_ablation_llm_calls} & \Use{tinyc_fourth_ablation_precision}  & \Use{tinyc_fourth_ablation_recall} & \Use{tinyc_fourth_ablation_f1} & \Use{tinyc_fourth_ablation_build_time} &            \Use{tinyc_fourth_ablation_oracle_calls} & \Use{tinyc_fourth_ablation_llm_calls}\\
\Use{c-500_programname} &
\Use{c-500_treevada_true_hdd_true_LLM_true_precision}  & 
\Use{c-500_treevada_true_hdd_true_LLM_true_recall} & 
\Use{c-500_treevada_true_hdd_true_LLM_true_f1} &
\Use{c-500_treevada_true_hdd_true_LLM_true_build_time} & 
\Use{c-500_third_ablation_precision}  & \Use{c-500_third_ablation_recall} & \Use{c-500_third_ablation_f1} & \Use{c-500_third_ablation_build_time} &            \Use{c-500_third_ablation_oracle_calls} & \Use{c-500_third_ablation_llm_calls} & \Use{c-500_fourth_ablation_precision}  & \Use{c-500_fourth_ablation_recall} & \Use{c-500_fourth_ablation_f1} & \Use{c-500_fourth_ablation_build_time} &            \Use{c-500_fourth_ablation_oracle_calls} & \Use{c-500_fourth_ablation_llm_calls}\\
\Use{tiny_programname} &
\Use{tiny_treevada_true_hdd_true_LLM_true_precision}  & 
\Use{tiny_treevada_true_hdd_true_LLM_true_recall} & 
\Use{tiny_treevada_true_hdd_true_LLM_true_f1} &
\Use{tiny_treevada_true_hdd_true_LLM_true_build_time} & 
\Use{tiny_third_ablation_precision}  & \Use{tiny_third_ablation_recall} & \Use{tiny_third_ablation_f1} & \Use{tiny_third_ablation_build_time} &            \Use{tiny_third_ablation_oracle_calls} & \Use{tiny_third_ablation_llm_calls} & \Use{tiny_fourth_ablation_precision}  & \Use{tiny_fourth_ablation_recall} & \Use{tiny_fourth_ablation_f1} & \Use{tiny_fourth_ablation_build_time} &            \Use{tiny_fourth_ablation_oracle_calls} & \Use{tiny_fourth_ablation_llm_calls}\\
\Use{minic_programname} &
\Use{minic_treevada_true_hdd_true_LLM_true_precision}  & 
\Use{minic_treevada_true_hdd_true_LLM_true_recall} & 
\Use{minic_treevada_true_hdd_true_LLM_true_f1} &
\Use{minic_treevada_true_hdd_true_LLM_true_build_time} & 
\Use{minic_third_ablation_precision}  & \Use{minic_third_ablation_recall} & \Use{minic_third_ablation_f1} & \Use{minic_third_ablation_build_time} &            \Use{minic_third_ablation_oracle_calls} & \Use{minic_third_ablation_llm_calls} & \Use{minic_fourth_ablation_precision}  & \Use{minic_fourth_ablation_recall} & \Use{minic_fourth_ablation_f1} & \Use{minic_fourth_ablation_build_time} &            \Use{minic_fourth_ablation_oracle_calls} & \Use{minic_fourth_ablation_llm_calls}\\
\Use{curl_programname} &
\Use{curl_treevada_true_hdd_true_LLM_true_precision}  & 
\Use{curl_treevada_true_hdd_true_LLM_true_recall} & 
\Use{curl_treevada_true_hdd_true_LLM_true_f1} &
\Use{curl_treevada_true_hdd_true_LLM_true_build_time} & 
\Use{curl_third_ablation_precision}  & \Use{curl_third_ablation_recall} & \Use{curl_third_ablation_f1} & \Use{curl_third_ablation_build_time} &            \Use{curl_third_ablation_oracle_calls} & \Use{curl_third_ablation_llm_calls} & \Use{curl_fourth_ablation_precision}  & \Use{curl_fourth_ablation_recall} & \Use{curl_fourth_ablation_f1} & \Use{curl_fourth_ablation_build_time} &            \Use{curl_fourth_ablation_oracle_calls} & \Use{curl_fourth_ablation_llm_calls}\\
\midrule
\Use{Avg-Small-Program_programname} &
\Use{Avg-Small-Program_treevada_true_hdd_true_LLM_true_precision}  & 
\Use{Avg-Small-Program_treevada_true_hdd_true_LLM_true_recall} & 
\Use{Avg-Small-Program_treevada_true_hdd_true_LLM_true_f1} &
\Use{Avg-Small-Program_treevada_true_hdd_true_LLM_true_build_time} & 
\Use{Avg-Small-Program_third_ablation_precision}  & \Use{Avg-Small-Program_third_ablation_recall} & \Use{Avg-Small-Program_third_ablation_f1} & \Use{Avg-Small-Program_third_ablation_build_time} &            \Use{Avg-Small-Program_third_ablation_oracle_calls} & \Use{Avg-Small-Program_third_ablation_llm_calls} & \Use{Avg-Small-Program_fourth_ablation_precision}  & \Use{Avg-Small-Program_fourth_ablation_recall} & \Use{Avg-Small-Program_fourth_ablation_f1} & \Use{Avg-Small-Program_fourth_ablation_build_time} &            \Use{Avg-Small-Program_fourth_ablation_oracle_calls} & \Use{Avg-Small-Program_fourth_ablation_llm_calls}\\
\midrule
\Use{lua_programname} &
\Use{lua_treevada_true_hdd_true_LLM_true_precision}  & 
\Use{lua_treevada_true_hdd_true_LLM_true_recall} & 
\Use{lua_treevada_true_hdd_true_LLM_true_f1} &
\Use{lua_treevada_true_hdd_true_LLM_true_build_time} & 
\Use{lua_third_ablation_precision}  & \Use{lua_third_ablation_recall} & \Use{lua_third_ablation_f1} & \Use{lua_third_ablation_build_time} &            \Use{lua_third_ablation_oracle_calls} & \Use{lua_third_ablation_llm_calls} & \Use{lua_fourth_ablation_precision}  & \Use{lua_fourth_ablation_recall} & \Use{lua_fourth_ablation_f1} & \Use{lua_fourth_ablation_build_time} &            \Use{lua_fourth_ablation_oracle_calls} & \Use{lua_fourth_ablation_llm_calls}\\
\Use{c_programname} &
\Use{c_treevada_true_hdd_true_LLM_true_precision}  & 
\Use{c_treevada_true_hdd_true_LLM_true_recall} & 
\Use{c_treevada_true_hdd_true_LLM_true_f1} &
\Use{c_treevada_true_hdd_true_LLM_true_build_time} & 
\Use{c_third_ablation_precision}  & \Use{c_third_ablation_recall} & \Use{c_third_ablation_f1} & \Use{c_third_ablation_build_time} &            \Use{c_third_ablation_oracle_calls} & \Use{c_third_ablation_llm_calls} & \Use{c_fourth_ablation_precision}  & \Use{c_fourth_ablation_recall} & \Use{c_fourth_ablation_f1} & \Use{c_fourth_ablation_build_time} &            \Use{c_fourth_ablation_oracle_calls} & \Use{c_fourth_ablation_llm_calls}\\
\Use{mysql_programname} & 
\Use{mysql_treevada_true_hdd_true_LLM_true_precision}  & 
\Use{mysql_treevada_true_hdd_true_LLM_true_recall} & 
\Use{mysql_treevada_true_hdd_true_LLM_true_f1} &
\Use{mysql_treevada_true_hdd_true_LLM_true_build_time} & 
\Use{mysql_third_ablation_precision}  & \Use{mysql_third_ablation_recall} & \Use{mysql_third_ablation_f1} & \Use{mysql_third_ablation_build_time} &            \Use{mysql_third_ablation_oracle_calls} & \Use{mysql_third_ablation_llm_calls} & \Use{mysql_fourth_ablation_precision}  & \Use{mysql_fourth_ablation_recall} & \Use{mysql_fourth_ablation_f1} & \Use{mysql_fourth_ablation_build_time} &            \Use{mysql_fourth_ablation_oracle_calls} & \Use{mysql_fourth_ablation_llm_calls}\\
\midrule
\Use{Avg-Big-Program_programname} &
\Use{Avg-Big-Program_treevada_true_hdd_true_LLM_true_precision}  & 
\Use{Avg-Big-Program_treevada_true_hdd_true_LLM_true_recall} & 
\Use{Avg-Big-Program_treevada_true_hdd_true_LLM_true_f1} &
\Use{Avg-Big-Program_treevada_true_hdd_true_LLM_true_build_time} & 
\Use{Avg-Big-Program_third_ablation_precision}  & \Use{Avg-Big-Program_third_ablation_recall} & \Use{Avg-Big-Program_third_ablation_f1} & \Use{Avg-Big-Program_third_ablation_build_time} &            \Use{Avg-Big-Program_third_ablation_oracle_calls} & \Use{Avg-Big-Program_third_ablation_llm_calls} & \Use{Avg-Big-Program_fourth_ablation_precision}  & \Use{Avg-Big-Program_fourth_ablation_recall} & \Use{Avg-Big-Program_fourth_ablation_f1} & \Use{Avg-Big-Program_fourth_ablation_build_time} &            \Use{Avg-Big-Program_fourth_ablation_oracle_calls} & \Use{Avg-Big-Program_fourth_ablation_llm_calls}\\

\bottomrule
\end{tabular}
}
\end{table}